\documentclass[aps,pre,reprint,superscriptaddress,longbibliography]{revtex4-1}

\usepackage{graphicx}
\usepackage{amsmath}
\begin{document}


\title{Spontaneous and stimulus-induced coherent states of critically balanced neuronal networks}


\author{Takashi Hayakawa}
\email[]{hayakawa.takashi@nihon-u.ac.jp}
\affiliation{RIKEN Center for Brain Science, Wako, Saitama, 3510198, Japan}
\affiliation{School of Medicine, Nihon University, Itabashi, Tokyo, 1738610, Japan}
\author{Tomoki Fukai}
\affiliation{RIKEN Center for Brain Science, Wako, Saitama, 3510198, Japan}


\date{\today}

\begin{abstract}
How the information microscopically processed by individual neurons is integrated and used in organizing the behavior of an animal is a central question in neuroscience. The coherence of neuronal dynamics over different scales has been suggested as a clue to the mechanisms underlying this integration. Balanced strong excitation and inhibition may amplify microscopic fluctuations to a macroscopic level, thus providing a mechanism for generating coherent multiscale neuronal dynamics. Previous theories of brain dynamics, however, were restricted to cases in which inhibition dominated excitation and suppressed fluctuations in the macroscopic population activity. In the present study, we investigate the dynamics of neuronal networks at a critical point between excitation-dominant and inhibition-dominant states. In these networks, the microscopic fluctuations in neuronal activities are amplified by the strong excitation and inhibition to drive the macroscopic dynamics, while the macroscopic dynamics determine the statistics of the microscopic fluctuations. Developing a novel type of mean-field theory applicable to this class of interscale interactions, for which an analytical approach has previously been unknown, we show that the amplification mechanism generates spontaneous, irregular macroscopic rhythms similar to those observed in the brain. Through the same mechanism, microscopic inputs to a small number of neurons effectively entrain the dynamics of the whole network. These network dynamics undergo a probabilistic transition to a coherent state, as the magnitude of either the balanced excitation and inhibition or the external inputs is increased. Our mean-field theory successfully predicts the behavior of this model. Furthermore, we numerically demonstrate that the coherent dynamics can be used for state-dependent read-out of information from the network. These results show a novel form of neuronal information processing that connects neuronal dynamics on different scales, advancing our understanding of the circuit mechanisms of brain computing.
\end{abstract}


\maketitle

\section{Introduction}
The cerebral cortex and hippocampus, the areas believed to be the origin of the versatile intelligent functionality of the mammalian brain, exhibit characteristic activities on two different scales. On the microscopic scale, neurons in these areas display various temporal patterns of firing activities in response to external stimuli or to being driven internally. These activities are correlated with fine features of the information the animal is processing \cite{fuster1971neuron, Miles:1993wy, romo1999neuronal, hahn2012spontaneous, Guo:2017df}. On the macroscopic scale, electroencephalograms (EEGs) and measurements of local-field potentials (LFPs) have revealed a diverse range of rhythmic activities. These vary in both frequency and amplitude, but they are clearly correlated with the behavioral states of the animal, such as its attention and arousal levels \cite{hobson1986neuronal, klimesch1999eeg, buzsaki2002theta, jensen2007human}. Furthermore, in recent years, increasing numbers of experimental results have suggested that {\it coherence} of activities on these two scales---namely, the degree of temporal cross-correlation among the activities---is finely controlled, reflecting the mechanisms underlying the binding of sensory stimuli, sensori-motor coordination, and learning in behavioral tasks \cite{gray1989stimulus, o1993phase, engel2001dynamic, Harris:bo, fries2007gamma, poulet2008internal, lisman2010working, Struber2014, gulati2014reactivation}.

How patterns emerge in multiscale dynamics in highly non-linear and non-equilibrium regimes has been a subject of active research in statistical physics. From this perspective, understanding the multiscale dynamics in the brain and their coherence can be considered as a challenge in statistical physics. Physicists have thus far constructed various models of the activities in the brain and have investigated those models both numerically and theoretically. In particular, a mean-field theory (MFT) of randomly connected neuronal networks (RNNs) has provided a solid theoretical foundation that allows us to investigate neuronal dynamics using analytical methods similar to those employed for spin-glass systems \cite{PhysRevLett.61.259}. To enhance its applicability, different versions of the theory have been developed for different models, ranging from simple networks of neurons described by one-dimensional firing-rate variables to structured networks of neurons described by binary spike variables or more realistic kinetic variables of biological membranes \cite{amari1972learning, Hopfield2554, amit_1989, PhysRevE.50.3171, doi:10.1162/089976698300017214, Brunel2000, faugeras2009constructive, renart2010asynchronous, 10.3389/fncom.2011.00028, PhysRevE.82.011903, PhysRevX.4.021039, PhysRevE.90.062710, PhysRevX.5.041030, PhysRevX.6.031024, PhysRevE.97.062314}. These studies have theoretically shown that their RNNs have dynamical phases with different characteristics, such as chaotic fluctuations in firing-rates, asynchronous irregular firing, and regular and irregular synchronized firing \cite{PhysRevLett.61.259, van1996chaos, Brunel2000, renart2010asynchronous}. Efficient computation that takes advantage of the dynamical properties of RNNs has also been investigated recently, in both biological and engineering contexts \cite{maass2002real, jaeger2004harnessing, verstraeten2005isolated, karmarkar2007timing, sussillo2009generating, Appeltant:2011jy, lukovsevivcius2012reservoir, Dominey:2013bg, PierreEnel:2016cv}.

A primary constraint in modeling neuronal dynamics of cortical areas is the fact that neurons in a local circuit densely form synapses on one another, and that these synapses obey {\it ``Dale's law''}, a principle that prohibits neurons from forming both excitatory and inhibitory synapses. Despite the knowledge about RNNs and MFTs accumulated over decades, only recently have researchers successfully begun to develop MFTs for networks under these two constraints. In RNNs comprising $N(\gg 1)$ neurons with these two constraints, $O(N)$ excitatory and inhibitory neurons are connected with a fixed $O(1)$ probability. As is usually the case for physical systems with random couplings, non-trivial dynamics are observed for synaptic strengths with $O(1/\sqrt{N})$ standard deviation. Then, Dale's law requires us to determine the means of the excitatory and inhibitory synaptic strengths to be $\pm O(1/\sqrt{N})$, respectively. As a result, neurons receive very strong excitatory and inhibitory recurrent inputs from other neurons. By extending a previous theory \cite{doi:10.1162/089976698300017214}, recent work showed that as observed experimentally\cite{Anderson2000, Shu_nature_2003, Haider4535, Okun_Lampl, Barral:2016er}, feedback inhibition balance strong excitatory recurrent inputs to neurons and stably produces an asynchronous and irregular firing state in such a model \cite{renart2010asynchronous}. Although this model did not have non-trivial population dynamics, two very recent studies \cite{Darshan:iy, Rosenbaum:tn} investigated spatially extended versions of such balanced RNNs in the presence of external inputs, reporting multiscale dynamics in which macroscopic spatiotemporal patterns and microscopic irregular firing of individual neurons coexisted.

Although these studies have successfully demonstrated the relationship between spatial structures and multiscale dynamics of balanced neuronal networks, there remains a fundamental issue that stems from a limitation inherent in these models. In the previous models, a change in the population activity caused by a small number of constituent neurons is quickly counter-balanced by the feedback-inhibition mechanism, resulting in only vanishingly small responses in the population dynamics [see further discussion in Sec.\ref{Discussion:C}]. Presumably, this is the reason why the previous studies required a network-wide application of external inputs---that is, an extrinsic origin---to induce multiscale dynamics. The vanishingly small responses of their models, however, contrast with recent experimental results that a weak stimulation of a small number of excitatory neurons effectively evokes a population response within the local circuit, suggesting an intrinsic origin of multiscale dynamics \cite{london2010sensitivity, Chettih:2019wm}. This discrepancy may imply a fundamental difference between the nature of the dynamics of the previous models and those of the neuronal circuits investigated experimentally. Theoretically, the weak effects of the stimulation of neurons in the previous models originate from the fact that a set of population statistics of neuronal activities follows equations that are closed among themselves \cite{Rosenbaum:tn}. This implies that the time evolution of those population statistics are independent of the microscopic fluctuations in the activities of individual neurons; namely, that there exists a separation of scales. From a general point of view in statistical mechanics, finding such a separation of scales is a common step in constructing an MFT. The description of intrinsically generated, multiscale neuronal dynamics, however, requires a theory without such a separation of scales. Although microscopic fluctuations are known to evoke very large responses in systems in critical states, the previous theories of critical phenomena are not of immediate use for this purpose, because the average of critical fluctuations over time and population are still vanishingly small in those theories \cite{altland2010condensed}. Thus, regardless of the observation of critical responses in the brain both on the microscopic and macroscopic scales \cite{COCCHI2017132}, it remains theoretically unclear how the critical responses of neurons are reflected in the population dynamics.

In this study, we present a solution to this fundamental issue by constructing a novel type of MFT for densely connected RNNs with Dale's law, for which mean synaptic weights are set to critical values between those for excitation-dominant and inhibition-dominant states. In this theory, unlike the previous theories of critical dynamics, we show that fluctuations in individual neuronal activities are amplified by the strong excitation and inhibition to provide stochastic driving forces for the population dynamics. We also show that external inputs to a $O(\sqrt{N})$ number of neurons effectively entrain the whole network, comprising $N$ excitatory and $N$ inhibitory neurons, through the same amplification mechanism. Then, we observe that the network dynamics undergo a transition from irregular dynamics to coherent dynamics as the magnitude of either the excitation and inhibition or that of the external inputs is increased. The transition to a coherent state is found to be strongly dependent on the configuration of the random connectivity. These phenomena are predicted by our MFT, which yields good quantitative agreement with direct numerical results. Numerical results further suggest that such coherent dynamics can be used for reading out information from the network in a state-dependent manner.  Although, for the sake of mathematical clarity, our theory is derived for a network of simplified neurons described by firing-rate variables, we confirm numerically that similar multiscale dynamics arise in a network of leaky integrate-and-fire (LIF) neurons.
\section{Model} \label{sec:section2}
\begin{figure}[h!]
\includegraphics[width=85mm]{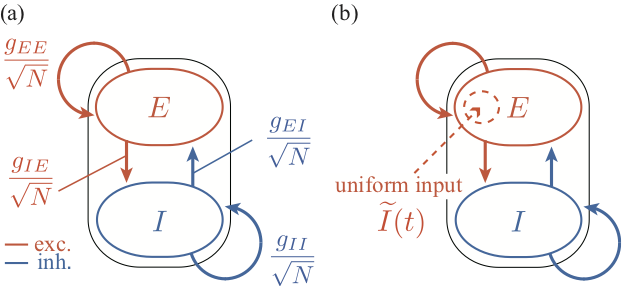}
\caption{A schematic illustration of the model. (a) The model network consists of a pair of excitatory and inhibitory populations (denoted by $E$ and $I$, respectively), each of which contains $N$ neurons. The neurons interact with one another through synaptic connections with (almost) {\it i.i.d.} quenched random weights. The mean strengths of the synaptic weights depend on the populations to which the presynaptic and postsynaptic neurons belong. They are parameterized as $g_{k\ell } /\sqrt{N}$ ($k,\ell =E,I$) as indicated in the panel. The variance of the synaptic weights is $\sigma _0^2/N$, with parameter $\sigma _0$. (b) External inputs are applied uniformly to only $\sqrt{N}$ neurons in the $E$ population [Eq.(\ref{modelling_sparse_input})]. } \label{model_schema}
\end{figure}
Our theory is formulated for a single pair of excitatory and inhibitory neuronal populations (denoted by index $k=E, I$, respectively), each of which consists of $N$ neurons [Fig.\ref{model_schema}(a)]. The $i$-th neuron of population $k$ is described by a single, real-valued dynamical variable $h_i^{(k)}$ that obeys the following dynamical equation:  
\begin{eqnarray}
\frac{\mathrm{d}}{\mathrm{d} t} h_i^{(k)}(t)&=&-h_i^{(k)}(t)+\hspace{-0.5cm}\sum _{\ell =E,I, 1\leq j\leq N}\hspace{-0.4cm}J _{k\ell }^{ij}\phi (h_j^{(\ell )}(t))+I_{i}^{(k)}(t), \label{dynamical_equation} \\
J_{k\ell }^{ij}&=&\frac{\sigma _{0}\mathcal{J} _{k\ell }^{ij}}{\sqrt{N}}+\frac{g_{k\ell }}{\sqrt{N}}. 
\label{basic_equation_nonzerosum}
\end{eqnarray}
In these equations, the function $\phi $ is the hyperbolic tangent function, which describes the sigmoidal response of the neurons. The quantities $h_i^{(k)}$ and $\phi (h_i^{(k)})$ represent the internal state and firing rate of the neuron. The variables $J_{k\ell }^{ij}$ and $I_{i}^{(k)}(t)$ denote the strengths of the recurrent synapses on, and the external input to, the neuron. The synaptic strengths are independently and identically distributed ({\it i.i.d.}) quenched random variables, and their means and standard deviations are parameterized as $g_{k\ell }/\sqrt{N}$ and $\sigma _0/\sqrt{N}$, respectively. Note that the means, but not the standard deviations, depend on the populations to which the presynaptic and postsynaptic neurons belong [Fig.\ref{model_schema}(a)]. To describe this randomness, we have used the ${\it i.i.d.}$ quenched random variables $\mathcal{J} _{k\ell }^{ij}$ with zero mean and unit variance in Eq.(\ref{basic_equation_nonzerosum}). Unless otherwise stated, we consider the following case throughout this study: 
\begin{eqnarray}
g_{EE}=g_{IE}=-g_{EI}=-g_{II}=g_0, \  \  g_0\geq 0.\label{pedagogical_g}
\end{eqnarray}
Equation (\ref{pedagogical_g}) tunes the model to the critical point at which neuronal fluctuations evoke large responses in the population dynamics. With suitable distributions for $\mathcal{J} _{k\ell }^{ij}$ [see Appendix \ref{appendix:D}], the model constrained by Eq.(\ref{pedagogical_g}) describes a densely connected network of excitatory and inhibitory neurons obeying the Dale's law. In the case with $g_0=0$ and $I_{i}^{(k)}(t)\equiv 0$, this model is equivalent to the classical model investigated by a previous study \cite{PhysRevLett.61.259}. For the case with $I_i^{(k)}(t)\neq 0$, we apply external inputs of the same strength, $\widetilde{I} (t)$, to only $\sqrt{N}$ neurons in the excitatory population [Fig.\ref{model_schema}(b)] as
\begin{eqnarray}
I_{i}^{(k)}(t)=\left \{ \begin{array}{cc} \widetilde{I} (t), & (k,i)\in \mathcal{S}, \\ 0, &  (k,i)\notin \mathcal{S}, \end{array} \right. \label{modelling_sparse_input}
\end{eqnarray}
where we define $\mathcal{S} \overset{\mathrm{def}}{=}\{ (k,i)| k=E, 1\leq i\leq \sqrt{N}\} $. 
Neurons in cortical areas have been thought to receive such sparse inputs \cite{Olshausen:1997cr, olshausen2004sparse}. Previous experiments showed that inputs to a small number of cortical neurons can drive the whole local circuit \cite{london2010sensitivity, Chettih:2019wm}. We model these neuronal responses.

In addition to the above model, we also study a model obeying the same dynamical equation as Eq.(\ref{dynamical_equation}) with the following additional tuning of the synaptic weights:  
\begin{eqnarray}
J_{k\ell }^{ij}=\frac{\sigma _{0}\widetilde{\mathcal{J} } _{k\ell }^{ij}}{\sqrt{N}}+\frac{g_{k\ell }}{\sqrt{N}},\  \  \  \widetilde{\mathcal{J}} _{k\ell }^{ij}=\mathcal{J} _{k\ell }^{ij}-\frac{1}{N}\sum _{1\leq j^{\prime }\leq N}\mathcal{J} _{k\ell }^{ij^{\prime }}.
\label{basic_equation}
\end{eqnarray}
Note that the sum of the random variations in the weights of the excitatory and inhibitory synapses on each neuron is finely tuned to zero; namely, $\sum _{1\leq j\leq N}\widetilde{\mathcal{J}} _{k\ell }^{ij}=0$. The weight matrix, $J_{k\ell }^{ij}$, is known to have a finite number of outlier eigenvalues for the untuned synaptic weights obeying Eq.(\ref{basic_equation_nonzerosum}), even in the limit of infinitely large $N$, but not for the finely tuned synaptic weights obeying Eq.(\ref{basic_equation}) \cite{PhysRevLett.97.188104, Tao2013}. We observe how this qualitative difference is reflected in the dynamics of the model.

\section{Mean-field theory} \label{sec:section3}
\subsection{Model with finely tuned synaptic weights} \label{subsec:3_A}
In this section, we formulate an MFT for the models described above. The detailed derivation is given in Appendix \ref{appendix:O}. The MFT is slightly simpler for the finely tuned model obeying Eqs.(\ref{dynamical_equation}), (\ref{pedagogical_g}), and (\ref{basic_equation}) with $I_{k,i}(t)\equiv 0$ than for the untuned model. Therefore, we first consider the dynamics of the finely tuned model.

Following a similar analysis to the previous one \cite{PhysRevX.5.041030}, we divide the dynamics of the model into macroscopic and microscopic parts. Let $m_k(t)$ be certain averages of neuronal variables $h_i^{(k)}(t)$ for $k=E, I$. The microscopic deviations from these averages are defined as
\begin{eqnarray}
\delta h_i^{(k)}(t)\overset{\mathrm{def}}{=} h_i^{(k)}(t)-m_k(t).
\end{eqnarray}
Similarly, we decompose the outputs of the neurons into macroscopic and microscopic parts as
\begin{eqnarray}
\phi (h_i^{(k)}(t))=\overline{\phi} _{k}(t)+\delta \phi _i^{(k)}(t).
\end{eqnarray}
The precise definitions of $m_k$ and $\overline{\phi } _k$ will be stated below. Here, we note that $m_k$ and $\overline{\phi } _k$ coincide with the averages of $h_i^{(k)}$ and $\phi (h_i^{(k)})$, respectively, over the population $k$ up to the leading order in $N$.

With these decompositions, we can rearrange the model equations into the following form in the large $N$ limit: 
\begin{eqnarray}
\frac{\mathrm{d}}{\mathrm{d} t} m_k(t)&=&-m_k(t)+\sum _{\ell, j}\frac{g_{k\ell }}{\sqrt{N} } \phi (h_j^{(\ell )}(t)), \label{macroequation} \\
\frac{\mathrm{d}}{\mathrm{d} t}\delta h_i^{(k)}(t)&=&-\delta h_i^{(k)}(t)+\sigma _{0}\sum _{\ell ,j}\frac{\widetilde{\mathcal{J}} _{k\ell }^{ij}}{\sqrt{N}} \delta \phi _j^{(\ell )}(t). \label{microequation}
\end{eqnarray}
The first of the above equations with a suitable initial condition defines $m_k$.

The configuration of the random synaptic weights does not change during time evolution. In the framework of MFT, however, we consider the distributions of the microscopic variables, $\delta h_i^{(k)}$ and $\delta \phi _j^{(\ell )}$, over a large ensemble of networks with different weight configurations. Therefore, the time evolution of these variables is stochastic. In the stochastic dynamics, the driving-force term in the right-hand side of Eq.(\ref{microequation}), 
\begin{eqnarray}
\eta _i^{(k)}(t)\overset{\mathrm{def}}{=}\sigma _0\sum _{\ell ,j}\frac{\widetilde{\mathcal{J}} _{k\ell }^{ij}}{\sqrt{N}} \delta \phi _j^{(\ell )}(t), \label{microfluctuation}
\end{eqnarray}
has the following property: given the entire time evolution of the mean activity $m_E$ and $m_I$, the conditional probability distribution for $(\eta _i^{(k)}(t_1), \eta _i^{(k)}(t_2), \cdots )$, for any finite set of time points, $t_1, t_2, \cdots $, is a zero-mean Gaussian that is {\it i.i.d.} with respect to the index $i$. This is intuitively justified by the central limit theorem applied to Eq.(\ref{microfluctuation}), under the assumption that the random synaptic weights and fluctuations in the neuronal outputs are almost independent [see Appendix \ref{appendix:O} for further justification]. Furthermore, since the driving-force term in the linear equation (\ref{microequation}) has a conditionally {\it i.i.d.} Gaussian distribution with zero mean, so also do fluctuations in $\{ \delta h_i^{(k)}\} _{i,k}$. Then, the following first- and second-order moments fully characterize these Gaussian fluctuations:
\begin{eqnarray}
C_k(t,s)&\overset{\mathrm{def}}{=}&\langle \delta \phi _i^{(k)}(t)\delta \phi  _i^{(k)}(s)\rangle ,\ \ \ \overline{\phi } _k(t)\overset{\mathrm{def}}{=}\langle \phi (h_i^{(k)}(t))\rangle ,\nonumber \\
D_k(t,s)&\overset{\mathrm{def}}{=}&\langle \delta h_i^{(k)}(t)\delta h_i^{(k)}(s)\rangle ,\ \ \ m_k(t)=\langle h_i^{(k)}(t)\rangle , \label{definition_of_statistics}
\end{eqnarray}
where the brackets denote averages over the Gaussian fluctuations. Note that $\overline{\phi } _k$ is defined in the above. The correlation function for $\eta _i^{(k)}$ is obtained from $C_{\ell }(t,s)$:
\begin{eqnarray}
\langle \eta _i^{(k)}(t)\eta _i^{(k)}(s)\rangle =\sigma _0^2\sum _{\ell =E,I}C_{\ell}(t,s). \label{correlation_of_micro_eta}
\end{eqnarray}
Below, we omit the population index $k$ because the excitatory and inhibitory populations have the same statistical properties in our model setting. Thus, we have 
\begin{eqnarray}
\overline{\phi }\overset{\mathrm{def}}{=}\overline{\phi }_E=\overline{\phi }_I , &\ \ \ & m\overset{\mathrm{def}}{=}m_E=m_I,\nonumber \\
C\overset{\mathrm{def}}{=}C_E=C_I,&\ \ \ &D\overset{\mathrm{def}}{=}D_E=D_I.
\end{eqnarray}

The statistics of the fluctuations defined above must satisfy certain consistency conditions. First, $m$ and $\overline{\phi}$ have been defined from dynamical variables that evolve under the influences of $m$ and $\overline{\phi }$ themselves [Eqs.(\ref{macroequation}), (\ref{microequation}), and (\ref{definition_of_statistics})]. Therefore, their values need to be determined in a self-consistent manner. Second, the two Gaussian fluctuations characterized by ($m$, $D$) and ($\overline{\phi }$, $C$) are related to each other because they originate from the same dynamical variables. Consistency among these statistics gives rise to self-consistent equations that determine the time evolution of $\overline{\phi }$, $C$, and $D$ for given values of $m$ and boundary conditions. Firstly, $\overline{\phi } $ and $C$ are represented as nonlinear functions of $m$ and $D$ [see Eqs.(\ref{recovering_C_again}) and (\ref{recovering_phi_bar_again}) in Appendix \ref{appendix:A} for the details]: 
\begin{eqnarray}
\overline{\phi } (t)&=&G_1(m(t), D(t,t)), \nonumber \\
C(t,s)&=&G_2(m(t),m(s),D(t,t),D(s,s),D(t,s)). \label{self_consistent_eq_1}
\end{eqnarray}
Secondly, the relation between $\eta _i^{(k)}$ and $\delta h_i^{(k)}$ results in the following dynamical equation: 
\begin{eqnarray}
(1+\partial _t)(1+\partial _s)D(t,s)=2\sigma _0^2C(t,s). \label{self_consistent_eq_2}
\end{eqnarray}

An important difference between our MFT and conventional MFTs lies in the macroscopic driving-force in Eq.(\ref{macroequation}). The right-hand side of this equation involves the summation of fluctuations in the outputs of individual neurons which can be considered as independent random quantities with correlation $C(t,s)$. Then, the sum of these quantities in Eq.(\ref{macroequation}), 
\begin{eqnarray}
\eta (t)&\overset{\mathrm{def}}{=}&\frac{1}{\sqrt{2} g_0}\sum _{\ell, j}\frac{g_{k\ell }}{\sqrt{N} }\phi (h_j^{(\ell )}(t)) \nonumber \\
&=&\frac{1}{\sqrt{2N}} \left \{ \sum _{i}\delta \phi _i^{(E)}(t)-\delta \phi _i^{(I)}(t)\right \} , \label{definition_eta}
\end{eqnarray}
is also a random quantity (which is equal for $k=E,I$). From the first line to the second line, we have used $\overline{\phi }_E(t)=\overline{\phi }_I(t)$. This stochasticity contrasts starkly with the deterministic macroscopic dynamics in conventional MFTs. The central limit theorem implies that $\eta $ obeys the following probability distribution: 
\begin{eqnarray}
p(\eta )\propto \exp \left (-\frac{1}{2}\eta ^TC^{-1}\eta -F\right ), \label{realization_probability_eta}
\end{eqnarray} 
where the normalization term $F$ is given by 
\begin{eqnarray}
F=\ln \mathrm{det} \left [ \left (1-\frac{\partial C^{1/2}}{\partial \eta}C^{-1/2}\eta \right )^{-1}C^{1/2}\right ]. \label{normalization_F}
\end{eqnarray}
Here, we have regarded $\eta $ and $C$ as a vector and a matrix, respectively, that consist of their values for infinitesimally discretized timesteps. With these stochastic dynamics for $\eta $, the macroscopic dynamics in Eq.(\ref{macroequation}) reads 
\begin{eqnarray}
\frac{\mathrm{d}}{\mathrm{d} t}m(t)=-m(t)+\sqrt{2} g_0\eta (t). \label{dynamics_of_m}
\end{eqnarray}
We note that, for a given history of $\eta $ and $m$ up to time $t$, the conditional distribution of $\eta (t+\Delta t)$ determined by Eq.(\ref{realization_probability_eta}) with small $\Delta t$ is approximately Gaussian. In this case, in the conditional distribution up to time $ t+\Delta t$,
\begin{eqnarray}
p(\eta (t+\Delta t)|\{ \eta (s)\} _{s\leq t})\propto \exp \left (-\frac{1}{2}\eta ^TC^{-1}\eta -F\right ), \label{numerical_MFE}
\end{eqnarray}
the correlation matrix $C$ and normalization term $F$ are independent of $\eta (t+\Delta t)$ up to the leading order in $\Delta t$. Therefore, the deviations of the conditional probability distribution from a Gaussian distribution are negligibly small. This fact enables us to solve the stochastic dynamics numerically for $\eta $, $m$, $\overline{\phi }$, $C$, and $D$ by iteratively updating their values with the Euler method [see Appendix \ref{appendix:A} for the details].

In sum, the microscopic fluctuations in the neuronal activities obey a Gaussian distribution that depends on the mean activity $m$ [Eqs.(\ref{self_consistent_eq_1}) and (\ref{self_consistent_eq_2})], and the probability of realizing the mean activity $m$ depends on the correlation matrix $C$ of the microscopic fluctuations [Eqs.(\ref{realization_probability_eta}) and (\ref{dynamics_of_m})]. Due to this strong link between the microscopic and macroscopic dynamics, the entire dynamics are, in general, non-Gaussian, even though the distribution of $\eta $ resembles a Gaussian distribution [Eq.(\ref{realization_probability_eta})]. This means that---unlike in conventional MFTs---a solution for the model cannot be completely determined by the first- and second-order moments. 
\subsection{Balance equations}
In Eq.(\ref{definition_eta}), we removed the population-averaged part by using $\overline{\phi } _E(t)=\overline{\phi }_I(t)$. Before using this relation, the macroscopic part of the mean-field equations reads, to leading order, 
\begin{eqnarray}
\frac{\mathrm{d}}{\mathrm{d} t}m_E(t)&\approx &-m_E(t)+\sqrt{N} (g_{EE}\overline{\phi }_E(t)+g_{EI}\overline{\phi }_I(t)), \nonumber \\
\frac{\mathrm{d}}{\mathrm{d} t}m_I(t)&\approx &-m_I(t)+\sqrt{N} (g_{IE}\overline{\phi }_E(t)+g_{II}\overline{\phi }_I(t)).
\end{eqnarray}
In these equations, the driving-force terms on the right-hand side are $O(\sqrt{N})$ and hence may diverge. Thus, the following condition must hold for stable dynamics: except for $O(1/\sqrt{N} )$ residuals, 
\begin{eqnarray}
&&g_{EE}\overline{\phi }  _{E}(t) +g_{EI}\overline{\phi }  _{I}(t)\approx 0,\nonumber \\
&&g_{IE}\overline{\phi }  _{E}(t) +g_{II}\overline{\phi }  _{I}(t)\approx 0. \label{balance_equation}
\end{eqnarray}
Eqs.(\ref{balance_equation}) are called {\it ``balance equations.''} In previous theories, the balance equations were often non-degenerate and determined unique values for $\overline{\phi } _E(t)$ and $\overline{\phi } _I(t)$. This implies that population-averaged activities exhibit only vanishingly small fluctuations if the entire dynamics are stable. In contrast, in our theory---for the values of $g_{k\ell}$ satisfying Eq.(\ref{pedagogical_g})---Eqs.(\ref{balance_equation}) are degenerate, and they are satisfied as long as equality $\overline{\phi } _{E}(t)=\overline{\phi } _{I}(t)$ holds. Furthermore, this equality is always ensured to hold as a consequence of the fact that the macroscopic equation, Eq.(\ref{macroequation}), has exactly the same driving-force term for the excitatory and inhibitory populations, and hence $m_E(t)=m_I(t)$. As a result, the average of the neuronal outputs $\overline{\phi } _k(t)$ are allowed to fluctuate strongly.

\subsection{Model with untuned synaptic weights} \label{subsec:2_A_2}
Next, we describe how the above theory is modified for the untuned model [Eqs.(\ref{dynamical_equation}) and (\ref{basic_equation_nonzerosum})]. In this case, the dynamical equation is divided into microscopic and macroscopic parts in a slightly different manner:
\begin{eqnarray}
\frac{\mathrm{d}}{\mathrm{d} t} m_{k}(t)&=&-m_k(t)+\sum _{\ell ,j}\frac{g_{k\ell }}{\sqrt{N} }\phi (h_j^{(\ell )}(t)), \label{macroequation_no_zerosum} \\
\frac{\mathrm{d}}{\mathrm{d} t}\delta h_i^{(k)}(t)&=&-\delta h_i^{(k)}(t)+\sum _{\ell ,j}\sigma _{0}\frac{\mathcal{J} _{k\ell }^{ij}}{\sqrt{N}} \phi (h_j^{(\ell )}(t)). \label{microequation_no_zerosum}
\end{eqnarray}
Note that, in Eq.(\ref{microequation_no_zerosum}), the fluctuation term $\delta \phi _j^{(\ell )}(t)$ in Eq.(\ref{microequation}) is replaced by the uncentered quantity $\phi (h_j^{(\ell )}(t))$. As a result, the microscopic driving-force terms, 
\begin{eqnarray}
\widetilde{\eta } ^{(k)}_{i}(t)\overset{\mathrm{def}}{=} \sigma _0\sum _{\ell, j}\frac{\mathcal{J} _{k\ell }^{ij}}{\sqrt{N}}\phi (h_{j}^{(\ell )}(t)), \label{gaussian_process_no_zerosum}
\end{eqnarray}
have the following correlation functions:
\begin{eqnarray}
\langle \widetilde{\eta } ^{(k)}_{i}(t)\widetilde{\eta } ^{(k)}_{i}(s)\rangle &=&2\sigma _0^2\widetilde{C} (t,s) \label{define_C_tilde} \\
\widetilde{C} (t,s)\overset{\mathrm{def}}{=}&&\hspace{-0.3cm}\langle \phi (h_i^{(k)}(t))\phi (h_i^{(k)}(s))\rangle =C(t,s)+\overline{\phi }(t)\overline{\phi }(s). \nonumber
\end{eqnarray}
This equation means that individual neurons receive additional synchronous inputs with random amplitudes, which can be represented as 
\begin{eqnarray}
\widetilde{\eta } _i^{(k)}=\eta _i^{(k)}(t)+\sqrt{2}\sigma _0\xi _i^{(k)}\overline{\phi }(t), \label{random_synchronous_input}
\end{eqnarray}
where $\{ \xi _i^{(k)}\} _{i,k}$ are {\it i.i.d.} quenched Gaussian variables with zero mean and unit variance. Then, the modified self-consistent equation, 
\begin{eqnarray}
(1+\partial _t)(1+\partial _s)D(t,s)&=&2\sigma _0^2\widetilde{C}(t,s), \label{self_consistent_eq_2_untune}
\end{eqnarray}
together with Eqs. (\ref{self_consistent_eq_1}) and (\ref{define_C_tilde}) determines the time evolution of $\overline{\phi }$, $\widetilde{C}$, and $D$ for a given orbit $m$. On the other hand, the macroscopic dynamics of $m$ are described by Eqs.(\ref{realization_probability_eta}) and (\ref{dynamics_of_m}). 
\subsection{Application of external inputs} \label{MFT_stimulus}
In Sec.\ref{subsection4B}, we apply external inputs of strength $\widetilde{I} (t)$ to $\sqrt{N}$ neurons in the $E$ population of the finely tuned model. The stimulus-driven dynamics are analyzed by an MFT that is slightly modified from the one introduced above. It is described by 
\begin{eqnarray}
\frac{\mathrm{d}}{\mathrm{d} t} m(t)&=&-m(t)+\sqrt{2} g_0\eta (t)+g_0\widetilde{\phi }(t), \nonumber \\
\frac{\mathrm{d}}{\mathrm{d} t}\delta h_i^{(k)}(t)&=&-\delta h_i^{(k)}(t)+\eta _i^{(k)}(t)+I_i^{(k)}(t), \nonumber \\
\widetilde{\phi }(t)\overset{\mathrm{def}}{=}&&\hspace{-0.5cm} \langle \phi (h_i^{(k)}(t))\rangle _{(k,i)\in \mathcal{S}} -\langle \phi (h_i^{(k)}(t))\rangle _{(k,i)\notin \mathcal{S}}, \label{MFT_ext}
\end{eqnarray}
where the microscopic and macroscopic stochastic driving-force terms, $\eta _i^{(k)}$ and $\eta $, are distributed according to the same equations as those for the autonomous case, namely, Eqs.(\ref{correlation_of_micro_eta}) and (\ref{realization_probability_eta}), with the same self-consistent equations, Eqs.(\ref{self_consistent_eq_1}) and (\ref{self_consistent_eq_2}). The difference, $\widetilde{\phi }(t)$, between the averages of the stimulus-driven and undriven neuronal variables over the Gaussian fluctuations gives an additional driving force term for the mean activity in Eq.(\ref{MFT_ext}). 

\begin{figure*}
\includegraphics[width=178mm]{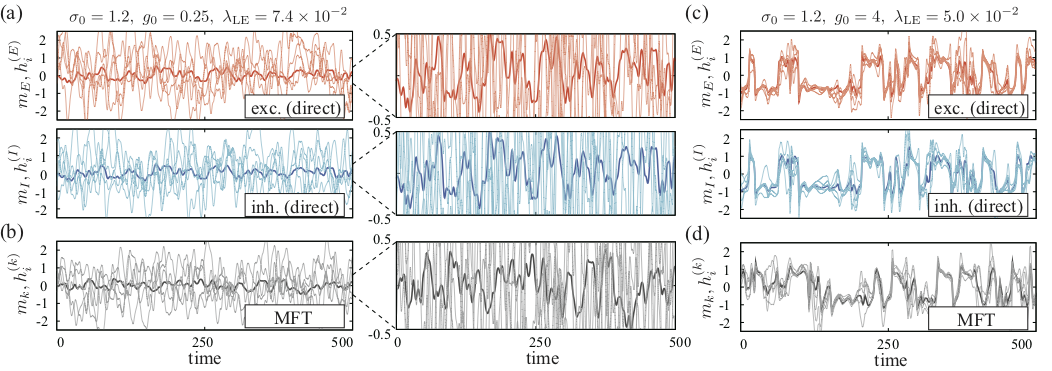}
\caption{Activity patterns of the finely tuned model for $\sigma _0=1.2$ and for two different values $g_0=0.25$ and $4$. In each plot, the thick and thin lines represent the mean activity $m_k$ and the activities $h_i^{(k)}$ of five representative neurons, respectively. For the direct simulations, the mean activity $m_k$ is approximated by the average of $h_i^{(k)}$ over the population $k$ comprising $N=10240$ neurons [see Appendix \ref{appendix:D} for the details of the simulations]. Solutions for the $E$ and $I$ populations from direct simulations (indicated by ``direct'') and solutions of the mean-field equations (indicated by ``MFT'') are depicted in red, blue, and gray, respectively. The numerically determined value of the largest Lyapunov exponent, $\lambda _{\mathrm{LE}}$, is shown above each panel. } \label{figure_single_EI_pair}
\end{figure*}

\section{Results} \label{sec:section4}
\subsection{Dynamics of the finely tuned model} \label{subsection4A}
\subsubsection{Fluctuations in the mean activity}
We first examine the finely tuned model described by Eqs.(\ref{dynamical_equation}) and (\ref{basic_equation}) without external inputs, for which the MFT takes the simplest form [Sec.\ref{subsec:3_A}]. For $g_0=0$, the MFT of this model is essentially equivalent to that studied previously \cite{PhysRevLett.61.259}. The previous theory showed that the present model with $g_0=0$ undergoes a transition from a trivial fixed point to a chaotic state at $\sigma _0=1/\sqrt{2}$, in which the mean activity, $m$, is constantly zero and individual neuronal activities, $h_i^{(k)}$, exhibit Gaussian fluctuations around it.

We are particularly interested in cases with non-zero values of $g_0$. We study these cases both by numerically solving the mean-field equations and by directly simulating the model for a large value of $N$. In these numerical simulations, we find only a trivial fixed-point solution for $\sigma _0\leq 1/\sqrt{2}$. In contrast, for $\sigma _0>1/\sqrt{2}$, we obtain non-trivial solutions. Since the repertoire of solutions is qualitatively the same for different values of $\sigma _0$, we show typical activity patterns only for $\sigma _0=1.2$ in Fig.\ref{figure_single_EI_pair}(a)--(d). In our MFT, the excitatory and inhibitory populations obey the same dynamical equations. Therefore, we plot only a single representative solution from MFT for each value of $g_0$. In fact, in the plots from the direct simulations, the mean activities of the excitatory and inhibitory populations are almost equal, and the individual neuronal activities in the two populations exhibit similar temporal patterns. Comparing the plots from the MFT and direct simulations, we observe similar amplitudes and temporal patterns for the mean activities and the microscopic fluctuations around them. These results suggest that our theory successfully predicts the behavior of the model. Below, we further evaluate this point quantitatively.

As the value of $g_0$ is increased from zero, the mean activity of the model starts to fluctuate with non-zero amplitudes. For relatively small values of $g_0$, the temporal profiles of the fluctuations, both in the mean and in the individual neuronal activities, are similar to the Gaussian fluctuations in individual neuronal activities at $g_0=0$ [Fig.\ref{figure_single_EI_pair}(a) and (b)]. This is expected from the MFT, which shows that the driving force for the mean activity is the summation of individual neuronal fluctuations scaled by $g_0/\sqrt{N}$ [Eq.(\ref{macroequation})]. With a further increase in the value of $g_0$, the model starts to show irregular, intermittent dynamics, varying between positive and negative values close to $\pm 1$, with patterns that are reminiscent of the UP-DOWN states observed in the brain \cite{hobson1986neuronal, poulet2008internal} [Fig.\ref{figure_single_EI_pair}(c) and (d)]. This bimodality in the mean activity indicates the non-Gaussianity of the dynamics and contrasts with the dynamics for small values of $g_0$. Numerically determined largest Lyapunov exponent [Fig.\ref{figure_single_EI_pair}] indicates that the both types of solutions described above are chaotic.

Increasing the value of $g_0$ still further, we occasionally observe stable fixed-points and regularly oscillating solutions, as well as irregular, chaotic solutions. Although these non-chaotic solutions are observed for networks with a fairly large number of neurons [Appendix \ref{system_size_dependence}], further theoretical analyses suggest that these solutions are due to finite-size effects and not stable in the thermodynamic limit [see the discussion in Sec.\ref{perturbative_analysis_fp} and Appendix \ref{perturbative_expansion_fixed_point}]
\begin{figure}
\includegraphics[width=88mm]{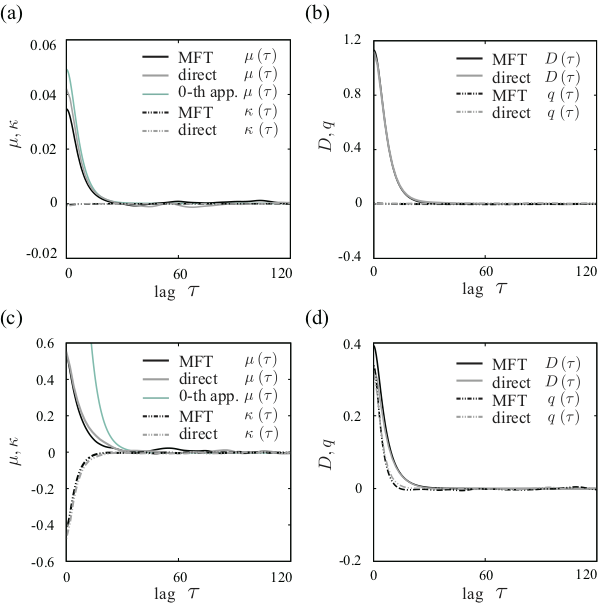}
\caption{Statistics of the network dynamics with the following values of the parameters: (a) and (b), $(\sigma _0,g_0)=(1.2,0.25)$; (c) and (d), $(\sigma _0,g_0)=(1.2,4)$. Panels (a) and (c) show the values of the autocorrelation functions of the mean activities $\mu (\tau )$ and of the neuronal activities $D(\tau )$. Panels (b) and (d) show the values of the fourth-order statistics of the mean activities $\kappa (\tau )$ and of the neuronal activities $q(\tau )$. These values are determined either from the direct simulations (labeled ``direct''), the simulations of the mean-field equations (labeled ``MFT''), or the zero-th order perturbative approximation (labeled ``0-th app.''), and they are indicated by different types of lines. In the simulations, these quantities are averaged over the time period $1000\leq t\leq 2200$ and over 15 random configurations of directly simulated networks with population size, $N=40960$ [(a) and (b)]; $N=20480$ [(c) and (d)]; or averaged over the same time period and 15 sequences of random numbers used for the simulations of the mean-field equations. } \label{figure_statistics_of_dynamics}
\end{figure}

To examine the extent to which the description provided by our MFT is accurate, we calculate statistics of the dynamics from numerical solutions of the mean-field equations and of the original model equations. We calculate the autocorrelation functions for the mean activity and for the individual neurons as 
\begin{eqnarray}
\mu (\tau )&\overset{\mathrm{def}}{=}&\langle m(t)m(t-\tau )\rangle ,  \label{autocorrelation_mean} \\
D(\tau )&\overset{\mathrm{def}}{=}&\langle \delta h_i^{(k)}(t)\delta h_i^{(k)}(t-\tau )\rangle . \label{autocorrelation_indiv}
\end{eqnarray}
Since we expect the dynamics to be non-Gaussian, we also calculate the fourth-order statistics defined by 
\begin{eqnarray}
\kappa (\tau )&=&\langle m(t)^2m(t-\tau )^2\rangle \nonumber \\
&&-\langle m(t)^2\rangle ^2-2\langle m(t)m(t-\tau )\rangle ^2, \label{fourth_moements_mean} \\
q(\tau )&=&\langle \delta h_i^{(k)}(t)^2\delta h_i^{(k)}(t-\tau )^2\rangle \nonumber \\
&&-\langle \delta h_i^{(k)}(t)^2\rangle ^2-2\langle \delta h_i^{(k)}(t)\delta h_i^{(k)}(t-\tau )\rangle ^2. \label{fourth_moments_indiv}
\end{eqnarray}
In these equations, bracketing indicates averaging over both time and configurations of the random connectivity. In Eqs.(\ref{autocorrelation_indiv}) and (\ref{fourth_moments_indiv}), we also take averages over population $k$ in direct simulations and averages over microscopic Gaussian fluctuations in the corresponding MFT. The fourth-order statistics defined above vanish if the dynamics are Gaussian. For the direct simulations, we show only the statistics of the $E$ population, because those of the $I$ population are essentially the same.

The panels in Fig.\ref{figure_statistics_of_dynamics} compare these calculated statistics, and they show good agreement between the theory and direct simulations. This indicates that our theory predicts the behavior of the model quantitatively, at least statistically. In this figure, we also observe large fourth-order statistics for networks with large values of $g_0$, which implies the highly non-Gaussian nature of the dynamics.

For the autocorrelation functions defined above, one would expect a perturbative expansion to provide a good analytical approximation, as it does for many physical systems. We can actually formulate such a method by expanding the dynamics around $g_0=0$. However, it is numerically intractable to carry out the calculation of even the first-order expansion [see Appendix \ref{perturbative_expansion_chaos}]. Here, we restrict ourselves to showing only the zero-th order term, $\mu (\tau )\approx g_0^2D_0(\tau)/\sigma _0^2$, of this perturbative expansion [Fig.\ref{figure_statistics_of_dynamics}(a) and (c)]. Here, $D_0(\tau )$ is the autocorrelation function of the microscopic variables $\delta h_i^{(k)}$ for $g_0=0$. This zero-th order approximation gives vanishing fourth-order statistics: $\kappa (\tau )=q(\tau )=0$. For small values of $g_0$, this solution shows relatively good agreement with the estimate obtained from direct simulations, while it does not do so for large values of $g_0$.

\subsubsection{Waveforms of the mean activity and the signature of time-reversal symmetry breaking}
For larger values of $g_0$, the analytical approach encounters another difficulty in addition to the computational problems mentioned above. Fig.\ref{figure_single_EI_pair}(c) and (d) show that the trajectories of the mean activity are observed with frequencies that are obviously asymmetric with respect to the time reversal of the trajectories. Note that the mean activity overshoots immediately after it makes an intermittent transition between positive and negative values, and that the temporal order of the transition and the overshooting is never reversed. Analytical approaches---such as a perturbative expansion around $g_0=0$---however, yield only symmetric solutions [see Appendix \ref{perturbative_expansion_chaos}]. This inconsistency suggests the possibility of symmetry breaking with respect to time reversal. If a symmetry is broken, one cannot expect a symmetry-breaking solution to be obtained from a series expansion around the symmetric solutions. In the following, we use a heuristic approach to seek clues to the occurrence and mechanism of such symmetry breaking and to an understanding of the waveform of the mean activities for large $g_0$.

\begin{figure}
\includegraphics[width=68mm]{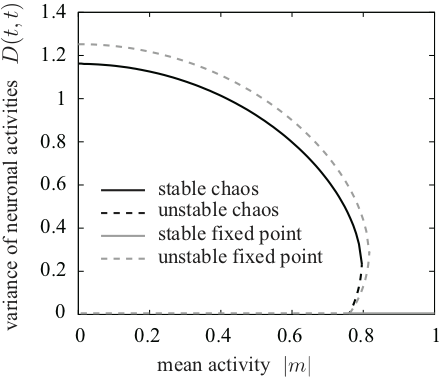}
\caption{Variances of neuronal activities in fixed-point and chaotic solutions and their stability, depicted for different absolute values of the constant mean activity, $|m|$. Connected pairs of branches of nontrivial fixed-point and chaotic solutions, as well as trivial fixed-point solutions, are shown.} \label{figure_fixed_point_chaos_diagram}
\end{figure}

In our MFT, the correlation function of the microscopic fluctuations is determined by Eqs.(\ref{self_consistent_eq_1}) and (\ref{self_consistent_eq_2}) for a given trajectory of the mean activity, which in turn determines the realization probability of the mean activity. Since this dependence is complicated, we first focus on the case with constant mean activities of different values, expecting the results to provide some clue to the dynamics with time-varying mean activities. Applying the previous theory \cite{PhysRevLett.61.259, PhysRevX.5.041030} to this analysis, we find multiple fixed-point solutions and chaotic solutions. Fig.\ref{figure_fixed_point_chaos_diagram} shows the variance of the neuronal activities of these solutions for different constant values of the mean activity. A branch of chaotic solutions [the black solid line in Fig.\ref{figure_fixed_point_chaos_diagram}] coincides with the solution examined in a previous study \cite{PhysRevLett.61.259} for $m(t)=0$. As the absolute value of the mean activity increases, the neuronal fluctuations in these solutions decrease. Another branch of chaotic solutions with smaller neuronal fluctuations [the black dotted line in Fig.\ref{figure_fixed_point_chaos_diagram}] emerges at the value satisfying $2\sigma _0^2\phi ^{\prime }(m)=1$, $|m|\approx 0.76$. The neuronal fluctuations in these solutions increase as the mean activity increases, and this branch eventually connects to that with larger neuronal fluctuations. From the numerical simulations, we find that the branch of chaotic solutions with larger fluctuations is stable, while that with smaller fluctuations is unstable. Fixed-point solutions and their stability can also be examined by applying the previous theory \cite{PhysRevX.5.041030} (or the method presented in Appendix \ref{perturbative_expansion_fixed_point}), and we find two connected branches of unstable fixed-point solutions as well as trivial fixed-point solutions [Fig.\ref{figure_fixed_point_chaos_diagram}]. The trivial fixed-point solution is stable for $|m|$ larger than the bifurcation point given by $2\sigma _0^2\phi ^{\prime }(|m|)=1$, while it is unstable below this point.    

Fig.\ref{figure_fixed_point_chaos_diagram} suggests the following explanation for the waveform of the mean activity and its time-reversal asymmetry observed in Fig.\ref{figure_single_EI_pair}(c) and (d). Let us assume that, for a time-varying mean activity, the instantaneous behavior of the neuronal fluctuations is the same as the above solution for the corresponding value of the constant mean activity. When the mean activity remains small for some time, the neuronal fluctuations increase. Since the neuronal fluctuations serve as a driving force for the mean activity, the mean activity is stochastically pushed to larger values. For larger values of the mean activity, the neuronal fluctuations decrease (along the black solid line in Fig.\ref{figure_fixed_point_chaos_diagram}), while still remaining chaotic. When the mean activity reaches a value for which there are no stable chaotic solutions, the neuronal fluctuations start to decay to the trivial fixed point. Then, the mean activity loses its driving force and decays to smaller values. In this descending part of the mean activity, the network state passes through the region with the unstable chaotic and fixed-point solutions (the lower branches of the non-trivial solutions in Fig.\ref{figure_fixed_point_chaos_diagram}). The profile of the neuronal fluctuations in this descending part is therefore different from that of the ascending part. Because of this passage through the region with unstable fixed points, both the neuronal fluctuations and the mean activity slow down, as we observe in Fig.\ref{figure_single_EI_pair}(c) and (d). We suggest that this hysteresis in the multiscale dynamics is the mechanism for the observed waveform of the mean activity and its time-reversal asymmetry.  
\subsubsection{Ferromagnetic effects and critical fluctuations} \label{section_ferromagnetic}
\begin{figure}
\includegraphics[width=90mm]{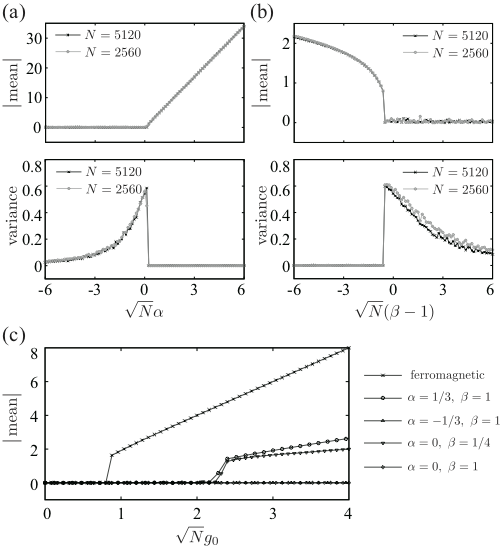}
\caption{Absolute values of the mean and variance of fluctuations in the mean activity of the excitatory population, calculated by taking long-time averages in numerical simulations with $(\sigma _0,g_0)=(1.2,4)$ and different values of (a) $\alpha $ and (b) $\beta $, respectively. We performed simulations with the two different system sizes indicated in the panels. (c) The absolute value of the long-time average of the mean activity of the excitatory population, calculated by taking long-time averages in numerical simulations with $\sigma _0=1.2$, different values of $g_0$, and the following values of $\alpha $ and $\beta $: $(\alpha ,\beta )=(\pm 1/3 ,0)$, $(0,1/4)$, $(0,0)$. Results for $g_{EE}=g_{IE}=g_{EI}=g_{II}=3g_0/2$ (indicated as ``ferromagnetic'') are also shown.} \label{figure_unbalanced}
\end{figure}

Thus far, we have examined balanced networks with parameter values satisfying the condition in Eq.(\ref{pedagogical_g}). In this section, we briefly mention what happens if this condition is not satisfied. As in a previous study \cite{PhysRevX.5.041030}, if the balance equation is not degenerate, the mean activities of the neuronal populations take a set of constant values uniquely determined by the balance equations, or else diverge. The remaining cases are described with two parameters $\alpha $ and $\beta $ as 
\begin{eqnarray}
\left (\begin{array}{cc} g_{EE} & g_{EI} \\ g_{IE} & g_{II} \end{array} \right )=g_0\left (\begin{array}{cc} 1+\alpha & -1+\alpha \\ \beta (1+\alpha) & \beta (-1+\alpha )\end{array} \right ). 
\end{eqnarray} 
By definition, the parameter $\alpha $ is interpreted as the magnitude of ferromagnetic interaction, while $\beta $ is interpreted as the relative gain of the synaptic input to inhibitory neurons. The case we have examined in the previous sections corresponds to $(\alpha ,\beta)=(0,1)$. We examine the fluctuations in the mean activity by calculating their mean and variance averaged over a long period of simulations and by observing how they change as the value of $\alpha $ or $\beta $ deviates from $(\alpha ,\beta)=(0,1)$ [Fig.\ref{figure_unbalanced}(a) and (b)]. We find that the mean activity diverges as $\alpha$ increases or $\beta $ decreases, while the variance of the fluctuations decays to zero as $\alpha $ decreases or $\beta $ increases. As shown in Fig.\ref{figure_unbalanced}(a) and (b), the rate of this divergence and decay is proportional to $\sqrt{N}$, which indicates that in the $N\rightarrow \infty$, the macroscopic dynamics of the network are divergent or trivial for $\alpha \neq 0$, $\beta =1$ and $\alpha =0$, $\beta \neq 1$. 

This behavior can be understood by first examining $O(1/\sqrt{N})$ values of $g_0$ and then taking the $g_0 \rightarrow \infty $ limit. As shown in a previous study \cite{PhysRevX.5.041030}, for $g_{k\ell }\propto 1/\sqrt{N}$, the dynamics of the mean activity are no longer subject to the balance between strong excitation and inhibition but instead are described by a simpler MFT:
\begin{eqnarray}
\frac{\mathrm{d}}{\mathrm{d} t} m_k(t)=-m_k(t)+\sum _{\ell =E,I} \sqrt{N} g_{k\ell } \check{\phi } _{\ell }(t), \label{mean_equation}
\end{eqnarray}
for $k=E$ and $I$, where $\check{\phi } _{\ell }(t)$ is the population average of $\phi (h_j^{(\ell )}(t))$. Fig.\ref{figure_unbalanced}(c) shows how the mean activity changes as $g_0$ increases for fixed values of $\alpha $ and $\beta $.  As $g_0$ increases for $\alpha >0$, $\beta =1$ or $\alpha =0$, $\beta <1$, the second-term on the right-hand side of Eq.(\ref{mean_equation}) starts to dominate, causing the trivial solution to bifurcate in a similar manner to a ferromagnetic transition [the result with $(\alpha ,\beta)=(1/3, 1)$, $(0, 1/4)$ and the result for a purely ferromagnetic interaction shown in Fig.\ref{figure_unbalanced}(c)]. For $\alpha \leq 0$, $\beta =1$ or $\alpha =0$, $\beta \geq 1$, we find that the mean activity remains at zero. For this case of strict inequality, the second term on the right-hand side of Eq.(\ref{mean_equation}) supplies feedback suppression to changes in the mean activity in a similar manner to anti-ferromagnetic effects. For $\alpha =0$, $\beta =1$, however, such a feedback mechanism does not work. These behaviors of the model with $O(1/\sqrt{N})$ values of $g_0$ account for the divergent or suppressed dynamics observed for $O(1)$ values of $g_0$ as the limit of the former. These results also suggest that the present model under the condition given by Eq.(\ref{pedagogical_g}) is at the critical point between the two states governed by extremely strong ferromagnetic and anti-ferromagnetic interactions.

\subsection{Dynamics of the untuned model} \label{subsection4A2}
\subsubsection{Qualitatively different solutions}
\begin{figure*}
\includegraphics[width=178mm]{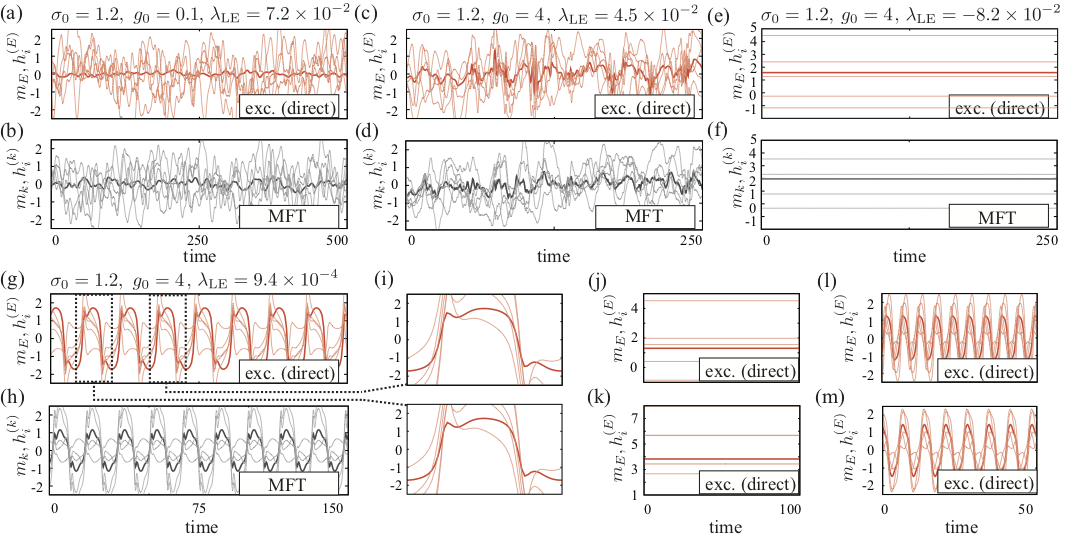}
\caption{Activity patterns of the untuned model. (a)--(h) Typical activity patterns for networks with the indicated parameter values are shown. In each plot, the thick and thin lines represent the mean activity $m_k$ and the activities $h_i^{(k)}$ of five representative neurons, respectively. For the direct simulations, the mean activity $m_k$ is approximated by the average of $h_i^{(k)}$ over the population $k$ comprising $N=10240$ neurons (see Appendix \ref{appendix:D} for the details of the simulations). Solutions for the $E$ populations from direct simulations (indicated by ``direct'') and solutions of the mean-field equations (indicated by ``MFT'') are depicted in red and gray, respectively. The plots for the $I$ population are omitted. For $(\sigma _0, g_0)=(1.2,4)$, we find three qualitatively different solutions: irregular solutions [(c) and (d)], fixed-point solutions [(e) and (f)], and regularly oscillating solutions [(g) and (h)]. Panel (i) shows magnified images of the regular oscillations, where the coherence between the mean activity and the activities of individual neurons can be seen. (j)--(m) For the fixed-point and regularly oscillating solutions, additional plots of the activity patterns illustrate the diversity of their values, waveforms, and frequencies. Which of the diverse solutions is observed for a given set of parameter values depends on the configuration of the random connectivity of the directly simulated networks or on the sequence of the random numbers used for the simulations of the mean-field equations. The numerically determined value of the largest Lyapunov exponent, $\lambda _{\mathrm{LE}}$, is shown above each panel. } \label{figure_single_EI_pair_nonzerosum}
\end{figure*}

The behavior of the untuned model, described by Eqs.(\ref{dynamical_equation}) and (\ref{basic_equation_nonzerosum}), is different from the results discussed above. We show plots of its activity patterns in Fig.\ref{figure_single_EI_pair_nonzerosum}(a)--(m). Simulations based on our MFT yield solutions with profiles similar to those from the direct simulations in this case, too. For smaller values of $g_0$, the network exhibits nearly Gaussian dynamics [Fig.\ref{figure_single_EI_pair_nonzerosum}(a) and (b)]. For larger values of $g_0$, it exhibits not only irregular dynamics [Fig.\ref{figure_single_EI_pair_nonzerosum}(c) and (d)] but also constant activities (fixed-point solutions) [Fig.\ref{figure_single_EI_pair_nonzerosum}(e) and (f)] and regularly oscillating dynamics (limit-cycle solutions) [Fig.\ref{figure_single_EI_pair_nonzerosum}(g) and (h)]. The values of the mean activities of the observed fixed-point solutions are widely distributed over positive and negative values [Fig.\ref{figure_single_EI_pair_nonzerosum}(j) and (k)]. Note that because of the symmetry of the model equations, fixed points are necessarily located symmetrically at two points with positive and negative mean activities of the same absolute value. The waveforms and frequencies of the observed regular oscillations are also diverse [Fig.\ref{figure_single_EI_pair_nonzerosum}(l) and (m)]. Which of these diverse solutions is observed for a given set of parameter values depends on the configuration of the random connectivity of the directly simulated networks or on the sequence of the random numbers used for the simulations of the mean-field equations. In the regularly oscillating solutions, we also observe that both the mean activity and the activities of individual neurons are coherent, which means that the activities of individual neurons have various waveforms but are all phase-locked to the same rhythm [Fig.\ref{figure_single_EI_pair_nonzerosum}(i)].

\subsubsection{Strong dependence on the configuration of the random connectivity}
\begin{figure}
\includegraphics[width=86mm]{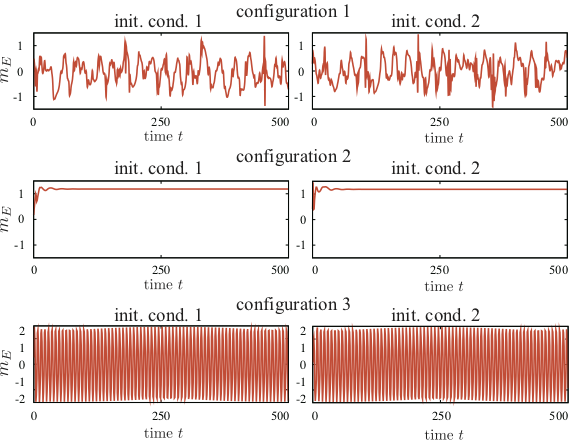}
\caption{Direct simulations of networks with three different configurations of the unadjusted random connectivity for the same parameter values, $(\sigma _0, g_0)=(1.2,4)$, for two different initial conditions. The mean activities of the excitatory populations of the three networks are plotted. The three networks consistently showed irregular, static, and regularly oscillating dynamics, respectively, in simulations from different initial conditions.} \label{figure_symmetry_breaking}
\end{figure}

Next, we examine in further detail the three qualitatively different solutions for larger values of $g_0$. Here, we emphasize that the type of the observed dynamics depends on the configuration of the random connectivity but not on the initial condition of the simulations. Fig.\ref{figure_symmetry_breaking} plots the activity patterns of three networks with the same parameter values but different configurations. We find that a network with the same configuration shows dynamics convergent to the same attractor when it is simulated from different initial conditions, while those with the same parameter values but different configurations show various dynamics. Such individuality among networks with different configurations was expected from the outlier eigenvalues of the synaptic weight matrices \cite{PhysRevLett.97.188104, Tao2013, Stern2016}, which we also confirm numerically [Appendix \ref{appendix:eigenvalues_of_weight_matrices}]. The outlier eigenvalues in the synaptic weight matrices indicate that the untuned model has a strong configuration dependence at the level of its dynamical equation.

Despite this obvious configuration dependence, our mean-field equations reproduce activities similar to those of the directly simulated networks. We therefore expect the MFT to give us further insights into the configuration-dependent dynamics, and we examine this point below.

\subsubsection{Fixed-point solutions and their stability} \label{perturbative_analysis_fp}
We first examine the observed fixed-point solutions. Suppose that the activity of the entire network is constant, with mean activity $m(t)\equiv m_{\mathrm{f}}$. Recall that the dynamics of the mean activity are described by Eq.(\ref{dynamics_of_m}), rewritten here as 
\begin{eqnarray}
\frac{\mathrm{d}}{\mathrm{d} t} m(t)&=&-m(t)+\sqrt{2}g_0\eta (t),\label{macroscopic_equation_fp_analysis}
\end{eqnarray}
where the fluctuation term $\eta (t)$ is generated according to Eq.(\ref{realization_probability_eta}). If the network state remains at a fixed-point, the correlation function of the microscopic fluctuations, $\widetilde{C}(t,s)$, is constant, and therefore, neuronal activities take normally distributed values that do not change temporally. Then, the fluctuation term $\eta (t)$ does not change temporally either, because it is the sum of the microscopic neuronal fluctuations [Eq.(\ref{realization_probability_eta})]. From Eq.(\ref{macroscopic_equation_fp_analysis}), we find that
\begin{eqnarray}
m(t)\equiv \sqrt{2}g_0\eta (t)\equiv m_{\mathrm{f}}, \label{macro_consistency_fp}
\end{eqnarray}
must hold in order for the network state to remain at the fixed point without requiring an external input. Applying the MFT, we find that there is a continuous band of values for $m_{\mathrm{f}}$ for which the stable solution for $\widetilde{C}(t,s)$ is constant. From that analysis, we expect that a solution satisfying Eq.(\ref{macro_consistency_fp}) exists with a non-zero probability [also see the discussion at the end of Appendix \ref{perturbative_expansion_fixed_point}]. The existence of this band suggests the mechanism for the appearance of fixed-point solutions as follows: When the mean activity stays in this band, $\widetilde{C}(t,s)$ tends to be constant, and hence, the microscopic fluctuations slow down. As a result, the mean activity receiving driving forces from the microscopic fluctuations also slows down. This leads to the convergence of the entire dynamics to an equilibrium point.

This scenario is justified by a perturbative stability analysis. In this analysis, we examine the response of the system around a fixed-point solution to a temporary external perturbative input. Suppose that Eq.(\ref{macro_consistency_fp}) holds and that the network state is set to a fixed-point solution with mean activity $m(t)=m_{\mathrm{f}}$ for a long time prior to $t=0$. Then, suppose that temporary external inputs, collectively denoted by $\mathbf{p}$, are applied in $t>0$. For $t\leq 0$, the self-consistent equation, Eq.(\ref{self_consistent_eq_2}), reads 
\begin{eqnarray}
D_0=2\sigma _0^2\widetilde{C}_0, \  \  \  
\widetilde{C}_0=\int \mathrm{d}\mathcal{N} (z)\phi (\sqrt{D_0}z+m_{\mathrm{f}})^2, \label{self-consistent_equation_fp}
\end{eqnarray}
with the variance of $\delta h_i^{(k)}$, denoted by $D_0$, and with the mean square of $\phi (h_i^{(k)})$, denoted by $\widetilde{C} _0$. Here, $\mathcal{N}(z)$ denotes a unit Gaussian distribution. The condition for the stability of a static solution to Eq.(\ref{self-consistent_equation_fp}) is given by
\begin{eqnarray}
1-a_1-2a_2>0&,& \ \ \ a_1<1,\ \ \ \mathrm{with} \nonumber \\
a_1=2\sigma _0^2\langle \phi ^{\prime }(h_i^{(k)})^2\rangle _0&,&\  \  a_2=\sigma _0^2\langle \phi ^{\prime \prime }(h_i^{(k)})\phi (h_i^{(k)})\rangle _0,  \label{microscopic_stability}
\end{eqnarray}
[see Appendix \ref{perturbative_expansion_fixed_point} for the derivation]. Here, the angle brackets with subscript $0$ denote averaging over the unperturbed dynamics with $m(t)\equiv m_{\mathrm{f}}$.

For $t\geq 0$, we perturbatively expand the dynamics around the fixed-point solution. We calculate how a change in the mean activity, $\delta m(t)=m(t)-m_{\mathrm{f}}$, evokes a response in the correlation $D(t,s)$ and how the evoked response in the correlation generates additional fluctuations in $\eta (t)$. We refer readers interested in the details of this analysis to Appendix \ref{perturbative_expansion_fixed_point}. From this analysis, we find that up to the first order, a self-consistent equation of the following form---with {\it i.i.d.} unit-Gaussian coefficients $\xi _{j\ell }$ and $\xi _{j\ell }^{\prime}$ determined by the configuration of the random connectivity---must be satisfied: 
\begin{eqnarray}
\delta m(t)&=&(1+\partial _t)^{-1}p_0(t)+m_{\mathrm{f}}d_1[\delta m](t)\nonumber \\
&&+\frac{g_0}{\sigma _0}\left \{ \sum _{1\leq j\leq 3,1\leq \ell <\infty }\xi _{j\ell }d_{2,j\ell }[\delta m](t)\right. \nonumber \\
&&\left. +\sum _{j,1\leq \ell <\infty }\xi _{j\ell }^{\prime }d_{3,j\ell }[\mathbf{p}](t)\right \} +O(|\mathbf{p}|^2). \label{series_solution_fp}
\end{eqnarray}
Here, the term $p_0$ is the component of the perturbative input that is uniformly applied to all neurons, and the terms $d_1[\delta m]$, $d_{2,j\ell }[\delta m]$, and $d_{3,j\ell}[\mathbf{p}]$ are certain linear transformations of $\delta m$ or $\mathbf{p}$, respectively. The operation denoted by $(1+\partial _t)^{-1}$ is defined as
\begin{eqnarray}
(1+\partial _t)^{-1}x(t)\overset{\mathrm{def}}{=}\int _{-\infty }^te^{-(t-\tau )}x(\tau )\mathrm{d} \tau .
\end{eqnarray}

The solution of this self-consistent equation can be obtained explicitly. From this solution, we find that if we have 
\begin{eqnarray}
m_{\mathrm{f}}\theta _1<1, \ \ \ a_1<1, \label{macroscopic_stability}
\end{eqnarray}
for constant $\theta _1$ calculated from the unperturbed dynamics [see Eq.(\ref{theta_1_definition}) for the details], $\delta m(t)\rightarrow 0$ holds with a non-zero probability as $t\rightarrow \infty $, depending on the values of $\xi _{j\ell }$, and hence, on the random weight configuration. This convergence of the mean activity, together with the microscopic stability given by Eq.(\ref{microscopic_stability}), implies the stability of the entire dynamics around the fixed point, and hence, justifies the scenario with the slowing-down of both the mean activity and the microscopic fluctuations. From the same analysis, we also see that, depending on the values of $\xi _{j\ell }$, the obtained solution converges for a short time after the application of the input but eventually diverges.

\begin{figure}
\includegraphics[width=65mm]{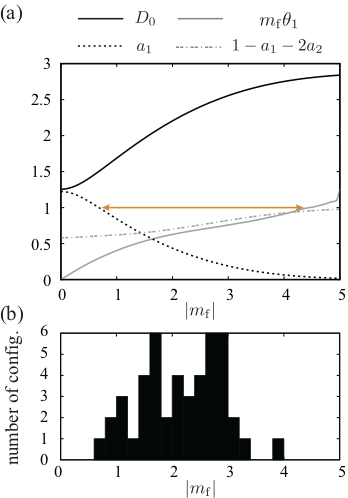}
\caption{(a)Values of the constants that appear in the stability condition for a fixed-point solution with $m(t)=m_{\mathrm{f}}$ and $\sigma _0=1.2$. The values of the constants do not depend on the value of $g_0$. The stability condition is satisfied for the range of $m_{\mathrm{f}}$ indicated by the double-ended orange arrow. (b)The histogram of (the absolute values of) the mean activity of the fixed-point solutions obtained by directly simulating one hundred networks in the same setting as Fig.\ref{figure_single_EI_pair_nonzerosum}(e) except for different weight configurations.} \label{figure_stability_condition_fp}
\end{figure}

In summary, we find that the model has fixed-point solutions for which the mean and variance of the neuronal activities are determined in a configuration-dependent manner [Eqs.(\ref{realization_probability_eta}), (\ref{macro_consistency_fp}), and (\ref{self-consistent_equation_fp})]. Then, even for fixed-point solutions with the same statistics of neuronal activities, we find that their stability depends on the individual configurations. For the fixed-point solution for which Eqs.(\ref{microscopic_stability}) and (\ref{macroscopic_stability}) are satisfied, configurations that yield stable fixed points exist with a non-zero probability. To check the stability of the numerically observed fixed points, we compute the values of $m_{\mathrm{f}}\theta _1$, $D_0$, $a_1$, and $1-a_1-2a_2$ for different values of $m_{\mathrm{f}}$ [Fig.\ref{figure_stability_condition_fp}(a)].  
We find that the conditions for the stability are actually satisfied for a certain range of $m_{\mathrm{f}}$ [orange arrows in Fig.\ref{figure_stability_condition_fp}(a)]. We find that all fixed points observed in numerical simulations, including those shown in Fig.\ref{figure_single_EI_pair_nonzerosum}, fall in this range of $m_{\mathrm{f}}$ [Fig.\ref{figure_stability_condition_fp}(b)]. This contrasts with fixed points that are occasionally observed for the finely tuned model with large $g_0$. Fixed points of that model never satisfy the corresponding stability condition. This implies that fixed points do not exist in the thermodynamic limit. In the above analysis, we find that the condition given by Eqs.(\ref{microscopic_stability}) and (\ref{macroscopic_stability}) itself does not depend on the values of $g_0$, while the probability of realizing stable fixed points does depend on it. For small $g_0$, the values of $\eta (t)$ that satisfy Eq.(\ref{macro_consistency_fp}) with a value of $m_{\mathrm{f}}$ within the range of stability is very large. The realization probability of such a large value of $\eta (t)$ is expected to be small from Eq.(\ref{realization_probability_eta}). This explains the reason that we do not observe fixed points for a very small $g_0$ and also suggests that stable fixed points still exist, albeit with a very small probability, for such small $g_0$.

\subsubsection{Regularly oscillating solutions and their stability} \label{perturbative_analysis_osc}
\begin{figure*} 
\includegraphics[width=178mm]{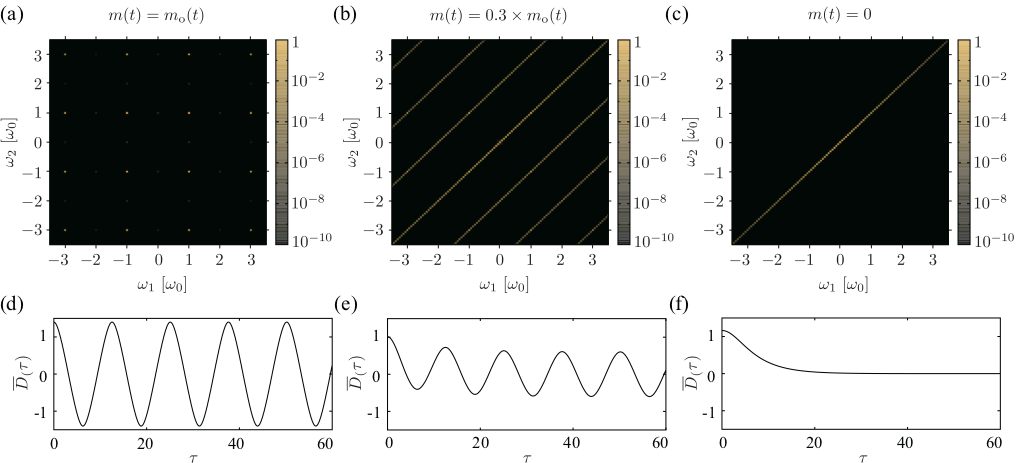}
\caption{(a)--(c) Fourier-transformed autocorrelation function $\widehat{D} (\omega _1,\omega _2)$ for the mean activity indicated above each panel. The times series $m_{\mathrm{o}}(t)$ is the  mean activity shown in Fig.\ref{figure_single_EI_pair_nonzerosum}(g), which oscillates regularly with the basic frequency $\omega _0$. The values of $\widehat{D} (\omega _1,\omega _2)$ are calculated over the frequency domain $-256\omega _0\leq \omega _1,\omega _2\leq 256\omega _0$ discretized into bins for which the frequency values are represented by multiples of $\omega _0/16$. We plot the absolute values of the calculated $\widehat{D} (\omega _1,\omega _2)$ for $-3.5\omega _0\leq \omega _1,\omega _2\leq 3.5\omega _0$ in a logarithmic color scale. The two argument frequencies, $\omega _1$ and $\omega _2$, are shown in units of the basic frequency $\omega _0$. (d)--(f) The averaged autocorrelation function $\overline{D} (\tau )$ for the three cases in the upper panels.} \label{figure_autocorrelation_oscillation}
\end{figure*}
Next, we examine the regularly oscillating solutions. If the entire network dynamics have stable oscillations, so do the microscopic parts of the dynamics. To find such microscopic oscillations for a given oscillatory orbit of the mean activity, $m(t)=m_{\mathrm{o}}(t)$, we solve the self-consistent equation, Eq.(\ref{self_consistent_eq_2_untune}), which we rewrite as
\begin{eqnarray}
(1+\partial _t)(1+\partial _s)D(t,s)=2\sigma _0^2\widetilde{C} (t,s). 
\end{eqnarray} 
This equation can be solved iteratively in the frequency domain [see Appendix \ref{perturbative_expansion_oscillation}]. Using the mean activity observed in Fig.\ref{figure_single_EI_pair_nonzerosum}(g) for $m_{\mathrm{o}}(t)$, we compute the autocorrelation $D(t,s)$ and show its magnitudes in the frequency domain [Fig.\ref{figure_autocorrelation_oscillation}(a)]. We find that the solution has a non-zero value only for multiples of the basic frequency $\omega _0$ of the mean activity, which indicates that the microscopic dynamics are completely entrained by the oscillatory mean activity. We also confirm this by checking the following averaged autocorrelation function calculated from $D(t,s)$: with $T_0=2\pi /\omega _0$,
\begin{eqnarray}
\overline{D} (\tau )\overset{\mathrm{def}}{=}\frac{1}{T_0}\int _{0}^{T_0}D(t,t+\tau )\mathrm{d} t.
\end{eqnarray}
The autocorrelation function in Fig.\ref{figure_autocorrelation_oscillation}(d) shows that the neuronal variables are periodic. In contrast, as is well known from a previous study \cite{PhysRevLett.61.259}, the autocorrelation function $D(t,s)$ for zero mean activity, $m(t)\equiv 0$, has the frequency representation, $\widehat{D} (\omega _1,\omega _2)=\widehat{D_0}(\omega _1)\delta (\omega _1-\omega _2)$, with a continuous function $\widehat{D} _0(\omega _1)$ [Fig.\ref{figure_autocorrelation_oscillation}(c)]. The averaged autocorrelation function for this case is unimodal and tends to zero as $\tau \rightarrow \infty $ [Fig.\ref{figure_autocorrelation_oscillation}(f)]. The qualitative difference between these two autocorrelation functions suggests the occurrence of a phase transition from one to the other. To check this, we decrease the amplitude of the mean activity, $m_{\mathrm{o}}(t)$, without changing its waveform, and we find that the solution starts to have a continuous spectrum extending over frequencies other than the multiples of $\omega _0$. After the transition, the autocorrelation function has the form of $\widehat{D} (\omega _1,\omega _2)=\sum _k\widehat{D} _k(\omega _1)\delta (\omega _1-\omega _2-k\omega _0)$ [Fig.\ref{figure_autocorrelation_oscillation}(b)] and the averaged autocorrelation function has both a periodic component and a component that vanishes at infinity [Fig.\ref{figure_autocorrelation_oscillation}(e)]. As the amplitude of the mean activity decreases, the periodic component in the autocorrelation function gradually decays, disappearing at $m(t)=0$. The analysis we perform in Appendix \ref{perturbative_expansion_oscillation} actually shows that these observed entrained dynamics are stably realized. The transition behavior observed above is qualitatively the same as that observed in a previous study \cite{PhysRevE.82.011903}, although that study used periodic inputs with random phases to induce the transition.

This microscopic transition gives an intuitive explanation for the mechanism of the observed oscillations. Recall that the dynamics of the mean activity is described by Eq.(\ref{dynamics_of_m}), rewritten here as 
\begin{eqnarray}
\frac{\mathrm{d}}{\mathrm{d} t} m(t)&=&-m(t)+\sqrt{2}g_0\eta (t).
\end{eqnarray}
Because the driving-force term, $\eta (t)$, is the sum of the microscopic fluctuations, it is entrained to the oscillation of the mean activity itself, if the mean activity oscillates with a sufficiently large amplitude, to entrain the microscopic fluctuations. We suggest this reverberation of entrainment between the mean activity and the microscopic fluctuations as the mechanism underlying the coherent oscillations we observe in Fig.\ref{figure_single_EI_pair_nonzerosum}(g) and (h).

The stability of this reverberation mechanism can be examined by using a perturbative method similar to that employed for fixed points: we derive a self-consistent equation with random coefficients that determines linear responses to external inputs, and construct its solution from which the condition for the stability of the reverberation can be examined numerically. From this analysis, we draw the same conclusion about the stability as that for the fixed points: regularly oscillating solutions for the untuned model, such as that observed in Fig.\ref{figure_single_EI_pair_nonzerosum}(g), are linearly stable with a non-zero probability; occasionally observed regular oscillations of the finely tuned model [Fig.\ref{zerosum_autonomous_oscillation}(c)], however, turn our to be unstable. These conclusions are consistent with the tendency observed in the results of direct simulations of networks with different system sizes [Fig.\ref{size_dependence}(a),(b)]. Since this analysis is complicated and essentially the same as that for fixed points, we omit its presentation here and refer interested readers to Appendix \ref{perturbative_expansion_oscillation}. We only note that we cannot examine exhaustively oscillatory orbits, and therefore we cannot completely exclude the existence of stable limit-cycle solutions for the finely tuned model. 
\subsection{External inputs to $\sqrt{N}$ excitatory neurons} \label{subsection4B}
\begin{figure*} 
\includegraphics[width=178mm]{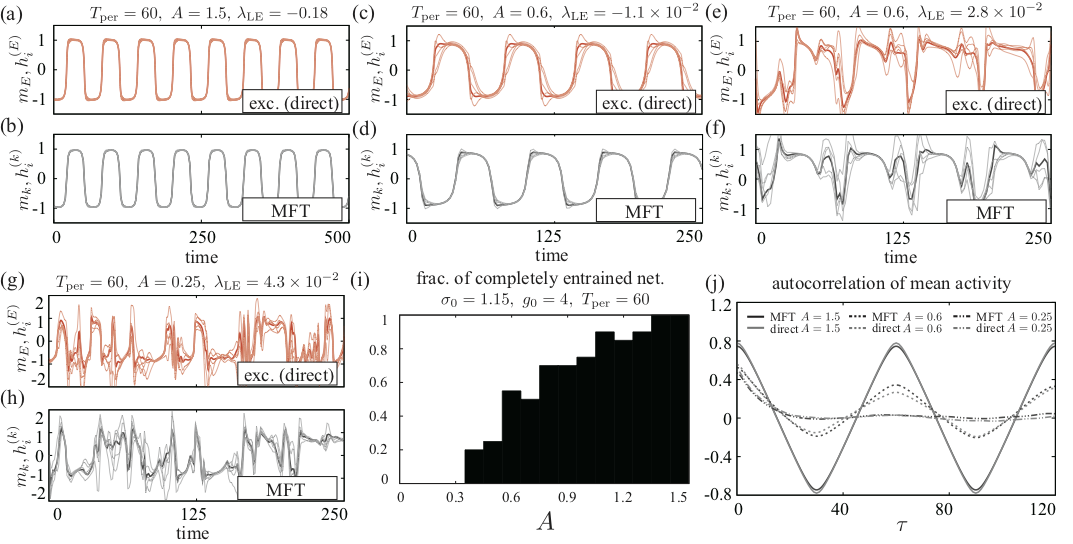}
\caption{(a)--(h) Typical activity patterns for the finely-tuned model with $(\sigma _0, g_0)=(1.15, 4)$ under sinusoidal external inputs with the indicated values of $A$ and $T_{\mathrm{per}}$. In each plot, the thick and thin lines represent the mean activity $m_k$ and the activities $h_i^{(k)}$ of five representative neurons, respectively. For the direct simulations, the mean activity $m_k$ is approximated by the average of $h_i^{(k)}$ over the population $k$ comprising $N=10240$ neurons (see Appendix \ref{appendix:D} for the details of the simulations). Solutions for the $E$ population in direct simulations and solutions of the mean-field equations are depicted in red and gray, respectively. The plots for the $I$ population are omitted. Qualitatively different dynamics are observed for the same parameter values [(d)--(g)], depending on the configuration of the random connectivity of the directly simulated networks or on the sequence of random numbers used for the simulations of the mean-field equations. (i) For each value of $A$, the percentage of twenty directly simulated networks with random weight configurations that synchronize with the sinusoidal inputs is plotted as a histogram. (j) Autocorrelation function of the mean activity $\mu _1(\tau )$ for $(\sigma _0, g_0)=(1.15, 4)$ and for the indicated values of the amplitude $A$ averaged over 15 different configurations or 15 series of random numbers and over the period $1000\leq t\leq 2500$ for both direct simulations and solutions of the mean-field equations.} \label{figure_external_input}
\end{figure*}

\subsubsection{Coherent states induced by sinusoidal inputs}
In this section, we apply sinusoidal inputs of amplitude $A$ and period $T_{\mathrm{per}}$ to $\sqrt{N}$ neurons in the $E$ population of the finely tuned model, namely, 
\begin{eqnarray}
\widetilde{I} (t)=A\sin (2\pi t/T_{\mathrm{per}}) \label{sinusoidal_input}
\end{eqnarray}
in Eq.(\ref{modelling_sparse_input}). In this model setting, we observe two qualitatively different types of behavior [Fig.\ref{figure_external_input}(a)--(h)]. We find that the activity patterns obtained from the direct simulations and from the MFT are quite similar, suggesting that our theory successfully predicts the behavior of the model for the case with external inputs as well. As we increase the amplitude $A$ for a fixed value of $g_0$, the solutions undergo a transition from irregular chaotic dynamics partially entrained by the input to regular non-chaotic dynamics synchronous with the input. This indicates that inputs to an $O(\sqrt{N})$ number of neurons can effectively entrain the whole network in this model.

We further observe that this transition occurs at different values of $A$, depending on the configuration, but not on the initial condition, similarly to Fig.\ref{figure_symmetry_breaking} (not shown). Fig.\ref{figure_external_input}(i) shows a histogram depicting the percentage of twenty networks with random configurations that synchronize with the inputs for each value of $A$. We see that the transition point is highly variable among networks with different configurations. Nevertheless, in the autocorrelation function $\mu (\tau )$ of the mean activity averaged over configurations and time according to the definition in Eq.(\ref{autocorrelation_mean}), we observe good agreement between direct simulations and MFT [Fig.\ref{figure_external_input}(j)].

\subsubsection{Stability of the entrained dynamics}
The stability of the regular, entrained dynamics can be examined using the same perturbative method as that for the regularly oscillating solutions of the model without external inputs. In this case, the results of the analysis indicate that the numerically observed oscillations of the model are linearly stable [see Appendix \ref{perturbative_expansion_oscillation}]. Consistent with this finding, in the direct numerical simulations, the induced coherent states are robustly observed for networks with large system sizes [see Appendix \ref{system_size_dependence}]. In contrast with the untuned model, in the present case, the synaptic weight matrix of the model does not have configuration-dependent outlier eigenvalues. We observe, however, that the linear variational equation around the attractors does have coefficient matrices with outlier eigenvalues [see Appendix \ref{appendix:eigenvalues_of_weight_matrices} for details]. These results suggest that the finely tuned model still shows strong configuration-dependence in its stimulus-driven dynamics.
\subsubsection{Reading out information from coherent dynamics} \label{section_readout}
\begin{figure}
\includegraphics[width=86mm]{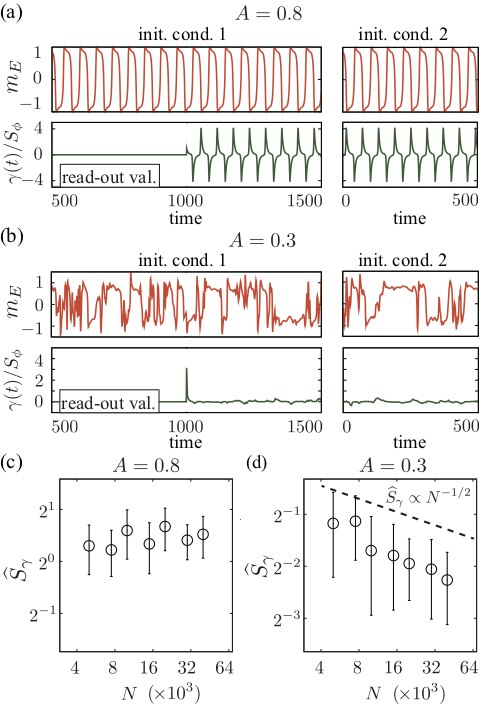}
\caption{(a) and (b) Mean activities of the excitatory population $m_E$ and normalized read-out values $\gamma (t)/S_{\phi }$, as calculated for networks with $(\sigma _0,g_0)=(1.15,4)$ that receive sinusoidal inputs with the indicated values of $A$ and $T_{\mathrm{per}}=60$. The weighting coefficients $\{ r_k^i\} _{k,i}$ are determined at $t_{\mathrm{init}}=1000$ in the simulations from initial condition 1. The mean activities and read-out values of networks from another random initial condition (initial condition 2) are also shown. (c) and (d) Scaling of the normalized standard deviation $\widehat{S} _{\gamma }$ of values read out from networks with the same parameter values as in (a),(b). The averages of $\widehat{S} _{\gamma }$ over fifteen different configurations, together with their standard errors, are plotted on logarithmic scales. Only networks with configurations giving rise to synchronous dynamics (c) or irregular dynamics (d) are analyzed. In (d), a straight line with slope $-1/2$, representing $\widehat{S} _{\gamma }\propto N^{-1/2}$, is also shown.} \label{figure_readout_with_regular}
\end{figure}

Fig.\ref{figure_external_input}(a)--(d) show that the mean activities and individual neuronal activities are coherent in dynamics synchronous with the inputs. This has a computational implication. Suppose that we read out microscopic fluctuations of these networks by taking a weighted average with $O(1/N)$ weighting coefficients, as described by the following equation:
\begin{eqnarray}
\gamma (t)=\sum _{k,i}\frac{r_k^i}{N} (\phi (h_i^{(k)}(t))-\check{\phi } _k(t)). \label{reading_out_equation}
\end{eqnarray}
Here, $\check{\phi } _k$ denotes the population average of $\phi (h_i^{(k)}(t))$ over the population $k$. In the above equation, the contribution of this population average is subtracted. This is because the population-averaged activity simply replicates the external inputs, and reading out this component does not have much computational value.

If the dynamics are coherent, we expect the read-out values, $\gamma (t)$, to be $O(1)$. In contrast, if the dynamics are chaotic, the ensemble of neuronal activities can be regarded as an incoherent Gaussian fluctuations, and therefore, values read out from them are expected to be $O(1/\sqrt{N} )$. We numerically test this hypothesis by examining the values read out with the following weighting coefficients:
\begin{eqnarray}
r_{k}^i=\left \{ 
\begin{array}{cc}
1 & \mathrm{if}\   \phi (h_i^{(k)}(t_{\mathrm{init}}))-\check{\phi } _k(t_{\mathrm{init}}) \geq 0\\
-1 & \mathrm{otherwise} 
\end{array}
\right. .  \label{determining_ro_coefficients}
\end{eqnarray}
In this equation, we set the coefficients to such values that the initial value of $\gamma (t)$ at time $t_{\mathrm{init}}$ is O(1). We show the network activities and typical read-out values obtained from them in Fig.\ref{figure_readout_with_regular}(a) and (b). The values read out from the coherent regular oscillation show a regular pattern of magnitudes comparable to the initial value, $\gamma (t_{\mathrm{init} })$, [Fig.\ref{figure_readout_with_regular}(a)], while those read out from the irregular activity decay rapidly from the initial value [Fig.\ref{figure_readout_with_regular}(b)]. This observation is consistent with the above argument. To evaluate the magnitudes of the read-out values further, we calculate the following normalized standard deviation, $\widehat{S} _{\gamma }$, of $\gamma (t)$ for networks of different system sizes: 
\begin{eqnarray}
&&\hspace{-0.3cm}S_{\gamma }\overset{\mathrm{def}}{=} \left \{ \frac{1}{T_{\mathrm{avg}}} \int _{t_{0}}^{t_0+T_{\mathrm{avg}}}\hspace{-0.5cm} (\gamma (t)-\overline{\gamma })^2 \mathrm{d} t\right \}^{1/2}\hspace{-0.5cm} ,\    \    \overline{\gamma }\overset{\mathrm{def}}{=}\frac{1}{T_{\mathrm{avg}}}\int _{t_{0}}^{t_{0}+T_{\mathrm{avg}}}\hspace{-0.5cm} \gamma (t)\mathrm{d} t. \nonumber \\
&&\hspace{-0.3cm} \widehat{S} _{\gamma }\overset{\mathrm{def}}{=} \frac{S_{\gamma } }{S_{\phi }} ,\    \    S_{\phi }\overset{\mathrm{def}}{=} \left \{ \frac{1}{T_{\mathrm{avg}}} \int _{t_0}^{t_0+T_{\mathrm{avg}}}\hspace{-0.5cm} \langle (\phi (h_i^{(k)}(t))-\check{\phi } _k)^2\rangle  \mathrm{d} t\right \}^{1/2}\hspace{-0.5cm}.
\end{eqnarray} 
The bracket in this equation indicates the average over the entire network. We also take the average over a long time, of length $T_{\mathrm{avg}}$, starting from a suitably chosen initial time $t_0$, for a simulation that starts from a random initial condition different from that of the simulation for which we have determined the weighting coefficient $r_k^i$. We also show the activity patterns obtained from this initial condition in Fig.\ref{figure_readout_with_regular}(a) and (b). In Fig.\ref{figure_readout_with_regular}(c) and (d), we show the calculated values of the normalized standard deviations on logarithmic scales, and we find that the values read out from regular oscillations do not depend much on the system size, while those read out from irregular dynamics are roughly proportional to $1/\sqrt{N}$, as we expect. These results suggest that the above mechanism for reading out $O(1)$ values only when the network dynamics are coherent enables neuronal networks to transmit information in a state-dependent manner. Note that we have repeatedly read out the same pattern from the coherent dynamics, regardless of the independent initial conditions [Fig.\ref{figure_readout_with_regular}(a)]. Regardless of the symmetry among the neurons that receive inputs through statistically the same set of synaptic weights, identical coherent dynamics---not coherent dynamics randomly reshuffled with respect to the neuronal indices---are always realized. Although the above coherent states are induced by artificial sinusoidal input, similar results are obtained for the case with irregular input [see Appendix \ref{appendix:irregular_iinput} for details]. 
\subsubsection{Remarks on the untuned model under external inputs}
The untuned model behaves in a qualitatively similar manner to the finely tuned model when both are driven by external inputs. The untuned model also shows transitions from irregular, partially entrained dynamics to regular dynamics that are completely synchronous with external inputs in a configuration-dependent manner. Thus, to avoid redundancy, we do not present the results for this model setting in this article. From a quantitative viewpoint, we note that the entrainment in this model setting is more complicated than that for the finely tuned model, presumably because the untuned model has inherent configuration-dependent rhythms, as observed in Fig.\ref{figure_single_EI_pair_nonzerosum}. 
\subsection{Multiscale dynamics of critically balanced networks of LIF neurons} \label{section_spiking}
\begin{figure*} 
\includegraphics[width=178mm]{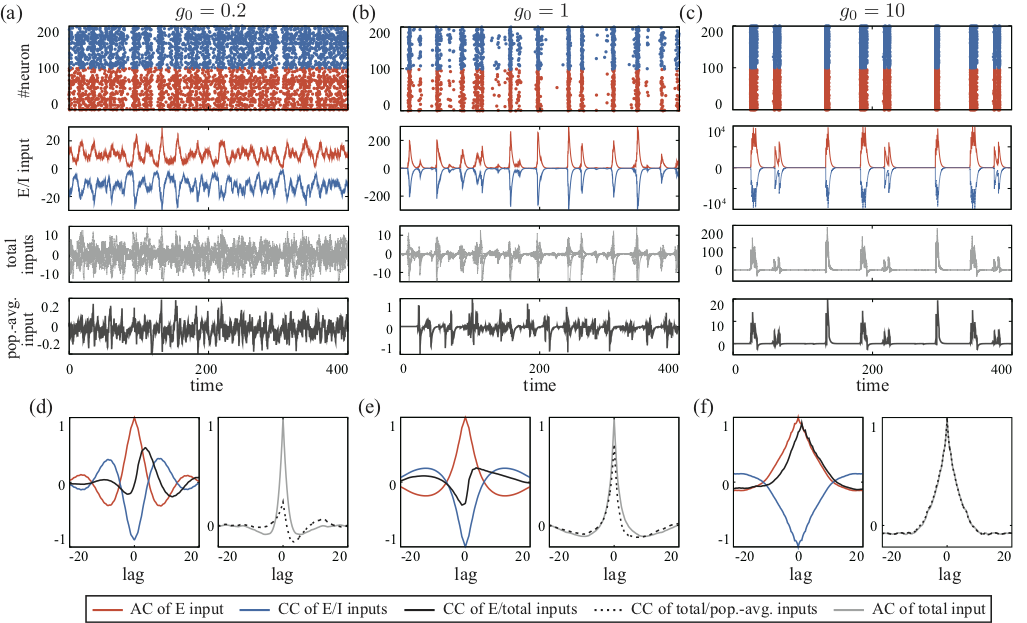}
\caption{(a)--(c): Multiscale dynamics obtained by numerically simulating a network of LIF neurons with $g_0=0.2, 1.0, 10.0$, respectively. The top panels are raster plots of firing activities of one hundred representative neurons from the excitatory (red) and inhibitory (blue) populations. The upper middle panels show the total excitatory (red) and inhibitory (blue) inputs to a single representative excitatory neuron. The lower middle panels show total inputs to five representative excitatory neurons (gray). The bottom panels show the population average of the inputs to all neurons. (d)--(f): The autocorrelation functions (ACs) of the excitatory input to a representative neuron and of the population average of the inputs to all neurons, and the cross-correlations functions (CCs) of those inputs and another fraction of the input to the representative neuron. The quantities shown are the AC of the excitatory input to the neuron (red); the CC of the excitatory and inhibitory inputs to the neuron (blue); the CC of the excitatory input to the neuron and the population-averaged input (black); the CC of the total input to the neuron and the population-averaged input (black dotted); and the AC of the total input to the neuron. The following parameter values are used for the simulations: $\tau _m=20, t_{\Delta }=0.5, t_{\mathrm{ref}}=0.55, V_{\mathrm{rest}}=10, V_{\mathrm{reset}}=10, V_{\mathrm{ths}}=20$.} \label{figure_spiking}
\end{figure*}

Thus far, we have focused on a highly simplified model with firing-rate variables. However, from way we have constructed our MFT, we expect that in principle a similar theoretical framework will hold for networks of spiking neurons (and hence we expect similar multiscale dynamics to those observed in the rate model). To demonstrate this, we numerically examine a commonly investigated network of leaky-integrate-and-fire (LIF) neurons (see e.g. \cite{ostojic2014two}) described by the following equation: for $k=E,I$, 
\begin{eqnarray}
\tau _m\frac{\mathrm{d}}{\mathrm{d} t}V_i^{(k)}(t)&=&-V_i^{(k)}(t) \nonumber \\
&+&\hspace{-0.5cm} \sum _{\ell=E,I, 1\leq j\leq N, n}\hspace{-0.5cm} J_{k\ell }^{ij}\delta (t-t_{j,n}^{(\ell)}-t_{\Delta}), \label{LIF_equation}
\end{eqnarray} 
In this equation, the variables $V_i^{(k)}$ and $t_{i,n}^{(k)}$ denote the membrane potential and the $n$-th spike time of the $i$-th neuron in the $k$ population, respectively. If $V_i^{(k)}$ exceeds a threshold potential $V_{\mathrm{ths}}$, a spike is emitted from the neuron, and $V_i^{(k)}$ is reset to $V_{\mathrm{reset}}$ and held for a time of length $t_{\mathrm{ref}}$. The constants $\tau _m$ and $t_{\Delta}$ represent the membrane time constant and the delay of synaptic transmissions, respectively. The synaptic weights $J_{k\ell} ^{ij}$ are given by Eq.(\ref{basic_equation_nonzerosum}) or (\ref{basic_equation}) subject to the condition in Eq.(\ref{pedagogical_g}). For this model, following the same argument as that for the simplified model [Sec.\ref{sec:section3}], fluctuations in the inputs to the neurons in the network are considered to be conditionally Gaussian, given the orbit of the population-averaged input to the neurons. Since the population-averaged input is given by the sum of the fluctuating inputs to individual neurons divided by $\sqrt{N}$, the population-averaged input is stochastic and its realization probability is determined through the correlation of the microscopic fluctuations. The neuron model used in the above, however, is highly nonlinear, and thus solving the mean-field equations demands much more intensive numerics than those we presented in the previous sections. Thus, we restrict ourselves to numerically simulating the model and examining whether similar multiscale dynamics arise intrinsically.

As the value of $g_0$ is increased from zero with the above model settings, we actually observe increasingly large fluctuations in the population-averaged activity [Fig.\ref{figure_spiking}]. Examining the autocorrelation and cross-correlation functions of individual and population-averaged inputs to the neurons, we observe that the population-averaged input is significantly correlated with the excitatory, inhibitory, and overall inputs to individual neurons [Fig.\ref{figure_spiking}(d)--(f)]. As we increase the value of $g_0$, these dynamics undergo a transition to coherent dynamics in a configuration-dependent manner [Fig.\ref{figure_spiking}(c) and (f)]. These transitions are observed robustly for different values of $N(\gg 1)$ [Fig.\ref{size_dependence} in Appendix \ref{system_size_dependence}]. Similar transitions are induced by external inputs to $\sqrt{N}$ neurons [Fig.\ref{stimulus-driven_LIF} in Appendix \ref{system_size_dependence}]. The successful prediction of the occurrence of multiscale dynamics indicates that the multiscale dynamics revealed by our MFT generally emerge in a variety of neuronal networks at critical parameter values. 

\section{Discussion} \label{sec:section5}
In the present study, we developed a novel type of MFT for RNNs consisting of a pair of excitatory and inhibitory populations of simplified neurons with finely-tuned or untuned synaptic weights that obey Dale's law. The mean strengths of the synaptic weights were assumed to take a set of critical values. In this theory, microscopic fluctuations in the neuronal activities that are amplified by the strong excitation and inhibition serve as driving forces for the macroscopic dynamics of the population activity, while the population activity constrains the statistics of the microscopic fluctuations. The investigated RNNs exhibited non-vanishing fluctuations in their population activity. When the magnitudes of excitation and inhibition were large, we found interesting dynamical properties in these fluctuations, such as high non-Gaussianity and asymmetry with respect to time reversal for the finely-tuned model, and strongly configuration-dependent transition to a static or oscillatory non-chaotic state for the untuned model. In the oscillatory state, neuronal activities have various waveforms while they are all phase locked to the rhythm of the population activity. Our theory successfully predicts these dynamical properties. We found that these multiscale dynamics occur at a critical point between extremely strong ferromagnetic and anti-ferromagnetic states.

In networks with external inputs, periodic inputs to an $O(\sqrt{N} )$ number of excitatory neurons effectively entrained the whole network. As the amplitude of the inputs was increased, the networks underwent a transition from irregular, partially entrained dynamics to regular dynamics synchronous with the input; the transition point again depended on the configuration. Unlike the autonomous case, the application of external inputs induced a transition to a coherent oscillation in the finely-tuned model, indicating that the fine-tuning of the synaptic weights reduces but does not remove the configuration dependence from the network dynamics. We also showed numerically that the induced coherent dynamics can be used as media for transmitting information in a state-dependent manner.

Furthermore, based on analogy, more biologically realistic networks of spiking neurons are expected to display similar multiscale dynamics. Although theoretical prediction of their dynamics is much more computationally demanding and beyond the scope of the current study, we numerically confirmed this hypothesis for networks of LIF neurons. 

\subsection{Closely related results}
The present study was largely inspired by a previous investigation of RNNs \cite{PhysRevX.5.041030} and by a couple of published and unpublished studies on the same model as ours \cite{PhysRevE.88.042824, Stern2016}. In the former study \cite{PhysRevX.5.041030}, the authors investigated a network with a balance between strong excitation and inhibition, providing a theoretical framework for dealing with a network with multiple neuronal populations and for analyzing its non-trivial fixed points and transitions to chaos. Our study used their theoretical framework as a starting point to analyze further the non-trivial population dynamics that they had not analyzed. In the latter published study \cite{PhysRevE.88.042824}, the authors numerically analyzed the same network as ours and found similar oscillatory dynamics. They further analyzed the dynamics with approximate reduced equations and related them to the eigenvalues of the connectivity matrices. Although these results were quite inspiring, their approach---focusing on the eigenvalues with the largest real parts---was not always sufficient to characterize behavior of the model. It is known that the eigenvalues of the connectivity matrices of their networks with finely tuned weights in the $N\rightarrow \infty $ limit are uniformly distributed over a disk \cite{Tao2013} [Fig.\ref{figure_outlier_eigenvalues_weight_matrix}(b)]. This implies that it is difficult to select a single eigenspace that effectively determines the dynamics in this limit, on which their theory relied. The unpublished study \cite{Stern2016} also took a similar approach for the same model and came to a pessimistic conclusion about the usefulness of an MFT in this setting. Sometime after we initially publicized the present results, a very recent work \cite{Landau_Sompolinsky} examined the case with $g_0\approx 0,\infty $ for our networks. Those authors developed a perturbative mean-field theory combined with the approach of \cite{PhysRevE.88.042824}. Although they gained insights into the behavior of the model analytically, their theory still largely relies on heuristic, approximate calculations, and its application is limited to nearly deterministic dynamics with small fluctuations. In contrast with these previous studies, we have here derived an MFT in a much more rigorous manner and have laid a foundation for further analysis. Our theory applies to the entire range of model parameters and gives accurate probabilistic descriptions of large macroscopic fluctuations.

\subsection{Stimulus-induced suppression of chaos and reservoir computing}
Prior to the present study, several authors have theoretically studied the externally driven, non-chaotic dynamics of neuronal networks without balanced excitation and inhibition \cite{PhysRevLett.69.3717, PhysRevE.82.011903, PhysRevX.8.041029}. These studies have shown that the chaoticity of the dynamics of RNNs is suppressed by external random inputs and that a non-trivial non-chaotic regime appears after a transition at some amplitude of the input. In particular, a seminal study \cite{PhysRevE.82.011903} showed that sinusoidal inputs with random phases induce a coherent state similar to ours. The transition to a coherent state in the present model is closely related to this suppression of chaos by stimulus, because the microscopic part of our model dynamics are statistically the same as the dynamics of a model without balanced excitation and inhibition that is suitably driven by a uniform external input, as shown by our MFT. In fact, the autocorrelation functions of neuronal activities of our networks [Fig.\ref{figure_autocorrelation_oscillation}] behave in a similar manner to those observed in the previous study \cite{PhysRevE.82.011903}. However, we note that the transition to this microscopic coherent state due to a uniform external input, not with random phases, has not been well studied to date. Besides the fact that the transition induced by uniform input is more difficult to analyze theoretically, the uniform application of an input often results in chaotic or trivial dynamics, and the transition to a coherent state is not found unless the waveform of the input is finely tuned. In the present model, the waveforms of the mean activity that induce a coherent state are determined by the network itself through the interactions between the microscopic and macroscopic dynamics, even in a case with external inputs. The main difference between the coherent states in the present study and those in previous studies lies in this spontaneity.

The spontaneously discovered coherent states discussed above may have implications for learning with RNNs. In previous studies, learning was first considered in the context of the {\it ``edge of chaos,''} where the variety and stability of network dynamics at the transition point to chaos were exploited in learning \cite{doi:10.1162/089976604323057443, legenstein2007makes, Boedecker2012}. More recent studies have focused on different non-chaotic dynamical phases induced by external or feedback inputs \cite{sussillo2009generating, Laje:2013bd, PhysRevX.8.041029}. In particular, the authors in a seminal work \cite{sussillo2009generating} stably reconstructed desired patterns from the coherent dynamics induced by randomly weighted strong feedback from read-out values to all of the neurons in the network. This strong random global feedback is expected to induce coherent dynamics by a mechanism similar to that studied in \cite{PhysRevE.82.011903} (see also \cite{PhysRevLett.118.018103,PhysRevX.8.041029} for a similar result with strong random global feed-forward input). This requirement for a strong global input, however, restricted the applicability of their framework to the supervised learning of a small number of temporal patterns. Our results suggest a new regime of dynamics, in which non-chaotic coherent dynamics emerge spontaneously and stably reproduce output patterns [Fig.\ref{figure_readout_with_regular}(b)], without being passively entrained by strong global inputs. Investigating learning based on the dynamical phase we have found is a worthy challenge for a future study.

\subsection{Population dynamics and critical fluctuations} \label{Discussion:C}
The relationship between population dynamics and individual neuronal activities has also been studied in previous models. In these studies, however, population dynamics and microscopic fluctuations in individual neurons were treated as statistically independent. Therefore, unlike our theory, none of the previously proposed theories for balanced networks could account for the experimentally observed strong impact of single neurons on the population activities \cite{london2010sensitivity, Chettih:2019wm}. 

In sparsely-connected balanced networks of spiking neurons, population dynamics are unaffected by fluctuations in the irregular firing of individual neurons. In fact. a previous study \cite{Brunel2000} showed that even when individual neurons fire irregularly, the population-averaged activities exhibit regular slow oscillations except for tiny fluctuations due to finite-size effects. This indicates the fact that the irregular firing of neurons exerts only negligible effects on the population dynamics of the sparsely connected networks.

In densely connected balanced networks, stimuli to a small number of excitatory spiking neurons also induce a vanishingly small response in the entire population. Two recent studies investigated the responses of such networks with spatial structures to correlated external inputs applied to a large number [$O(N)$] of neurons \cite{Darshan:iy, Rosenbaum:tn}. It was shown analytically that a non-negligible population response can be induced only when the spatial extent of the input correlation is narrower than that of recurrent connections from a single neuron \cite{Rosenbaum:tn}. This theoretical result should also hold when stimuli are given to a small number [$O(\sqrt{N})$] of excitatory neurons because such stimuli generate correlated internal inputs to the surrounding neurons that are connected with the stimulated neurons. In this case, the previous theory indicates that the population response is negligibly small.

The difference in the impact of single neurons on the population dynamics between the previous models and our model can be understood from the strong ferromagnetic effects examined in Sec.\ref{section_ferromagnetic}. Previous models focused on the dynamical regime in which the network activity is stabilized by strong inhibitory feedback that suppresses excessive excitations. This regime corresponds to the anti-ferromagnetic state we observed in Sec.\ref{section_ferromagnetic}. The anti-ferromagnetic effects strongly suppress the responses of the neuronal population when a small number of neurons are stimulated. In contrast, we have focused on the dynamical regime emerging at the critical point between the ferromagnetic and anti-ferromagnetic states. Activated spontaneously or driven by stimuli to a small number of neurons, our model displays strong macroscopic fluctuations at the critical point. Remarkably, our MFT precisely describes the probabilistic behavior of these critical fluctuations , which was, to our knowledge, difficult for any of the previous MFTs developed in the statistical mechanics of disordered systems. Although the parameter values yielding the critical point are not generic, experiments have shown evidence for self-organized critical dynamics in the brain \cite{COCCHI2017132}, and thus we can reasonably expect some adaptive mechanisms to finely tune the system to the critical point.

The intrinsic origin of the multiscale dynamics may also be supported by the experimentally observed large cross-correlations of EEG/LFPs and individual neuronal activities \cite{poulet2008internal}. EEG and LFPs are considered to reflect mainly a collective excitatory component of synaptic inputs to (apical dendrites of) neurons in the local circuit \cite{ebersole2003current}, and thus they should reflect the waveforms of the very large excitatory inputs to neurons. When strong excitatory and inhibitory inputs to neurons cancel out, the remaining fluctuations do not need to be strongly correlated with the original excitatory and inhibitory inputs, especially if the main driving-force for the population dynamics are extrinsic. In fact, from the multiscale dynamics of the previous model, small cross-correlations of population dynamics and recurrent excitatory input are expected (cf. Fig.1 of \cite{Rosenbaum:tn}). In contrast, our model shows large cross-correlations [Fig.\ref{figure_spiking}(d)--(f)]. From the fact that cortical activity displays switching behavior between states with small and large cross-correlations of EEG/LFPs and neuronal activities \cite{poulet2008internal}, our theory and previous theories are suggested to separately model two different operating regimes of the same cortical circuits.

In our theory, the dynamical nature of the critical fluctuations depends strongly on the detailed configuration of the synaptic connections. On the other hand, accumulating evidence suggests that the connectivity of local cortical circuits is rapidly remodeled \cite{trachtenberg2002long}. Whether the fine connectivity structure has a strong impact on critical fluctuations in cortical network dynamics---and what functional implications such an impact has---need to be further clarified. 

\subsection{Limitations and future extensions of the theory}
Despite the advantages of our theory mentioned above, it is fair to say that the validity of the theory is still restricted by the simplicity of the model settings. One of the most important steps for widening the applicability of our theory is to extend the theory to networks of spiking neurons. In the present study, we gave priority to the analytical tractability and simplicity in numerical simulations, and we restricted ourselves mainly to networks of firing-rate model neurons. However, the extension of our theory to networks of more realistic model neurons should be straightforward. This can be done by regarding balanced inputs to individual neurons as Gaussian fluctuations and by determining the related statistics to ensure consistency with the nonlinear dynamics of single neurons. In this calculation, we can use our method in combination with a previous method \cite{PhysRevE.90.062710} to describe the network dynamics. In the previous study, the mean-field equations for a simple RNN of nonlinear firing-rate units without balanced excitation and inhibition was solved by using the statistics calculated from extensive numerical simulations of single units driven by random forces. Applying the combined method to a critically balanced network of spiking neurons would be computationally demanding but in principle doable.

Although we leave this challenge for future study, we note that the computational costs associated with the approach of \cite{PhysRevE.90.062710} cannot be reduced by commonly employed approximate treatments such as white Gaussian approximation of inputs to neurons \cite{Brunel2000, ostojic2014two}. This is because we must take into account the time-dependent nonlinear interactions between the microscopic neuronal fluctuations and macroscopic population activity underlying the critical multiscale dynamics. These interactions cannot be handled by the approximate treatments. This distinction between a full treatment and an approximate treatment may be related to the recent controversial argument about the transition in networks of spiking neurons from a state with irregular spiking at a constant rate to a state with irregular firing-rate fluctuations, as the mean strength of the synaptic connections is increased \cite{ostojic2014two}. Although the interpretation of this observation based on an approximate description was controversial \cite{Engelken:2016ea}, a full treatment is expected to give an accurate description.

The other simplified aspects of the model include the neglect of different cellular properties of excitatory and inhibitory neurons and of spatial structure of cortical circuits. Although our theory can be extended to include these elements, substantial works will be needed for that. For example, for models with different membrane constants of excitatory and inhibitory neurons, automatic cancellation of excitatory and inhibitory inputs with $\overline{\phi } _E=\overline{\phi } _I$ is not ensured from the condition in Eq.(\ref{pedagogical_g}). However, it is plausible that some feedback mechanism dynamically clamp the population activity to satisfy $\overline{\phi } _E=\overline{\phi } _I$. Then, similar critical multiscale dynamics to those we have observed are expected to emerge. In experimental studies, characteristic spatial responses have been observed when a small number of neurons were stimulated \cite{Russell706010}. Thus, it is an important open challenge to understand how critical multiscale dynamics emerge in a spatially extended model and to examine whether those dynamics are consistent with the experimentally observed spatial patterns.

From the theoretical point of view, another question that remains unanswered concerns the way qualitatively different solutions bifurcate in our model when we increase the magnitudes of excitation and inhibition. Theoretical analysis of this bifurcation is hard due to the fact that the MFT is constructed based on an averaging over network configurations while the bifurcation point depends strongly on the individual configurations. Regarding this point, a recent study \cite{PhysRevX.6.031024} developed a theory of linear dynamics for disordered systems with individual configurations. The stochastic linear response theory shown in Appendices \ref{perturbative_expansion_fixed_point} and \ref{perturbative_expansion_oscillation} also allows us to analyze the response dynamics around fixed points and regular coherent oscillations for individual configurations. However, to identify the type of a bifurcation, information is needed about the lowest order nonlinear term relevant to the bifurcation. We expect higher order corrections to the lowest order result to provide nonlinear response dynamics valid for individual configurations and information about the bifurcations.

\begin{acknowledgements}
We thank Dr. Shun Ogawa, Dr. Naoki Hiratani, Dr. Yasuhiro Tanaka, Dr. Chigaku Itoi, Dr. Maximilian Schmidt, Dr. Takuma Tanaka, Ms. Milena M. Carvalho, Dr. Tomoki Kurikawa, and Mr. Toshitake Asabuki for their kind help with fruitful discussions and advice. We thank Mr. Keita Watanabe for his general advice on numerical simulations. We also thank two of our colleagues, Dr. Miho Itoi and Dr. Ryotaro Inoue, for their kind support to the work environment for the revision of the manuscript. This research is supported by the Brain Mapping by Integrated Neurotechnologies for Disease Studies (Brain/MINDS) from Japan Agency for Medical Research and development (AMED), Grant-in-Aid for Young Scientists (B) Number JP16K16121 from Japan Society for the Promotion of Science (JSPS) and Grants-in-Aid for Scientific Research on Innovative Areas Number JP16H01289 and JP17H06036 from the Ministry of Education, Culture, Sports, Science and Technology (MEXT).  
\end{acknowledgements}


\newpage
\clearpage
\newpage
\appendix
\section{Simulations of neuronal networks that do not violate Dale's law} \label{appendix:D}
In the main text, we simulate the model equations, Eq.(\ref{dynamical_equation}) or (\ref{LIF_equation}) together with synaptic weights described by Eq.(\ref{basic_equation_nonzerosum}) or (\ref{basic_equation}), directly. In this section, we describe the details of the simulations. We first describe the random variables $\mathcal{J} _{k\ell }^{ij}$. As indicated in \cite{Stern2016}, we can choose random variables for the connectivity, so that the model does not violate Dale's law, a rule that prohibits neurons from feeding both excitatory and inhibitory connections. We use the following random variables with zero mean and unit variance:
\begin{eqnarray}
\pm \mathcal{J} _{k\ell }^{ij}=\left \{ 
\begin{array}{cc}
\sqrt{\frac{1-p}{p} }& \  prob.\    p\nonumber \\
-\sqrt{\frac{p}{1-p} }& \  otherwise.
\end{array}
\right.
\end{eqnarray} 
The sign before $\mathcal{J} _{k\ell }^{ij}$ is positive if the population $\ell $ is excitatory, and negative otherwise. With the same usage of sign $\pm$ for $J_{k\ell }^{ij}$, these random variables give, for both Eqs.(\ref{basic_equation_nonzerosum}) and (\ref{basic_equation}),
\begin{eqnarray}
\pm \sqrt{N} J_{k\ell }^{ij}=\left \{ 
\begin{array}{cc}
\sigma _{0}\sqrt{\frac{1-p}{p} }+g_0+O_p(1/\sqrt{N})& \  prob.\    p\nonumber \\
-\sigma _{0}\sqrt{\frac{p}{1-p} }+g_0+O_p(1/\sqrt{N})& \  otherwise.
\end{array}
\right.
\end{eqnarray} 
Note that the effect of the adjustment in the second of Eq.(\ref{basic_equation}) is $O_p(1/\sqrt{N})$. For any finite value of $g_{0}$, we can thus choose $p$ such that the value on the right-hand side of the above equation converges to positive values in distribution. In practice, for finite values of $N$, values of $p$ that are too small reduce the reproducibility of the numerical results. Thus, we use $p=0.2$ or $0.4$ in all simulations, although fixing $p$ violates Dale's law for small values of $g_0$.

With these random synaptic weights, we integrate the model equations using the fourth-order Runge-Kutta algorithm \cite{galassi2002gnu} with discrete timesteps of size $\Delta t=0.05$ for Eq.(\ref{dynamical_equation}), and $\Delta t=0.02$ for Eq.(\ref{LIF_equation}). We use $N=10240$ for most of the results, except that we use $N=40960$ in Fig.\ref{figure_statistics_of_dynamics}(a),(b), $N=20480$ in Fig.\ref{figure_statistics_of_dynamics}(c),(d), $N=10000$ in Fig.\ref{figure_spiking} and \ref{stimulus-driven_LIF}, and different values of $N$ in Fig.\ref{figure_readout_with_regular}(c),(d), Fig.\ref{size_dependence}, and Fig.\ref{figure_irregular_input}(c),(d). For smaller values of $N$($<8000$), we use $p=0.4$ to enhance the stability of the results. Each direct simulation is performed using 16-cores of recent versions of Intel Xeon processors in parallel and takes a few hours--a few days.

 \section{Eigenvalue spectra of synaptic weight matrices and local stability matrices} \label{appendix:eigenvalues_of_weight_matrices}
\begin{figure}
\includegraphics[width=87mm]{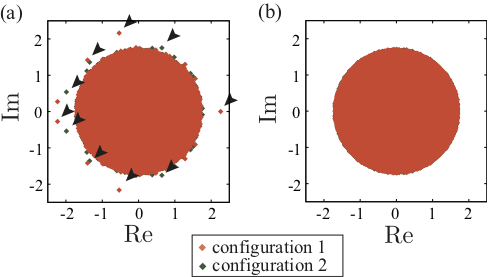}
\caption{The entire eigenvalue spectra of synaptic weight matrices for (a) the untuned model and (b) the finely tuned model. The spectra for two different configurations are calculated from networks with 10240 neurons for each population and plotted in different colors. For both panels, we use $(\sigma _0,g_0)=(1.2,4)$. In the spectra for the untuned model, outlier eigenvalues are observed [arrowheads in the panel (a)], while most of the eigenvalues are distributed over a common disk. Such outlier eigenvalues are not observed for the finely tuned model. In both panels, most of the eigenvalues for configuration 2 in the common disk are hidden behind those for configuration 1, although the distribution in the common disk is quite similar between the two configurations.} \label{figure_outlier_eigenvalues_weight_matrix}
\end{figure}

We show the entire eigenvalue spectra of the synaptic weight matrices of the untuned and finely tuned models in Fig.\ref{figure_outlier_eigenvalues_weight_matrix}(a) and (b). As pointed out in previous studies \cite{PhysRevLett.97.188104, Tao2013, Stern2016}, the synaptic weight matrices of the untuned model have configuration-dependent outliers while those of the finely tuned model do not.   

\begin{figure}
\includegraphics[width=87mm]{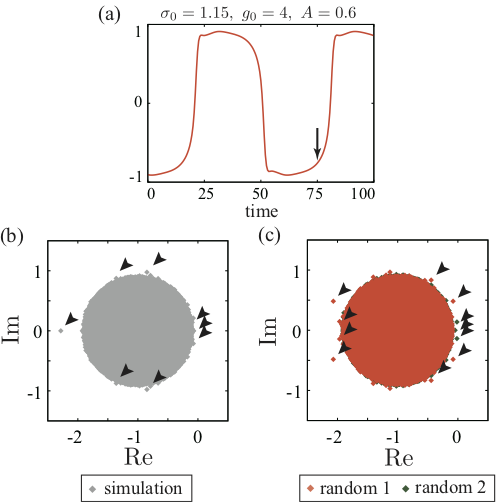}
\caption{(a) A simulation in the same setting as Fig.\ref{figure_external_input}(c) is performed and the mean activity of the excitatory population is shown. The values of the model parameters are indicated in the panel. (b) The entire eigenvalue spectrum of the coefficient matrix $B$ of the linear variational equation (\ref{linear_variational_eq}), calculated for the indicated time point of the numerical simulation [panel (a) arrow]. (c) The eigenvalue spectra of $B$ are also calculated by using two sets of independent random weight configurations and independent random neuronal activities of the same first- and second-order moments as those of the simulated activities used in (b) (indicated as ``random 1'' and ``random 2'', respectively). In these spectra, outlier eigenvalues are observed [arrowheads in (b) and (c)], while most of the eigenvalues are distributed over a common disk. In (c), most of the eigenvalues for ``random 2'' in the common disk are hidden behind those for ``random 1'', although the distributions in the common disk are quite similar in the two settings.} \label{figure_outlier_eigenvalues_jacobian}
\end{figure}

Fig.\ref{figure_outlier_eigenvalues_jacobian}(b) shows the entire eigenvalue spectrum of the coefficient matrix $B_{k\ell }^{ij}(t)$ of the following linear variational equation: 
\begin{eqnarray}
\hspace{-1cm}\frac{\mathrm{d}}{\mathrm{d}t} u_{i}^{(k)}(t)&=&\sum _{\ell, j}B_{k\ell}^{ij}(t)u_{j}^{(\ell)}(t),\label{linear_variational_eq} \\
B_{k\ell}^{ij}(t)&\overset{\mathrm{def}}=&-\delta _{ij}\delta _{k\ell }+\sigma _0J _{k\ell }^{ij}\phi ^{\prime }(h_{j}^{(\ell )}(t)). \nonumber
\end{eqnarray}
The variational equation (\ref{linear_variational_eq}) describes how an infinitesimal variation in $h_i^{(k)}$, denoted by $u_{i}^{(k)}$, evolves over time around the observed dynamics. The spectrum of $B_{k\ell}^{ij}(t)$ is calculated with $h_{i}^{(k)}(t)$ at an arbitrarily chosen time point $t$ in the regular orbit of the finely tuned model with external inputs [the indicated point of Fig.\ref{figure_outlier_eigenvalues_jacobian}(a)]. Since the sum of a row of $B$ is not finely tuned to a fixed value even if $J$ is the synaptic weight matrix of the finely tuned model, a previous result suggests that $B$ has configuration-dependent outlier eigenvalues \cite{PhysRevLett.97.188104}. We observe this in Fig.\ref{figure_outlier_eigenvalues_jacobian}(b). In Fig.\ref{figure_outlier_eigenvalues_jacobian}(c), we also show the eigenvalue spectrum of the same random coefficient matrix, $B_{k\ell }^{ij}=J_{k\ell }^{ij}\phi ^{\prime }(h_j^{(\ell )})$, but for which $J_{k\ell }^{ij}$ and $\phi ^{\prime }(h_j^{(\ell )})$ are generated independently in such a manner that they have the same first- and second-order moments as those used for Fig.\ref{figure_outlier_eigenvalues_jacobian}(b). These spectra have a common disk-form distribution of eigenvalues and outlier eigenvalues at different positions, as expected from the previous study \cite{PhysRevLett.97.188104}. A previous study \cite{PhysRevX.8.041029} also calculated the spectra of the coefficient matrices of the linear variational equations around dynamics of RNNs and showed that they agreed with those estimated from random matrix theory. The results also agreed with the largest Lyapunov exponent calculated from the linear response theory based on an MFT. By analogy with this, our stochastic linear response theory derived in Sec.\ref{perturbative_analysis_fp}, Sec.\ref{perturbative_analysis_osc}, Appendix \ref{perturbative_expansion_fixed_point}, and Appendix \ref{perturbative_expansion_oscillation} is expected to allow further quantitative evaluation of the agreement between random matrix theory with outliers and our stochastic MFT. We leave this as a challenge for the future.

\section{Variety of network dynamics for different system sizes} \label{system_size_dependence}
Here, we investigate how often we observe each of the qualitatively different solutions in simulations of the network models with different system sizes. We do this by conducting simulations of the models with random weight configurations. We use the same model equation and parameter values as for Figs.\ref{figure_single_EI_pair_nonzerosum}(g), \ref{figure_external_input}(c), and \ref{figure_spiking}(c), and for occasionally observed fixed-point and limit-cycle solutions for autonomous networks with finely tuned synaptic weights [Fig.\ref{zerosum_autonomous_oscillation}]. We also examine stimulus-driven dynamics of networks of LIF neurons, by applying sinusoidal inputs to $\sqrt{N}$ excitatory neurons as described by Eqs.(\ref{modelling_sparse_input}) and (\ref{sinusoidal_input}). The input term in these equations are added to the right-hand side of Eq.(\ref{LIF_equation}). As mentioned in the main text, the entire network gets entrained to the sinusoidal input in a configuration-dependent manner, as we increase the amplitude of the input [Fig.\ref{stimulus-driven_LIF}].

\begin{figure}
\includegraphics[width=87mm]{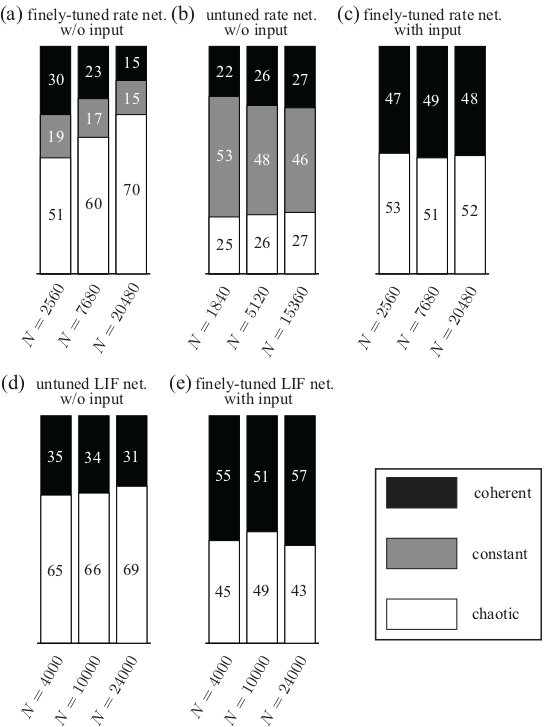}
\caption{(a)--(e) The numbers of the configurations for which each of the qualitatively different dynamics is observed are counted and summarized in a graph for one hundred direct simulations of (a) the finely-tuned rate-neuron network without external inputs, (b) the untuned rate-neuron network without external inputs, (c) the finely-tuned rate-neuron network with sinusoidal external inputs, (d) the untuned LIF-neuron network without external inputs, and (e) the finely-tuned LIF-neuron network with sinusoidal external inputs, with the indicated number of neurons for each population. For (a)--(e), we use the same settings as for Figs.\ref{zerosum_autonomous_oscillation}, \ref{figure_single_EI_pair_nonzerosum}(g), \ref{figure_external_input}(c), \ref{figure_spiking}(c), and \ref{stimulus-driven_LIF}(b), respectively, except for the random weight configurations.} \label{size_dependence}
\end{figure}

\begin{figure}
\includegraphics[width=86mm]{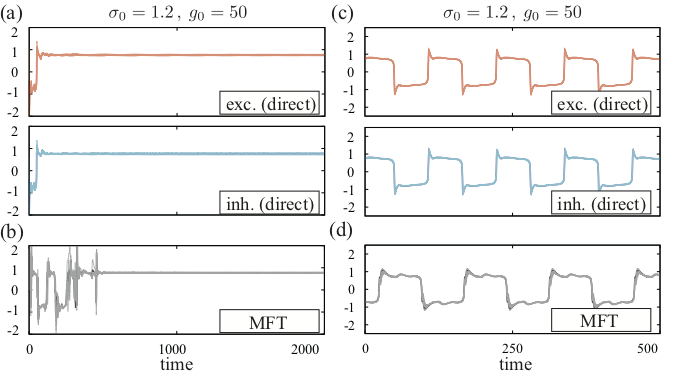}
\caption{Static states [(a) and (b)] and coherent oscillations [(c) and (d)] observed in direct simulations and simulations of the mean-field equations for the model with finely-tuned synaptic weights. The simulations are conducted for the same condition as Fig.\ref{figure_single_EI_pair}, except for the random numbers used for the simulations and the parameters $(\sigma _0, g_0)=(1.2, 50)$ used here. Activity patterns of the network are plotted in the same manner as in Fig.\ref{figure_single_EI_pair}. These non-chaotic solutions are occasionally observed, depending on the configuration of the random connectivity or on the sequence of random numbers used for solving the mean-field equations.} \label{zerosum_autonomous_oscillation}
\end{figure}

\begin{figure}
\includegraphics[width=55mm]{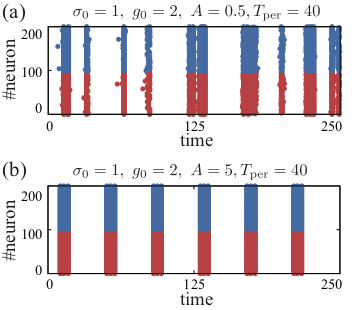}
\caption{The raster plots of one hundred representative excitatory and inhibitory neurons of the LIF model under uniform, sinusoidal external inputs. The parameter values used for the simulations are indicated above the panels. As we increase the amplitude $A$ of the sinusoidal inputs, the network undergoes a transition from (a) partially entrained dynamics with irregular firing to (b) coherent dynamics completely entrained to the input. The values of the other model parameters are the same as for Fig.\ref{figure_spiking}, but the synaptic weights are finely tuned.} \label{stimulus-driven_LIF}
\end{figure}

In the model setting for Figs.\ref{figure_single_EI_pair_nonzerosum}(g), \ref{figure_external_input}(c), \ref{figure_spiking}(c), and \ref{stimulus-driven_LIF}(b), we find that the frequency with which each type of solution is observed does not depend much on the system size [Fig.\ref{size_dependence}(b)--(e)], but it does in the model setting for Fig.\ref{zerosum_autonomous_oscillation} [Fig.\ref{size_dependence}(a)]. These findings are consistent with the results of the stability analysis we perform in Sec.\ref{perturbative_analysis_fp}, Sec.\ref{perturbative_analysis_osc}, Appendix \ref{perturbative_expansion_fixed_point}, and Appendix \ref{perturbative_expansion_oscillation}. 
\section{Application of irregular external inputs} \label{appendix:irregular_iinput}
\begin{figure}[h!]
\includegraphics[width=86mm]{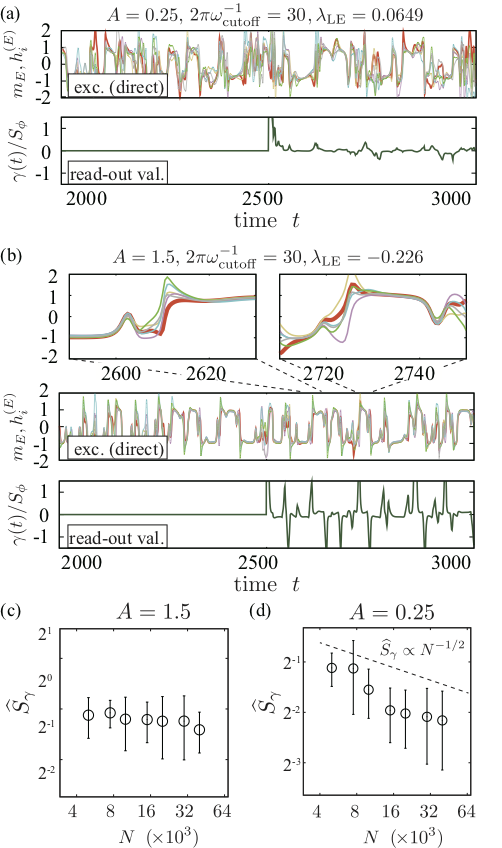}
\caption{(a), (b) Typical mean activities of the excitatory population (thick red line) and typical activities of five representative excitatory neurons (thin lines in different colors) in networks with the indicated parameter values and the irregular external input described by Eq.(\ref{irregular_input}) are shown in the upper panels. The largest Lyapunov exponents of the dynamics are shown above these panels. The values read out from the networks in the same manner as in Fig.\ref{figure_readout_with_regular}(a) and (b) ($t_{\mathrm{init}}=2500$) are shown in the lower panels. Two magnified images of the indicated parts are also shown. (c),(d) Normalized standard deviations of the read-out values from networks with $(\sigma _0, g_0)=(1.15, 4)$ and $A=1.5$ or $0.25$. The averages of $\widehat{S} _{\gamma }$ over fifteen different configurations, together with their standard errors, are plotted on logarithmic scales. In panel (d), a straight line with slope $-1/2$, representing $\widehat{S} _{\gamma }\propto N^{-1/2}$, is also shown. } \label{figure_irregular_input}
\end{figure}

A different type of external input of particular interest is an irregular input with no periodicity. As an example of such an input, we use the following filtered noise:
\begin{eqnarray}
&&I_{k,i}(t)=\sqrt{c_{0}}A\mathcal{F} ^{-1}[\widehat{I} _{\mathrm{cutoff}}](t), \   \   (k=E,1\leq i\leq \sqrt{N} ), \nonumber \\
&&\widehat{I} _{\mathrm{cutoff} }(\omega )=\Theta (\omega _{\mathrm{cutoff}}-|\omega |) \mathcal{F} [I_{\mathrm{WG}}] (\omega ), \nonumber \\
&&c_0\overset{\mathrm{def}}{=}\int _{\omega _{\min}}^{\omega _{\max}}\hspace{-0.3cm}|\mathcal{F} [I_{\mathrm{WG}}](\omega )|^2\mathrm{d} \omega / \int _{\omega _{\min}}^{\omega _{\max}}\hspace{-0.3cm}|\widehat{I} _{\mathrm{cutoff}}(\omega )|^2\mathrm{d} \omega , \label{irregular_input}
\end{eqnarray}
where $\mathcal{F}$ and $\mathcal{F} ^{-1}$ indicate Fourier and inverse-Fourier transforms, respectively. We define white Gaussian noise with unit variance, $I_{\mathrm{WG}}$, with $\langle I_{\mathrm{WG}}(t)I_{\mathrm{WG}}(t-\tau )\rangle =\delta (\tau )$. The domain of integration is between $\omega _{\min }$ and $\omega _{\max }$, which are given suitably in the discrete Fourier transforms we use for the numerical calculations below \cite{galassi2002gnu}. To filter out the high frequency components, we use the step function $\Theta (\omega )=1$ for $\omega \geq 0$ and $\Theta (\omega )=0$ for $\omega <0$.

Examining the properties of the network dynamics resulting from these irregular inputs, we find that the largest Lyapunov exponent decreases and becomes negative as we increase the amplitude $A$ of the input. By analogy with the results shown in Sec.\ref{section_readout}, we hypothesize that neuronal activities are coherent in the dynamics with the negative largest Lyapunov exponents. In Fig.\ref{figure_irregular_input}(a) and (b), we show typical activity patterns for the networks with the negative and positive largest Lyapunov exponents, given these irregular inputs. Coherence among the neurons is not seen in the activity patterns. To further test the hypothesis, we read out values linearly from these networks according to Eqs.(\ref{reading_out_equation}) and (\ref{determining_ro_coefficients}) and examine the scaling of the read-out values in the same manner as for Fig.\ref{figure_readout_with_regular}(c) and (d). Fig.\ref{figure_irregular_input}(c) and (d) show that the normalized standard deviations of the values read out from networks with negative exponents do not depend much on the system size, while those calculated from networks with positive exponents are roughly proportional to $1/\sqrt{N}$. This suggests that the networks undergo a transition to coherent dynamics as we increase the amplitude of the irregular external input, even though coherence among the neurons is not obvious from their activity patterns. 
\clearpage
\begin{widetext}
\section{Path-integral representation of the dynamics} \label{appendix:O}
In this section, we derive the MFT described in Sec.\ref{sec:section3} based on a path-integral representation of dynamics. The approach based on path-integral representations has recently become increasingly popular in the analysis of RNNs and other disordered systems \cite{PhysRevE.75.051919, 10.1371/journal.pcbi.1002872, chow2015path, Hertz:2016, PhysRevX.8.041029, PhysRevE.97.062314, Schuecker:2016uj}. Our argument and notation follow \cite{Schuecker:2016uj}. We refer readers to the first two chapters of \cite{Schuecker:2016uj} for precise definitions and notations for the path integral we use below.

We first analyze the autonomous dynamics of the model with finely-tuned synaptic weights described in Sec.\ref{subsec:3_A}. The moment-generating functional for the dynamics of the present model from an initial condition $a$ is given by 
\begin{eqnarray}
Z[\mathrm{j}]=\lim _{\Delta t\rightarrow +0}\prod _{\alpha =1}^{T/\Delta t}\left \{ \int _{-\infty }^{\infty }\mathrm{d} h_{\alpha }\exp (\mathrm{j} _{\alpha }^Th_{\alpha }\Delta t)\right \} p(h_1, h_2, \cdots , h_{T/\Delta t}|h_0=a).
\end{eqnarray}
In the above, following \cite{Schuecker:2016uj}, we collectively denote $\{ h_{i}^{(k)}\} _{k,i}$ by $h$ as a single column vector, and add the subscript $\alpha$ as the time index. The superscript $T$ denotes transposition. The probability density over the sample paths is denoted by $p$. Hereafter, we sometimes use the same sort of collective notation without mentioning it.  
The main step in deriving the MFT for the present model is to transform the above functional into an integral with respect to the sample paths for the mean activity. 
Putting the right-hand side of the model equation equal to $f(h)$, and using the following Fourier representation of the Dirac delta functional: 
\begin{eqnarray}
\delta (h)=\frac{1}{(2\pi \mathrm{i})^{2N}} \int _{-\mathrm{i}\infty}^{\mathrm{i}\infty }\mathrm{d} \widetilde{h} \exp (\widetilde{h} ^Th),
\end{eqnarray} 
the above generating functional is transformed as follows:
\begin{eqnarray}
Z[\mathrm{j}, \widetilde{\mathrm{j}}]&=&\lim _{\mathrm{D}\rightarrow +0}\lim _{\Delta t\rightarrow +0}\prod _{\alpha =1}^{T/\Delta t}\left \{ \int _{-\infty }^{\infty }\mathrm{d} h_{\alpha }\right \} \prod _{\alpha ^{\prime }=0}^{T/\Delta t-1}\left \{ \int _{-\mathrm{i}\infty }^{\mathrm{i}\infty}\frac{\mathrm{d} \widetilde{h} _{\alpha ^{\prime }}}{(2\pi \mathrm{i})^{2N}} \right \} \nonumber \\
&\times &\exp \left (\sum _{\alpha ^{\prime \prime }=0}^{T/\Delta t-1}\widetilde{h} _{\alpha ^{\prime \prime }}^T(h_{\alpha ^{\prime \prime }+1}-h_{\alpha ^{\prime \prime }}-f(h_{\alpha ^{\prime \prime }})\Delta t-a\delta _{\alpha ^{\prime \prime }, 0}+\frac{\mathrm{D}}{2}\widetilde{h} _{\alpha ^{\prime \prime}})+\mathrm{j}_{\alpha ^{\prime \prime }+1}^Th_{\alpha ^{\prime \prime }+1}\Delta t+\widetilde{\mathrm{j}} _{\alpha ^{\prime \prime }}^T\widetilde{h} _{\alpha ^{\prime \prime }}\Delta t \right ) \nonumber \\
&\overset{\mathrm{def}}{=}&\int \mathcal{D} h\mathcal{D} \widetilde{h} \exp \left (\int _{-\infty }^{\infty }\widetilde{h} (t)^T(\partial _th(t)-f(h(t))-a\delta (t))+\mathrm{j} (t)^Th(t)+\widetilde{\mathrm{j} }(t)^T\widetilde{h}(t)\mathrm{d}t\right ).
\end{eqnarray}
In the above, we have introduced an auxiliary field, $\widetilde{\mathrm{j}}$, for calculating the response function (see \cite{Schuecker:2016uj}). In the above definition, we adopt the Ito convention and take the noiseless limit in defining the path integrals for the dynamics. In this section, we assume that the noiseless limit, thermodynamic limit, and stepsize limit all commute with one another. Note that we have
\begin{eqnarray}
\left. Z[\mathrm{j} , \widetilde{\mathrm{j}}]\right | _{\mathrm{j}\equiv0}=1, \  \  (^{\forall }\widetilde{\mathrm{j}}), \label{normalisation_of_pathintegral}
\end{eqnarray}
because this quantity is the limit of integrals of proper probability densities. 
In the following, we represent inner products with respect to time in the $L^2$ sense, using a vectorial notation such as
\begin{eqnarray}
\widetilde{h}^Th\equiv \int _{-\infty }^{\infty }\widetilde{h} (t)^Th(t)\mathrm{d} t.
\end{eqnarray}
Rewriting the above with a concrete form for $f$ and making the dependence on configurations explicit, we obtain
\begin{eqnarray}
Z[\mathrm{j}, \widetilde{\mathrm{j}}|J]&=&\int \mathcal{D} h\mathcal{D} \widetilde{h} \exp \left [\sum _kS[h^{(k)}, \widetilde{h} ^{(k)}]-\sum _k\widetilde{h} ^{(k)T}\left \{ \sum _{\ell }\frac{\sigma _{0}}{\sqrt{N}} \widetilde{\mathcal{J}} _{k\ell }\phi (h^{(\ell )})+\sum _{\ell }\frac{g_{k\ell }}{\sqrt{N}}M_1\phi (h^{(\ell )})\right \} \right. \nonumber \\
&&\left. +\sum _k\mathrm{j} ^{(k)T}h^{(k)}+\sum _k\widetilde{\mathrm{j}} ^{(k)T}\widetilde{h} ^{(k)}\right ] \nonumber \\
&=&\int \mathcal{D} \theta \mathcal{D} \widetilde{m}\mathcal{D} h\mathcal{D} \widetilde{h}\exp \left [ \sum _kS[h^{(k)}, \widetilde{h} ^{(k)}]-\sum _k\widetilde{h} ^{(k)T}\sum _{\ell }\frac{\sigma _{0}}{\sqrt{N}} \widetilde{\mathcal{J}} _{k\ell }\phi (h^{(\ell )})\right. \nonumber \\
&&-\left. \sum _k\widetilde{m} _k^T\sum _{\ell }\frac{g_{k\ell }}{\sqrt{N}} \mathrm{1} ^T(\phi (h^{(\ell )})-\overline{\phi } _\ell \mathrm{1})+\sum _k(\widetilde{m} _k-\mathrm{1} ^T\widetilde{h} ^{(k)})^T\theta _k+\sum _k\mathrm{j} ^{(k)T}h^{(k)}+\sum _k\widetilde{\mathrm{j}} ^{(k)T}\widetilde{h} ^{(k)}\right ]. \label{explicit_form}
\end{eqnarray}
Here, we define the action:
\begin{eqnarray}
S[h^{(k)},\widetilde{h} ^{(k)}]\overset{\mathrm{def}}{=}\widetilde{h} ^{(k)T}(\partial _t+1)h^{(k)}.
\end{eqnarray}
We define the vector and matrix for which all entries are unity at each timestep as $\mathrm{1}$ and $M_1$, respectively. The left action of $M_1$ is thus given by
\begin{eqnarray}
M_1: \{ \phi (h_j^{(\ell )}(t))\} _{\ell, j, t}\rightarrow \{ \sum _{j^{\prime } }\phi (h_{j^{\prime }}^{(\ell )}(t))\} _{\ell, j, t}.
\end{eqnarray}
The column vector which consists of $\phi (h_i^{(k)})$ has been denoted by $\phi (h^{(k)})$. 
From the first line to the second line of Eq.(\ref{explicit_form}), we have inserted the Dirac delta functional equating $\widetilde{m} _k$ and $\mathrm{1} ^T\widetilde{h} ^{(k)}$ for each $t$, where $\mathrm{1} ^T$ denotes the following operation:
\begin{eqnarray}
\mathrm{1} ^T: \{ \widetilde{h} _i^{(k)}(t)\} _{i,t}\rightarrow \left \{ \sum _i\widetilde{h} _i^{(k)}(t)\right \} _t.
\end{eqnarray}
Note that, under the condition in Eq.(\ref{pedagogical_g}), the insertion of any value of $\overline{\phi } _{\ell }$ does not affect the value of the integrand in the last line of Eq.(\ref{explicit_form}) as long as $\overline{\phi }_E=\overline{\phi }_I$ and hence $\sum _{\ell =E,I}g_{k\ell }\overline{\phi } _{\ell }=0$ hold. We will determine the precise value of $\overline{\phi } _k$ below.

Next, we take the configurational average of the integrand of the above equation. We consider unit Gaussian measures for $\mathcal{J} _{k\ell }^{ij}$ denoted by $\mathcal{N} (\mathcal{J} _{k\ell }^{ij})$. Other distributions for these random variables can be analyzed in essentially the same manner, where distributions with zero mean and unit variance give the same result (to see this, expand the exponential in the integrand in terms of small values). Focusing on the term involving $\widetilde{\mathcal{J}} _{k\ell }^{ij}$, we have
\begin{eqnarray}
&&\int \mathrm{d} \mathcal{N} (\mathcal{J} _{k\ell }^{ij})\exp \left (-\widetilde{h} _i^{(k)T}\frac{\sigma _{0} }{\sqrt{N}} \mathcal{J} _{k\ell }^{ij}\phi (h_{j}^{(\ell )})+\sum _{j^{\prime }} \widetilde{h}_i^{(k)T}\frac{\sigma _{0} }{\sqrt{N}} \frac{\mathcal{J}  _{k\ell }^{ij} }{N} \phi (h_{j^{\prime }}^{(\ell )})\right ) \nonumber \\
&=&\exp \left [\frac{\sigma _{0}^2}{2N} \left \{ \int \widetilde{h} _i^{(k)}\phi (h_{j}^{(\ell )})-\sum _{j^{\prime }}\frac{1}{N} \widetilde{h} _{i}^{(k)}\phi (h_{j^{\prime }}^{(\ell )})\mathrm{d} t\right \} ^2\right ].
\end{eqnarray}
Taking the product of the second line over $k,\ell, i,j$, we obtain the part of the integrand involving $\{ \mathcal{J} _{k\ell }^{ij}\} _{k,\ell, i,j}$ as
\begin{eqnarray}
\exp \left [\sum _{k,\ell}\frac{\sigma _{0}^2}{2N}\left \{ \sum _{i,j} \int \mathrm{d} s\hspace{0.05cm} \mathrm{d} t\hspace{0.05cm} \widetilde{h} _{i}^{(k)}(s)\widetilde{h} _{i}^{(k)}(t)(\phi (h_j^{(\ell )}(s))-\check{\phi }_{\ell }(s))(\phi (h_j^{(\ell )}(t))-\check{\phi } _{\ell }(t))\right \} \right ].  \label{after_integrating_J}
\end{eqnarray}
Here, $\check{\phi }_{\ell }$ denotes the average of $\phi (h_j^{(\ell )})$ over the $\ell $ population. We now introduce an auxiliary field by inserting the following Dirac delta functionals: 
\begin{eqnarray}
&&\delta \left (-NQ_{\ell ,1}(s,t)+\sum _j(\phi (h_j^{(\ell )}(s))-\check{\phi }_{\ell }(s))(\phi (h_j^{(\ell )}(t))-\check{\phi } _{\ell }(t))\right ) \nonumber \\
&=&\int \mathcal{D} Q_{\ell, 2}\exp \left [\int \mathrm{d} s\hspace{0.05cm} \mathrm{d} t\hspace{0.05cm} Q_{\ell,2}(s,t)\left \{ -NQ_{\ell, 1}(s,t)+\sum _j(\phi (h_j^{(\ell )}(s))-\check{\phi }_{\ell }(s))(\phi (h_j^{(\ell )}(t))-\check{\phi } _{\ell }(t))\right \} \right ] , \label{introducing_auxiliary_fields} \\
&&\delta \left (-N\check{\phi } _{\ell }(t)+\sum _{j} \phi (h_j^{(\ell )}(t))\right ) \nonumber \\
&=&\int \mathcal{D} \widetilde{\psi } _{\ell } \exp \left [\widetilde{\psi } _{\ell }^T\left \{ -N\check{\phi } _{\ell }+\sum _{j} \phi (h_j^{(\ell )})\right \} \right ]. \label{introducing_mean_fields}
\end{eqnarray}
Following the convention in \cite{Schuecker:2016uj}, we regard $Q_{\ell ,i}$ $(i=1,2)$ as matrices and use the following notation:
\begin{eqnarray}
Q_{\ell, 1}^TQ_{\ell, 2}&=&\int \mathrm{d} s\hspace{0.05cm} \mathrm{d} t\hspace{0.05cm} Q_{\ell, 1}(s,t)Q_{\ell, 2}(s,t), \nonumber \\
\widetilde{h} _i^{(k)T}Q_{\ell, 1}\widetilde{h} _i^{(k)}&=&\int \mathrm{d} s\hspace{0.05cm} \mathrm{d} t\hspace{0.05cm} \widetilde{h} _i^{(k)}(s)Q_{\ell ,1}(s,t)\widetilde{h} _i^{(k)}(t).
\end{eqnarray}
Using Eqs.(\ref{after_integrating_J})--(\ref{introducing_mean_fields}), we take the average of the moment-generating functional over the probability distribution $P_J(J)$ of $J_{k\ell }^{ij}$ as follows:
\begin{eqnarray}
\overline{Z} [\mathrm{j}, \widetilde{\mathrm{j}}]&=&\int \mathrm{d} P_J(J)Z[\mathrm{j}, \widetilde{\mathrm{j}}|J] \nonumber \\
&=&\int \mathcal{D} \theta \mathcal{D} \widetilde{m} \mathcal{D}Q_1 \mathcal{D} Q_2 \mathcal{D} \widetilde{\psi } \mathcal{D} \check{\phi } \mathcal{D} h \mathcal{D} \widetilde{h} \exp \left (-N\sum _{\ell }Q_{\ell ,1}^TQ_{\ell ,2}+\sum _{k,i}S[h_i^{(k)},\widetilde{h} _i^{(k)}]+\sum _{k,\ell ,i}\frac{\sigma _{0}^2}{2} \widetilde{h} _i^{(k)T}Q_{\ell ,1}\widetilde{h} _i^{(k)}-\sum _{k,i}\widetilde{h} _i^{(k)T}\theta _k\right. \nonumber \\
&&\left. +\sum _{\ell,j}\Delta \phi _j^{(\ell ) T}Q_{\ell ,2}\Delta \phi _j^{(\ell )}+\sum _k\mathrm{j}^{(k)T}h^{(k)}+\sum _k\widetilde{\mathrm{j} }^{(k)T}\widetilde{h} ^{(k)}+\sum _{\ell ,j}\widetilde{\psi } _{\ell }^T\phi (h_j^{(\ell )})-N\sum _{\ell}\widetilde{\psi } _{\ell}^T\check{\phi } _{\ell } \right ) \nonumber \\
&&\times \exp \left( -\sum _k\widetilde{m} _k^T\sum _{\ell }\frac{g_{k\ell }}{\sqrt{N}} \mathrm{1}^T(\phi (h^{(\ell )})-\overline{\phi }_\ell \mathrm{1})+\sum _{k}\widetilde{m} _k^T\theta _k\right ). \label{expanded_integrand}
\end{eqnarray}
Here, $\phi (h_j^{(\ell )})-\check{\phi } _{\ell }$ is denoted by $\Delta \phi _j^{(\ell )}$. Using this representation, we note that the argument of the first exponential takes the form of an independent interaction between each neuron and the auxiliary fields if the values of $\theta _k$, $\check{\phi } _{\ell }$, $\widetilde{\psi } _{\ell }$, $Q_{\ell ,1}$, and $Q_{\ell, 2}$ are fixed.

In the above equation, we notice that, for fixed sample paths for $\theta $ and $\widetilde{m}$, the integral of the first exponential function with respect to $Q_1,Q_2,\widetilde{\psi }, \check{\phi }, h$, and $\widetilde{h}$ gives the generating functional for the following dynamics of a fictitious RNN with a uniform external input $\theta _k(t)$:
\begin{eqnarray}
\frac{\mathrm{d}}{\mathrm{d} t}h_i^{(k)}(t)=-h_i^{(k)}(t)+\sigma _0\sum _{j,\ell }\frac{\widetilde{\mathcal{J}} _{k\ell }^{ij}}{\sqrt{N}} \phi (h_j^{(\ell )}(t))+\theta _k(t). \label{fictitious_equation}
\end{eqnarray}
The dynamics of this fictitious RNN are no longer under the effects of balanced excitation and inhibition, and therefore solved by a conventional MFT. We first rewrite the averaged generating functional for these dynamics as  
\begin{eqnarray}
\overline{Z} ^*[\mathrm{j}, \widetilde{\mathrm{j}}|\theta ]&=&\int \mathcal{D}Q_1 \mathcal{D} Q_2 \mathcal{D} \widetilde{\psi } \mathcal{D} \check{\phi } \mathcal{D} h \mathcal{D} \widetilde{h} \exp \left (-N\sum _{\ell }Q_{\ell ,1}^TQ_{\ell ,2}+\sum _{k,i}S[h_i^{(k)},\widetilde{h} _i^{(k)}]+\sum _{k,\ell,i}\frac{\sigma _{0} ^2}{2} \widetilde{h} _i^{(k)T}Q_{\ell ,1}\widetilde{h} _i^{(k)}-\sum _{k,i}\widetilde{h} _i^{(k)T}\theta _k\right. \nonumber \\
&&+\left. \sum _{\ell ,j}\Delta \phi _j^{(\ell ) T}Q_{\ell ,2}\Delta \phi _j^{(\ell )}+\sum _k\mathrm{j} ^{(k)T}h^{(k)}+\sum _k\widetilde{\mathrm{j} }^{(k)T}\widetilde{h} ^{(k)}+\sum _{\ell ,j}\widetilde{\psi } _{\ell }^T\phi (h_j^{(\ell )})-N\sum _{\ell}\widetilde{\psi } _{\ell}^T\check{\phi } _{\ell }\right ). \label{gf_fictitious_system}
\end{eqnarray}
The method for obtaining the dynamics described by the above moment-generating functional has been developed in previous studies \cite{PhysRevX.5.041030, Schuecker:2016uj}. First, the above functional is rewritten as
\begin{eqnarray}
\overline{Z} ^*[\mathrm{j}, \widetilde{\mathrm{j}}|\theta ]&=&\int \mathcal{D}Q_1 \mathcal{D} Q_2 \mathcal{D} \widetilde{\psi } \mathcal{D} \check{\phi } \exp \left (-N\sum _{\ell }Q_{\ell ,1}^TQ_{\ell ,2}+N\sum _k\ln \mathcal{Z} _k[Q_1,Q_2,\widetilde{\psi }, \check{\phi }, \mathrm{j} ,\widetilde{\mathrm{j}} |\theta ]-N\sum _{\ell}\widetilde{\psi } _{\ell}^T\check{\phi } _{\ell }\right ), \nonumber \\
\mathcal{Z} _k[Q_1,Q_2,\widetilde{\psi }, \check{\phi }, \mathrm{j} ,\widetilde{\mathrm{j}} |\theta ]&=&\int \mathcal{D} h\mathcal{D} \widetilde{h} \mathcal{D} P_{\mathrm{j}, \widetilde{\mathrm{j}}}(\mathrm{l} ,\widetilde{\mathrm{l}}) \exp \left ( \widetilde{h} ^{(k)T}(\partial _t+1)h^{(k)}+\sum _{\ell}\frac{\sigma _{0} ^2}{2} \widetilde{h} ^{(k)T}Q_{\ell ,1}\widetilde{h} ^{(k)}-\widetilde{h} ^{(k)T}\theta _k +\mathrm{l}^{(k)T}h^{(k)}+\widetilde{\mathrm{l}} ^{(k)T}\widetilde{h}^{(k)} \right. \nonumber \\
&&\left. +(\phi (h^{(k)})-\check{\phi } _k)^TQ_{k,2}(\phi (h^{(k)})-\check{\phi } _k)+\widetilde{\psi } _{k}^T\phi (h^{(k)})\right ). \label{path_integral_for_one_body}
\end{eqnarray}
In the second equation, $h^{(k)}$ and $\widetilde{h} ^{(k)}$ are no longer collections of variables corresponding to individual neurons but instead are one-dimensional variables for a representative neuron feeling the mean-fields. Similarly, $\mathrm{l} ^{(k)}$ and $\widetilde{\mathrm{l}} ^{(k)}$ are one-dimensional variables that take values randomly drawn from the  measure $P_{\mathrm{j}, \widetilde{\mathrm{j}} }$ corresponding to $\mathrm{j}$ and $\widetilde{\mathrm{j}} $:
\begin{eqnarray}
P_{\mathrm{j}, \widetilde{\mathrm{j}}}=\otimes _k\left (\frac{1}{N}\sum _i \delta _{\mathrm{j} _i^{(k)}}\delta _{\widetilde{\mathrm{j} }_i^{(k)}}\right ).
\end{eqnarray}
Applying the saddle-point method, we find that the entire probability mass of the path integral of the dynamics at $\mathrm{j}=\widetilde{\mathrm{j}}=0$ concentrates at the values of $\check{\phi } _{\ell }$, $\widetilde{\psi } _{\ell }$, $Q_{\ell ,1}$, and $Q_{\ell, 2}$ that maximize the integrand of the first of Eq.(\ref{path_integral_for_one_body}). Taking the functional derivatives and examining the stationarity conditions, we obtain these optimal values as 
\begin{eqnarray}
Q_{\ell ,2}^*=\sum _k\frac{\sigma _{0}^2}{2}\langle \widetilde{h} ^{(k)}\widetilde{h} ^{(k)T}\rangle &=&0, \    Q_{\ell ,1}^*=\langle (\phi (h^{(\ell )})-\check{\phi } _{\ell })(\phi (h^{(\ell )})-\check{\phi } _{\ell })^T\rangle ,\nonumber \\
 \widetilde{\psi } _{\ell }^*=-2\langle Q_{\ell, 2}(\phi (h^{(\ell )})-\check{\phi } _{\ell })\rangle &=&0,\    \check{\phi } _{\ell }^*=\langle \phi (h^{(\ell )})\rangle . \label{deriving_updates_of_fictitious_system_again}
\end{eqnarray}
Here, the bracket denotes the expected value of the argument for all possible sample paths weighted according to the path-integral representation of the fictitious dynamics.
The zeros for $Q_{\ell ,2}^*$ and $\widetilde{\psi } _{\ell }^*$ are obtained by taking directional functional derivative with respect to $\widetilde{\mathrm{j} } _i^{(k)}(t)=\alpha ^{(k)}(t)$ at $\alpha =0$ and $\mathrm{j}=0$ and using the normalization of the probability density (see also the arguments in \cite{PhysRevB.25.6860, Schuecker:2016uj}). We actually have
\begin{eqnarray}
\left. \frac{\delta \overline{Z} ^*[0,\{ \widetilde{\mathrm{j} }_i^{(k)}(t)=\alpha ^{(k)}(t)\} _{k,i,t}|\theta ]}{\delta \alpha ^{(k)}\delta \alpha ^{(k)T}}\right |_{\alpha =0}\hspace{-0.5cm}=N\langle \widetilde{h} ^{(k)}\widetilde{h} ^{(k)T}\rangle =0, \   \   \left. \frac{\delta \overline{Z} ^*[0,\{ \widetilde{\mathrm{j} }_i^{(k)}(t)=\alpha ^{(k)}(t)\} _{k,i,t}|\theta ]}{\delta \alpha ^{(k)}}\right |_{\alpha =0}\hspace{-0.5cm}=N\langle \widetilde{h} ^{(k)}\rangle =0,
\end{eqnarray}
because we have $\overline{Z} ^*[0, \widetilde{\mathrm{j}}|\theta ]=1$ for any $\widetilde{\mathrm{j}}$, similarly to Eq.(\ref{normalisation_of_pathintegral}).
Then, following the argument in \cite{Schuecker:2016uj}, we can regard the dynamics of individual neurons, as described by the second of Eq.(\ref{path_integral_for_one_body}), as linear dynamics driven by a set of {\it i.i.d.} Gaussian noise with correlation $Q_{\ell ,1}^*$ and drift term $\theta _k$. Thus, we obtain 
\begin{eqnarray}
(1+\partial _t)(1+\partial _s)\langle \delta h^{(k)}(t)\delta h^{(k)}(s)\rangle &=&\sum _\ell \sigma _{0}^2\langle (\phi (h^{(\ell )}(t))-\langle \phi (h^{(\ell )}(t))\rangle )(\phi (h^{(\ell )}(s))-\langle \phi (h^{(\ell )}(s))\rangle )\rangle , \nonumber \\
(1+\partial _t)\langle h^{(k)}(t)\rangle &=&\theta _k(t). \label{deriving_updates_of_fictitious_system_again_2}
\end{eqnarray}
This is the solution for the fictitious RNN described by Eq.(\ref{fictitious_equation}).

We now return back to the generating functional for the original dynamics in Eq.(\ref{expanded_integrand}). Note that the integrand of Eq.(\ref{gf_fictitious_system}) gives a proper probability density over the sample paths for the fields. Then, because of the independence among neurons for fixed values of $\theta _k$ and $\widetilde{m} _k$, the integrand in Eq.(\ref{expanded_integrand}) can be rearranged into the following form with the aid of the central limit theorem:
\begin{eqnarray}
\overline{Z} [0, 0]=\int \mathcal{D} \theta \mathcal{D} \widetilde{m} \mathcal{D} \eta \exp \left (-\frac{1}{2}\sum _{\ell }\eta _{\ell }^TQ_{\ell ,1}^{*-1}\eta _{\ell }-\sum _{\ell }F_{\ell }+\sum _k\widetilde{m} _k^T(\theta _k-\sum _\ell g_{k\ell }\eta _{\ell }) -const. \right ). \label{deriving_updates_of_eta_again}
\end{eqnarray}
Here, we have set the values of $\overline{\phi }_E=\overline{\phi }_I$ to $\langle \phi (h_j^{(\ell)})\rangle $, the configuration and population average of $\phi (h_j^{(\ell )})$ over the fictitious dynamics for each sample path for $\theta _k$. We have defined $\eta _{\ell }\overset{\mathrm{def}}{=} \frac{1}{\sqrt{N}} 1^T(\phi (h^{(\ell )})-\overline{\phi } _{\ell }1)$.
We determine the above normalization term, $F_{\ell }$, below. Noting that the auxiliary field $\widetilde{m} _k$ is imposing equality between $\theta _k$ and $\sum _\ell g_{k\ell }\eta _{\ell }$, the above equation determines the probabilities with which the sample paths for $\eta _{\ell }$ and $\theta _k$ are realized, and the density for each value of $\eta _{\ell }$ is given by 
\begin{eqnarray}
p_{\{\eta \}}(\{ \eta _{\ell }\} _{\ell })\propto \exp \left (-\frac{1}{2}\sum _{\ell }\eta _{\ell }^TQ_{\ell ,1}^{*-1}\eta _{\ell }-\sum _{\ell }F_{\ell} \right ).  \label{probability_density_eta}
\end{eqnarray}
Note that this density is at most O(1) in terms of the neuron count, and therefore, does not affect the values of $Q_{\ell, 1}$ and $Q_{\ell ,2}$, which have most of the probability mass.

Because of the condition, $g_{EE}=g_{IE}=g_0$ and $g_{EI}=g_{II}=-g_0$, we have $\theta _E=\theta _I\overset{\mathrm{def}}{=}\theta $ from $\theta _k=\sum _{\ell }g_{k\ell }\eta _{\ell }$. Since the right-hand sides of Eq.(\ref{deriving_updates_of_fictitious_system_again_2}) take the same values for $k=E,I$, we have 
\begin{eqnarray}
\langle \delta h^{(E)}(t)\delta h^{(E)}(s)\rangle =\langle \delta h^{(I)}(t)\delta h^{(I)}(s)\rangle &\overset{\mathrm{def}}{=}&D(t,s), \\
\langle h^{(E)}(t)\rangle =\langle h^{(I)}(t)\rangle &\overset{\mathrm{def}}{=}&m(t), \\
\langle (\phi (h^{(E)}(t))-\langle \phi (h^{(E)}(t))\rangle )(\phi (h^{(E)}(s))-\langle \phi (h^{(E)}(s))\rangle )\rangle &&\nonumber \\
=\langle (\phi (h^{(I)}(t))-\langle \phi (h^{(I)}(t))\rangle )(\phi (h^{(I)}(s))-\langle \phi (h^{(I)}(s))\rangle )\rangle &\overset{\mathrm{def}}{=}&C(t,s). \label{C_from_path_integral}
\end{eqnarray}
We have $Q_{E,1}^*=Q_{I,1}^*=C$ from Eq.(\ref{deriving_updates_of_fictitious_system_again}), and hence $F_E=F_I\overset{\mathrm{def}}{=}F$. Defining $\eta =\frac{\sqrt{2}}{2}(\eta _E-\eta _I)$ and $\overline{\eta }=\frac{\sqrt{2}}{2}(\eta _E+\eta _I)$ and combining these with Eqs.(\ref{deriving_updates_of_fictitious_system_again_2}) and (\ref{probability_density_eta}) and the equality, $\theta _E=\theta _I=g_0(\eta _{E}-\eta _I)$, 
 we obtain the realization probability of $m(t)$ given by 
\begin{eqnarray}
(1+\partial _t)m(t)=\sqrt{2} g_0\eta (t), \  \  \  
p_{\eta }(\eta )\propto \exp \left (-\frac{1}{2}\eta ^TC\eta-\frac{1}{2}\overline{\eta }^TC\overline{\eta }-2F -const. \right ),  \label{probability_density_simplified_eta}
\end{eqnarray}
which is equivalent to Eqs.(\ref{realization_probability_eta}) and (\ref{dynamics_of_m}). We also obtain the dynamical equation for the correlation matrix for the microscopic fluctuations from Eqs.(\ref{deriving_updates_of_fictitious_system_again_2}) and (\ref{C_from_path_integral}) as 
\begin{eqnarray}
(1+\partial _t)(1+\partial _s)D(t,s)=2\sigma _0^2C(t,s), \label{final_linear_equation_of_C_D}
\end{eqnarray}  
which is identical to Eq.(\ref{self_consistent_eq_2}).

Finally, we identify the normalization term $F$. Let us consider the following representation, 
\begin{eqnarray}
\eta =H\xi , \  \  \  HH^T=C. 
\end{eqnarray} 
If $H$ is independent of $\eta $, this transformation, together with the probability density in Eq.(\ref{probability_density_simplified_eta}), yields {\it i.i.d.} unit Gaussian variables $\xi $. In this case, the normalization term is given by $\exp (-F)=\exp (-\frac{1}{2}\ln |C|)=|C|^{-1/2}$. This is the Jacobian accompanying the transformation of probability densities over $\eta $ and $\xi $. However, in the present setting, $H$ depends on $\eta $. For this case,  taking the nonlinear deformation of the coordinates spanned by $\xi$ into account, we expect that the normalization term $F$ is given by the Jacobian $\ln |H+\frac{\partial H}{\partial \xi }\xi |$ and hence that the transformation $\eta =H\xi $ still gives unit Gaussian variables. Here, the Jacobian matrix is rearranged as 
\begin{eqnarray}
H_{ij}+\sum _{\alpha }\frac{\partial H_{i\alpha }}{\partial \xi _j}\xi _\alpha =H_{ij}+\sum _{\alpha ,\beta }\frac{\partial H_{i\alpha }}{\partial \eta _{\beta}}\frac{\partial \eta _{\beta}}{\partial \xi _j}(H^{-1}\eta )_\alpha =\frac{\partial \eta _i}{\partial \xi _j}\ \Rightarrow \ \frac{\partial \eta }{\partial \xi }=\left (1-\frac{\partial H}{\partial \eta }H^{-1}\eta \right )^{-1}H,
\end{eqnarray}
which yields Eq.(\ref{normalization_F}). Although rigorous treatment of this point encounters the mathematical difficulty in rigorously dealing with path-integrals, the validity of the above normalization term is supported by the fact that it yields the proper normalization of the path integral consistently with Eq.(\ref{normalisation_of_pathintegral}). This normalization needs to be kept in mind as we perform perturbative expansion below.

Below, we describe the outline of how the above theory is modified for the case with external input to $\sqrt{N}$ excitatory neurons and for the case with untuned synaptic weights.
For the case with $\widetilde{I} (t)\neq 0$ in Eq.(\ref{modelling_sparse_input}), the following $O(\sqrt{N})$ correction is introduced to Eq.(\ref{path_integral_for_one_body}):  
\begin{eqnarray}
\overline{Z} ^*[\mathrm{j}, \widetilde{\mathrm{j}}|\theta ]&=&\int \mathcal{D}Q_1 \mathcal{D} Q_2 \mathcal{D} \widetilde{\psi } \mathcal{D} \check{\phi } \exp \left (-N\sum _{\ell }Q_{\ell ,1}^TQ_{\ell ,2}+N\sum _k\ln \mathcal{Z} _k[Q_1,Q_2,\widetilde{\psi }, \check{\phi }, \mathrm{j} ,\widetilde{\mathrm{j}} |\theta ] \right. \nonumber \\
&+&\left. \sqrt{N}(\ln \widetilde{\mathcal{Z} } _E[Q_1,Q_2,\widetilde{\psi }, \check{\phi }, \mathrm{j} ,\widetilde{\mathrm{j}} |\theta ]-\ln \mathcal{Z} _E[Q_1,Q_2,\widetilde{\psi }, \check{\phi }, \mathrm{j} ,\widetilde{\mathrm{j}} |\theta ])-N\sum _{\ell}\widetilde{\psi } _{\ell}^T\check{\phi } _{\ell }\right ), \nonumber \\
\widetilde{\mathcal{Z} } _E[Q_1,Q_2,\widetilde{\psi }, \check{\phi }, \mathrm{j} ,\widetilde{\mathrm{j}} |\theta ]&\overset{\mathrm{def}}{=}&\int \mathcal{D} h\mathcal{D} \widetilde{h} \mathcal{D} P_{\mathrm{j}, \widetilde{\mathrm{j}}}(\mathrm{l} ,\widetilde{\mathrm{l}}) \exp \left ( \widetilde{h} ^{(E)T}(\partial _t+1)h^{(E)}+\sum _{\ell}\frac{\sigma _{0} ^2}{2} \widetilde{h} ^{(E)T}Q_{\ell ,1}\widetilde{h} ^{(E)}-\widetilde{h} ^{(E)T}\theta _E -\widetilde{I}^{T}\widetilde{h} ^{(E)} \right. \nonumber \\
&&\left. +\mathrm{l}^{(E)T}h^{(E)}+\widetilde{\mathrm{l}} ^{(E)T}\widetilde{h}^{(E)} +(\phi (h^{(E)})-\check{\phi } _E)^TQ_{E,2}(\phi (h^{(E)})-\check{\phi } _E)+\widetilde{\psi } _{E}^T\phi (h^{(E)})\right ). \label{path_integral_for_one_body_ext}
\end{eqnarray}
In the subsequent saddle point method, however, we notice that the $O(\sqrt{N})$ correction does not affect the values of $Q_{\ell ,j}^*$, $\widetilde{\psi }_{\ell }^*$, and $\check{\phi }_{\ell }^*$ in the leading order. Then, the neurons feel the same Gaussian fields as those described by Eqs.(\ref{deriving_updates_of_fictitious_system_again}). Then, the second exponential in Eq.(\ref{expanded_integrand}) is modified as
\begin{eqnarray}
\exp \left [ -\sum _k\widetilde{m _k}^T\left \{ \sum _{\ell }\frac{g_{k\ell }}{\sqrt{N}} \mathrm{1} _{\setminus \mathcal{S}}^T(\phi (h^{(\ell )})-\overline{\phi }_\ell \mathrm{1})+\frac{g_0}{\sqrt{N}}\mathrm{1} _{\mathcal{S} }^T(\phi (h^{(\ell )})-\overline{\phi } _{\ell })-\theta _k\right \}\right ], 
\end{eqnarray} 
, with $\overline{\phi }_{\ell }=\langle \phi (h_j^{(\ell )})\rangle _{j\notin \mathcal{S}}$. In the above equation, $1_{\mathcal{S}}$ ($1_{\setminus \mathcal{S}}$) denotes a vector whose elements are one for indices belonging to $\mathcal{S}$, and zero otherwise. We easily find that the second term in the above exponential gives the driving-force term, $\widetilde{\phi }$, for the mean dynamics [Eq.(\ref{MFT_ext})]. We also find that the exclusion of the sum of the $\sqrt{N}$ random values in the first term followed by the division by $\sqrt{N}$ does not affect the values of the driving-force term $\eta $ in the leading order.  Thus, with $\eta $ given by Eq.(\ref{probability_density_simplified_eta}), we obtain the mean-field equations for the stimulus-driven dynamics as presented in Sec.\ref{MFT_stimulus}.

For the untuned model, we easily see that the averaging with respect to the untuned random synaptic weights leads to the replacement of $\Delta \phi _j^{(\ell ) T}Q_{\ell ,2}\Delta \phi _j^{(\ell )}$ in Eq.(\ref{expanded_integrand}) by $\phi (h_j^{(\ell )})^TQ_{\ell , 2}\phi (h_j^{(\ell )})$. By a straightforward application of the same argument as above, we obtain the results presented in Sec.\ref{subsec:2_A_2}.
\end{widetext}

\section{Efficient method for solving the mean-field equations} \label{appendix:A}
In this section, we show how the stochastic mean-field equations are numerically solved by recursively updating the statistics characterizing the microscopic and macroscopic dynamics of the system. In the main text and Appendix \ref{appendix:O}, we found that self-consistent equations (\ref{self_consistent_eq_1}) and (\ref{self_consistent_eq_2}) determine the time evolution of the statistics of the microscopic fluctuations. These equations are rewritten as
\begin{eqnarray}
(1+\partial _{t_1})(1+\partial _{t_2})D(t_1,t_2)=2\sigma _0^2C(t_1,t_2). \label{filter_C}
\end{eqnarray}
Dealing with the above matrix requires us to retain large matrices. To reduce the computational cost, we recursively update matrices of smaller size by using the following auxiliary matrix that retains the effect of the boundary conditions:
\begin{eqnarray}
R\overset{\mathrm{def}}{=}2\sigma _0^2(1+\partial _{t_1})^{-1} C, \label{define_R}
\end{eqnarray}
which means $R$ is the solution of the ordinary differential equation, $(1+\partial _{t_1})R=2\sigma _0^2C$, from suitable initial conditions, and the initial conditions do not affect the values of $A$ at time indices distant from those of the initial conditions because of the decay with the unit decay constant. For the dynamics of the untuned model, matrix $C$ should be suitably replaced by $\widetilde{C}=C+\overline{\phi }\overline{\phi }^T$.

Suppose that we have values of $m(t_1)$, $\eta (t_1)$, $\overline{\phi } (t_1)$, $C(t_1, t_2)$, $D(t_1,t_2)$, and $R(t_1, t_3)$ for timesteps $t\geq t_1,t_2\geq t-T_1$, and $t-\Delta t\geq t_3\geq t-T_1$. For the model with external inputs, also suppose that we have values of $\widetilde{\phi } (t_1)$. The length of the time interval $T_1$ should not be too large. In what follows, we obtain the values of these vectors and matrices for a time index one step beyond the currently available entries.

We first update the value of $m$, which is obtained from the values of $\eta (t)$, $\overline{\phi } (t)$, and $\widetilde{\phi } (t)$ by applying the Euler method to Eq.(\ref{dynamics_of_m}) or (\ref{MFT_ext}).

Next, we update $R$. The values of $R(t+\Delta t,t_3)$ for $t-\Delta t\geq t_3\geq t-T_1$ are obtained by solving Eq.(\ref{define_R}) from the initial values, $R(t,t_3)$, by using the Euler method. For the values of $R(t_1,t)$, we need to solve Eq.(\ref{define_R}) from the unknown initial value, $R(t-T_1,t)$. We obtain the initial value by applying a discrete Fourier transform to Eq.(\ref{define_R}):
\begin{eqnarray}
(1+\mathrm{i}\omega )\widehat{R} (\omega ,t)=2\sigma _0^2\widehat{C} (\omega ,t). \label{fourier_initial}
\end{eqnarray} 
This implicitly assumes that the regularity of the correlation function $2\sigma _0^2C$ is captured by the discrete Fourier transform. Actually, if the microscopic dynamics of the model are chaotic and have a relatively short correlation time, then, $C(t_1,t)\approx 0$ for $t_1\leq t-T_1$ holds and we can use the initial condition, $R(t-T_1,t)\approx 0$, which is given also by Eq.(\ref{fourier_initial}). If the entire dynamics have periodic components with a relatively short period, the Fourier transform detects the corresponding peak of the correlation function in the frequency domain and Eq.(\ref{fourier_initial}) is expected to provide an accurate initial condition. Either one of these two situations is almost always the case if we use a sufficiently large value of $T_1$. The exception among the cases we have investigated in this study is one in which the external input is irregular and the network dynamics retain finite irregularly varying correlation over a very long time. In this case, the above two assumptions do not hold and we do not attempt to obtain a numerical solution from the MFT.

We use the initial value of the solution obtained by solving Eq.(\ref{fourier_initial}) and then applying the inverse discrete Fourier transform. Note that, although the discrete Fourier transform captures decaying and regular patterns in the autocorrelation function, it does not necessarily provide a good approximation to the solution over the entire time domain. Thus, we solve the ordinary differential equation (\ref{define_R}) again from the initial values thus obtained, not directly using the solution obtained from the inverse discrete Fourier transform.

Once we obtain the updated values of $R$, those of $D$ are obtained straightforwardly with the Euler method by noting that Eqs.(\ref{filter_C}) and (\ref{define_R}) yield
\begin{eqnarray}
\left (1+\partial _{t_2} \right )D(s,t_2)=R(s,t_2). \label{recovering_D_from_R}
\end{eqnarray}
We are not concerned about the initial conditions for this equation; for the values of $D(s,t+\Delta t)$ with $s\leq t$, we can use the values of $D(s,t)$ as initial conditions. The value of $D(t+\Delta t, t+\Delta t)$ is almost independent of the initial value, $D(t+\Delta t, t-T_1)$, since the variation at time $t_2=t-T_1$ converges exponentially with unit time constant.

Next, we obtain the updated values of $C(t+\Delta t, s)$, $\overline{\phi } (t+\Delta t)$, and $\widetilde{\phi } (t+\Delta t)$ from $D$ and $m$. They are computed by using Eqs.(\ref{definition_of_statistics}), (\ref{self_consistent_eq_1}), and (\ref{MFT_ext}), which can be written more precisely as
\begin{widetext}
\begin{eqnarray}
C(t_1,t_2)&=&\langle (\phi (z_1)-\overline{\phi } (t_1))(\phi (z_2)-\overline{\phi } (t_2))\rangle \nonumber \\
&&\hspace{-2.7cm} =\int \mathrm{d} \mathcal{N} (w)\hspace{-0.1cm}\int \mathrm{d} \mathcal{N} (y_1)\left \{ \phi \left (D_{11}^{1/2}\left (\sqrt{1-\frac{|D_{12}|}{(D_{11}D_{22})^{1/2}}}y_1+\mathrm{sgn}D_{12}\sqrt{\frac{|D_{12}|}{(D_{11}D_{22})^{1/2}}}w\right )+m_{1}\right ) -\overline{\phi } (t_1)\right \} \nonumber \\
&&\hspace{-1cm} \times \int \mathrm{d} \mathcal{N} (y_2)\hspace{-0.1cm}\left \{ \phi \left (D_{22}^{1/2}\left (\sqrt{1-\frac{|D_{12}|}{(D_{11}D_{22})^{1/2}}}y_2+\sqrt{\frac{|D_{12}|}{(D_{11}D_{22})^{1/2}}}w\right )+m_{2}\right )-\overline{\phi } (t_2)\right \} , \label{recovering_C_again} \\
\overline{\phi } (t_1)&=&\langle \phi (z_1)\rangle \nonumber \\
&&\hspace{-1.5cm}=\int \mathrm{d} \mathcal{N} (w)\hspace{-0.1cm}\int \mathrm{d} \mathcal{N} (y_1)\phi \left (D_{11}^{1/2}\left (\sqrt{1-\frac{|D_{12}|}{(D_{11}D_{22})^{1/2}}}y_1+\mathrm{sgn}D_{12}\sqrt{\frac{|D_{12}|}{(D_{11}D_{22})^{1/2}}}w\right )+m_{1}\right ), \label{recovering_phi_bar_again} \\
\widetilde{\phi } (t_1)+\overline{\phi } (t_1) &=& \langle \phi (z_1+\delta v(t_1))\rangle \nonumber \\
&&\hspace{-3.0cm}=\int \mathrm{d} \mathcal{N} (w)\int \mathrm{d} \mathcal{N} (y_1)\hspace{-0cm}\phi \left (D_{11}^{1/2}\left (\sqrt{1-\frac{|D_{12}|}{(D_{11}D_{22})^{1/2}}}y_1+\mathrm{sgn}D_{12}\sqrt{\frac{|D_{12}|}{(D_{11}D_{22})^{1/2}}}w\right )+m_{1}+\delta v(t_1)\right ). \label{recovering_phi_tilde_again}
\end{eqnarray}
\end{widetext}
In the above, $y_1,y_2$ and $w$ are independent unit Gaussian variables integrated with $\mathrm{d} \mathcal{N} (w)=\exp(-w^2/2)\mathrm{d}w/\sqrt{2\pi }$. We have used the abbreviations $D_{\alpha \beta }=D(t_\alpha ,t_\beta )$ and $m_{\alpha }=m(t_\alpha )$ for $\alpha ,\beta =1,2$. In the calculation of $\widetilde{\phi}$, we use $\delta v(t)\overset{\mathrm{def}}{=}(1+\partial _t)^{-1}\widetilde{I}(t)$. To calculate the double integral in the above efficiently, we use a table that retains the values obtained by performing one of the two integrations with respect to $y_1$ or $y_2$ for a fixed value of $w$. More precisely, we prepare a table consisting of the values of the integral
\begin{eqnarray}
\int \mathcal{D} y\phi (\alpha y+\beta ),
\end{eqnarray}
for different values of $(\alpha ,\beta )$. In the calculation of Eq.(\ref{recovering_C_again}), we perform the double integration by interpolating the values in the table and integrating them with respect to $w$.

Finally, we obtain a realization of the random variable, $\eta (t+\Delta t)$. Recall that the realization probability of $\eta $ is give by Eq.(\ref{numerical_MFE}), which is rewritten as
\begin{eqnarray}
\hspace{-0.7cm} P(\eta (t+\Delta t)|\{ \eta (t-s)\} _{s\geq 0})\propto  \exp \left (-\frac{1}{2} \eta ^TC^{-1}\eta \right ). 
\end{eqnarray}
The above conditional Gaussian distribution has mean $\mu _{t+\Delta t}$ and variance $\nu _{ t+\Delta t}$, with 
\begin{eqnarray}
\mu _{t+\Delta t}&=&c_{t+\Delta t}^TC_{t:t-T_2}^{-1}\eta, \label{equation_for_mu} \\
\nu _{t+\Delta t}&=&C(t+\Delta t,t+\Delta t) \nonumber \\
&&-c_{t+\Delta t}^TC_{t:t-T_2}^{-1}c_{t+\Delta t}. \label{equation_for_nu}
\end{eqnarray}
In these equations, we define the column vector $c_{t+\Delta t}(s_1)\overset{\mathrm{def}}{=} C(s_1,t+\Delta t)$ and matrix $C_{t:t-T_2}(s_1,s_2)\overset{\mathrm{def}}{=} C(s_1,s_2)$ for the restricted range of time indices $t-T_2\leq s_1,s_2\leq t$. Since approximation errors may be larger at the boundary of the time domain, we use a smaller value of $T_2(<T_1)$. From the above, we obtain a realization of $\eta (t+\Delta t)$ using independent unit Gaussian random variable $\xi _{t+\Delta t}$ as
\begin{eqnarray}
\eta (t+\Delta t)=\mu _{t+\Delta t}+\sqrt{\nu _{t+\Delta t}}\xi _{t+\Delta t}. \label{realisation_eta_bar}
\end{eqnarray}
In this way, we obtain all necessary updated values.

For the computation in Eq.(\ref{realisation_eta_bar}), we need to calculate the inverse matrix, $C_{t:t-T_2}^{-1}$. Since our computation of matrix $C$ is based on numerical integration, small amounts of errors are inevitable. When the inverse matrix is computed, the effects of small errors in the small eigenvalues of the matrix can be large. Thus we introduce a small ridge, computing $(C_{t:t-T_2}+\epsilon \mathrm{diag} (C_{t:t-T_2}))^{-1}$ instead of $C_{t:t-T_2}^{-1}$, where the diagonal matrix $\mathrm{diag} (C_{t:t-T_2})$ consists of the diagonal elements of $C_{t:t-T_2}$. This amounts to ignoring the small eigenvalues of $C$. Also, since taking matrix inverse at every timestep is inefficient, we update the inverse matrix using the formula,
\begin{eqnarray}
(&A&+pq^T+qp^T)^{-1}=A^{-1}-\frac{1+a_{pq}}{(1+a_{pq})^2-a_{pp}a_{qq}} \nonumber \\
&&\hspace{-0.7cm}\times \left \{ b_qb_p^T+b_pb_q^T-\frac{a_{pp}}{1+a_{pq}}b_qb_q^T-\frac{a_{qq}}{1+a_{pq}}b_pb_p^T\right \} ,
\end{eqnarray}
for a square symmetric matrix $A$ and column vectors $p$ and $q$ of the corresponding size. 
In the above, we define
\begin{eqnarray}
&&a_{pp}=p^TA^{-1}p,\    a_{pq}=p^TA^{-1}q,\    a_{qq}=q^TA^{-1}q, \nonumber \\
&&b_p=A^{-1}p,\    b_q=A^{-1}q.
\end{eqnarray}
Applying this formula with $A=C_{t:t-T_2}$, $p(t-T_2)=-1$, $p(t-s_1)=0$ for $s_1\neq T_2$, $q(t-T_2)=0$ and $q(s_2)=C(t-T_2,s_2)$ for $t-T_2<s_2\leq t$ gives the inverse matrix of $C_{t:t-T_2+\Delta t}$ in the upper left part of the output matrix. Applying this formula with $A(t+\Delta t,t+\Delta t)=C(t+\Delta t,t+\Delta t)$, $A(s_1,t+\Delta t)=A(t+\Delta t,s_1)=0$, $A(s_1,s_2)=C(s_1,s_2)$, $p(t+\Delta t)=1$, $p(s_1)=0$, $q(t+\Delta t)=0$, and $q(s_1)=C_{\ell }(t+\Delta t, s_1)$ for $t-T_2<s_1,s_2\leq t$ then gives $C_{\ell ,t+\Delta t:t-T_2+\Delta t}^{-1}$.
To avoid the accumulation of numerical errors, we directly compute the inverse matrix every $500$ timesteps. In the main text, we use the following values of the parameters: $T_1=480, T_2=T_1/2$ for Fig.\ref{figure_single_EI_pair_nonzerosum} and Fig.\ref{figure_external_input}, and $T_1=960, T_2=T_1/2$ for Fig.\ref{figure_single_EI_pair}. We use $\epsilon=1.0\times 10^{-6}$ for $g_0=0.25$ in Fig.\ref{figure_single_EI_pair}, $\epsilon =1.0\times 10^{-3}$ for Fig.\ref{figure_single_EI_pair_nonzerosum} and Fig.\ref{figure_external_input}, and $\epsilon =1.0\times 10^{-4}$ for the rest. We use the discrete timesteps with stepsize $\Delta t=15/128$. Each simulation is performed using 16 cores of recent versions of Intel Xeon processors in parallel for a few hours--a few days.
\begin{widetext}

\section{Perturbative expansion for chaotic solutions with mean activities of small amplitudes} \label{perturbative_expansion_chaos}
In Sec.\ref{subsection4A}, we have mentioned a perturbative expansion around $g_0=0$ for the calculation of the moments of the mean activity. This perturbative calculation scheme, however, turns out to not work well. We briefly explain how it is performed and why it does not work well. According to our MFT, the probability distribution over sample paths for the mean activity $m(t)$ is given by $(1+\partial _t)m(t)=\sqrt{2} g_0\eta (t)$ with
\begin{eqnarray}
p(\eta )=\exp \left (-\frac{1}{2} \eta ^{T}C[\eta ]\eta -F[\eta ]-const. \right ).
\end{eqnarray}
In this section, we make explicit the dependence of $C$ and $F$ on $\eta $. Although the above equation is derived for sample paths from a fixed initial condition, we assume that it holds on the entire time axis. This is justified by the intuition that the mixing property of chaotic dynamics keep the calculated moments from being severely affected by the boundary values.

In what follows, for illustration, we focus on the calculation of autocorrelation 
\begin{eqnarray}
\langle m(t)m(t+\tau )\rangle =\frac{g_0^2}{\pi }\int _{\mathbf{R} ^2}e^{\mathrm{i}p_1t-\mathrm{i}p_2(t-\tau )}\frac{1}{(1+\mathrm{i}p_1)(1-\mathrm{i}p_2)}\langle \widehat{\eta } (p_1)\overline{\widehat{\eta }}(p_2)\rangle \mathrm{d} p_1\mathrm{d} p_2,
\end{eqnarray}
in the frequency domain, which is easier to carry out than that in the time domain. Here, $\widehat{\cdot }$ denotes the Fourier transform. Our objective is to compute the following moment: 
\begin{eqnarray}
\langle \widehat{\eta } (p_1)\overline{\widehat{\eta }}(p_2)\rangle =\int \mathcal{D} \widehat{\eta }\mathcal{D} \overline{\widehat{\eta }} \widehat{\eta } (p_1)\overline{\widehat{\eta }}(p_2)\frac{1}{Z}\exp \left (-\frac{1}{2} \overline{\widehat{\eta } }^{T}\widehat{C} [\widehat{\eta }] \widehat{\eta }-\widehat{F} [\widehat{\eta }]\right ), 
\end{eqnarray}
where we define the normalization constant $Z$ and the Fourier transform of the autocorrelation $\widehat{C} [\widehat{\eta }](\omega _1,\omega _2)\overset{\mathrm{def}}{=}\frac{1}{2\pi }\int _{\mathbf{R} ^2}e^{-\mathrm{i}\omega _1t+\mathrm{i} \omega _2s}C[\eta ](t,s)\mathrm{d} t\mathrm{d} s$. We also define $\widehat{F} [\widehat{\eta }]=F[\eta ]$.

For this calculation, we need to compute the perturbed correlation matrix, which can be carried out by differentiating Eq.(\ref{final_linear_equation_of_C_D}) with respect to $\eta $. The first-order response in $D(t,s)$ then obeys the following equation:  
\begin{eqnarray}
(1+\partial _t)(1+\partial _s)\delta D^{(1)}(t,s)&=&a_1(t-s)\delta D^{(1)}(t,s)+a_2(t-s)\delta D^{(1)}(t,t)+a_2(s-t)\delta D^{(1)}(s,s)\nonumber \\
&&\hspace{4cm}+a_3(t-s)\delta m(t)+a_3(s-t)\delta m(s), \label{first_order_response_dynamics_chaos}
\end{eqnarray}
where we define $a_1(t-s)=2\sigma _0^2\langle \phi ^{\prime }(h(t))\phi ^{\prime}(h(s))\rangle _{0}$, $a_2(t,s)=\sigma _0^2\{\langle \phi ^{\prime \prime }(h(t))\phi (h(s))\rangle _{0}-\overline{\phi ^{\prime \prime } _0}\overline{\phi _0}\} $, and $a_3(t-s)=2\sigma _0^2\{\langle \phi ^{\prime}(h(t))\phi (h(s))\rangle _{0}-\overline{\phi _0^{\prime }}\overline{\phi _0}\}$. Here, the average over the unperturbed dynamics with $m(t)\equiv 0$ is denoted by the angle bracket with subscript $0$, and we use abbreviations such as $\overline{\phi _0}\overset{\mathrm{def}}{=}\langle \phi (h(t))\rangle _0$. These coefficients originate from the differentiation of $C(t,s)$ with respect to $D(t,s)$ and $m(t)$, as summarized in Appendix \ref{differentiation_of_C}. Since $\phi (h(t))-\overline{\phi  _0}$ and $\phi ^{\prime }(h(t))-\overline{\phi ^{\prime }_0}$ are odd and even functions of $h(t)$, and since the dynamical variable $h(t)$ is distributed symmetrically around $h(t)=0$ in the unperturbed dynamics, we find $a_3(t,s)=0$. Thus, there is no driving force in the above equation, and we obtain $\delta D^{(1)}(t,s)\equiv 0$.

Next, we examine the second-order response in $D(t,s)$, which obeys the following non-homogeneous linear equation: 
\begin{eqnarray}
(1+\partial _t)(1+\partial _s)\delta D^{(2)}(t,s)&=&a_1(t-s)\delta D^{(2)}(t,s)+a_2(t-s)\delta D^{(2)}(t,t)+a_2(s-t)\delta D^{(2)}(s,s)\nonumber \\
&&+b_1(t-s)\delta m(t)^2+b_1(s-t)\delta m(s)^2+b_2(t-s)\delta m(t)\delta m(s),\label{second_order_response_dynamics_chaos}
\end{eqnarray}
where we define $b_1(t-s)=2\sigma _0^2\{ \langle \phi ^{\prime \prime }(h(t))\phi (h(s))\rangle _{0}-\overline{\phi ^{\prime \prime }_0}\overline{\phi _0}\} $ and $b_2(t-s)=2\sigma _0^2\{ \langle \phi ^{\prime}(h(t))\phi ^{\prime }(h(s))\rangle _{0}-\overline{\phi ^{\prime }_0}^2\} $.

To obtain a solution to this equation, we first compute an approximate solution that satisfies
\begin{eqnarray}
(1+\partial _t)(1+\partial _s)\delta D_0^{(2)}(t,s)=b_2(t-s)\delta m(t)\delta m(s)+\cdots ,
\end{eqnarray}
by ignoring the first three terms of the right-hand side of Eq.(\ref{second_order_response_dynamics_chaos}). 
This equation is solved in the frequency domain as 
\begin{eqnarray}
\widehat{\delta D} _0^{(2)}(\omega _1,\omega _2)=\frac{2g_0^2}{\sqrt{2\pi }(1+\mathrm{i} \omega _1)(1-\mathrm{i} \omega _2)} \int _{\mathbf{R}} \frac{\widehat{b_2}(\omega _3)\widehat{\eta }(\omega _1-\omega _3)\overline{\widehat{\eta }}(\omega _2-\omega _3)}{(1+\mathrm{i} (\omega _1-\omega _3))(1+\mathrm{i}(\omega _2-\omega _3))} \mathrm{d} \omega _3+\cdots 
\end{eqnarray}
The above solution leaves a residual error on the right-hand side of Eq.(\ref{second_order_response_dynamics_chaos}). Then, we recursively make corrections to the solution by considering the residual error as a new non-homogeneous term and solving
\begin{eqnarray}
(1+\partial _t)(1+\partial _s)\delta D_{k+1}^{(2)}(t,s)=a_1(t-s)\delta D_k^{(2)}(t,s)+a_2(t-s)\delta D_k^{(2)}(t,t)+a_2(s-t)\delta D_k^{(2)}(s,s), \label{recursion_relation}
\end{eqnarray}
in the frequency domain.

Assuming that the series thus obtained is convergent, we obtain the second-order response in the form $\delta D^{(2)}(t,s)=\sum _{k=0}^{\infty }\delta D_k^{(2)}(t,s)$. Using this response, we obtain the response in $C(t,s)$ as $\widehat{\delta C}^{(2)}(\omega _1,\omega _2)=(1+\mathrm{i}\omega _1)(1-\mathrm{i}\omega _2)\widehat{\delta D}^{(2)}(\omega _1,\omega _2)/(2\sigma _0^2)$.

Now, we obtain
\begin{eqnarray}
\langle \widehat{\eta }(p_1)\overline{\widehat{\eta }}(p_2)\rangle =\langle \widehat{\eta }(p_1)\overline{\widehat{\eta }}(p_2)\rangle _0+\left \langle  \widehat{\eta }(p_1)\overline{\widehat{\eta }}(p_2) \left (\frac{1}{2}\overline{\widehat{\eta }} ^{T}\widehat{C} [0]^{-1} \widehat{\delta C}^{(2)} \widehat{C} [0]^{-1} \widehat{\eta } -\frac{1}{2}\widehat{C} [0]^{-1}\odot \widehat{\delta C}^{(2)} +\cdots \right )\right \rangle _{\mathrm{c}}.
\end{eqnarray}
Here, the second angle bracket denotes average of just the connected contribution between the external legs and the interaction vertex, as is conventional in diagrammatic calculations \cite{altland2010condensed}. The symbol $\odot $ denotes the element-wise product of two matrices followed by integration with respect to the two argument variables. We omit to write down the terms originating from the complicated dependence of $F$ on $\eta $. In principle, we can calculate the desired moment from the above expansion.

However, this calculation scheme turns out to be difficult to carry out. The difficulty originates from the slow convergence of the sum of the series. Because of this, we need to compute $\delta D_k^{(2)}$ up to a large $k$. However, we find that the diagrammatic calculations of $\delta D_k^{(2)}$ up to a large $k$ requires us to carry out multiple integrals explicitly, and this is computationally intractable. This happens because of the lack of interchangeability between the diagrammatic averaging and the recursion in Eq.(\ref{recursion_relation}). In the main text, because of this difficulty, we restrict ourselves just to the autocorrelation function calculated using the crudest approximation, $\langle \widehat{\eta }(p_1)\overline{\widehat{\eta }}(p_2)\rangle \approx \langle \widehat{\eta }(p_1)\overline{\widehat{\eta }}(p_2)\rangle _0$ [Fig.\ref{figure_statistics_of_dynamics}(a) and (c)].

\section{Perturbative stability analysis of fixed points} \label{perturbative_expansion_fixed_point}
In this section, we analyze the stability of fixed-point solutions observed in Sec.\ref{perturbative_analysis_fp} of the main text. We consider the case in which the mean activity initially takes a constant value, $m(t)=m_{\mathrm{f}}$ for $t\leq 0$. According to the MFT, the mean activity is determined by 
\begin{eqnarray}
(1+\partial _t)\delta m(t)=\sqrt{2} g_0\eta (t)-m_{\mathrm{f}}, \label{mean_equation_fixed_point}
\end{eqnarray}
where we define, $\delta m(t)\overset{\mathrm{def}}{=}m(t)-m_{\mathrm{f}}$. To analyze the dynamics of the mean activity, we first need to examine the associated microscopic Gaussian fluctuations determined by the past values of $m(t)$. For a certain range of constant values of $m(t)$, these Gaussian fluctuations have a stable fixed-point solution. For these fixed points, the correlation matrices $D(t,s)$ and $\widetilde{C}(t,s)$ take constant values, $D_0$ and $\widetilde{C}_0$, which are obtained by solving the self-consistent equation, Eq.(\ref{self-consistent_equation_fp}). As described in the main text, we assume that $\eta (t)=m_{\mathrm{f}}/\sqrt{2}g_0$ holds for $t\leq 0$. Then, the network state stays at the fixed point without requiring external inputs, and we have $\delta m(t)=0$ for $t>0$ if there is no external input in $t>0$. This condition is expected to be satisfied for some value of $m_{\mathrm{f}}$ with a non-zero probability (see the discussion at the end of this section as well). We then examine the linear response for $t>0$ to temporary external perturbative inputs, collectively denoted by $\mathbf{p}$.

We first examine responses in the correlation matrices up to the second order in $\mathbf{p}$, which determines $O(|\mathbf{p}|)$ response in $\eta (t)$. By simply differentiating the dynamical equation for the autocorrelation function, Eq.(\ref{final_linear_equation_of_C_D}), we find that the first-order response in $D(t,s)$, which is denoted by $\delta D^{(1)}(t,s)$, obeys 
\begin{eqnarray}
\hspace{-1cm} (1+\partial _t)(1+\partial _s)\delta D^{(1)}(t,s)&=&a_1\delta D^{(1)}(t,s)+a_2(\delta D^{(1)}(t,t)+\delta D^{(1)}(s,s))+a_3(\delta m(t)+\delta m(s))+p_1(t)+p_1(s), \label{1st_order_linear_response}
\end{eqnarray} 
where the term, $p_1(t)$, represent the effects of the input component that is correlated with the unperturbed neuronal fluctuations, and we define $a_1=2\sigma _0^2\langle \phi ^{\prime }(h)^2\rangle _0$, $a_2=\sigma _0^2\langle \phi ^{\prime \prime }(h)\phi (h)\rangle _0$ and $a_3=2\sigma _0^2\langle \phi ^{\prime }(h)\phi (h)\rangle _0$. These coefficients originate from the differentiation of $\widetilde{C}(t,s)$ with respect to $D(t,s)$ and $m(t)$, which is summarized in Appendix \ref{differentiation_of_C}. Note that these coefficients must be appropriately replaced by the values of centered statistics, such as $\sigma _0^2\langle (\phi ^{\prime \prime }(h)-\overline{\phi _0^{\prime \prime }})(\phi (h)-\overline{\phi _0})\rangle _0$, for the case with finely-tuned synaptic weights. Here, we use abbreviations such as $\overline{\phi _0^{\prime \prime }}=\langle \phi ^{\prime \prime }(h)\rangle _0$ for the statistics averaged over the unperturbed dynamics. 
The solution for the above equation, with boundary condition, $\delta D^{(1)}(t,s)=0$ for $t,s\leq 0$, satisfies 
\begin{eqnarray}
\delta D^{(1)}(t,s)&=&\int _{0}^te^{-(1-a_1)(t-\tau )}(a_2\delta D^{(1)}(\tau ,\tau )+a_3\delta m(\tau )+p_1(\tau ))\mathrm{d} \tau \nonumber \\
&&\hspace{2cm}+\int _0^se^{-(1-a_1)(s-\tau )}(a_2\delta D^{(1)}(\tau ,\tau )+a_3\delta m(\tau )+p_1(\tau ))\mathrm{d} \tau . \label{1st_order_perturbative_self_consistent_equation}
\end{eqnarray}
Putting $s=t$ and differentiating both sides of the above equation with respect to $t$, we have
\begin{eqnarray}
((1-a_1-2a_2)+\partial _t)\delta D^{(1)}(t,t)=2(a_3\delta m(t)+p_1(t)), 
\end{eqnarray} 
which then gives 
\begin{eqnarray}
\delta D^{(1)}(t,t)=2\int _0^te^{-(1-a_1-2a_2)(t-\tau )}(a_3\delta m(\tau )+p_1(\tau ))\mathrm{d} \tau . \label{1st_order_perturbation_in_D}
\end{eqnarray}
Comparing the above solution with Eq.(\ref{1st_order_perturbative_self_consistent_equation}) gives the solution for $\delta D^{(1)}(t,s)$. This response is bounded if $1-a_1-2a_2>0$ and if $\delta m$ and $p_{1}$ are bounded.

From the above analysis, by putting $\delta D^{(1)}(t,s)=r(t)+r(s)$, we obtain 
\begin{eqnarray}
D(t,s)\approx D_0+r(t)+r(s)+O(|\mathbf{p}|^2). \label{first_order_representation_of_D_given_delta_m}
\end{eqnarray}
Noting that the first-order response in $\overline{\phi }(t)\overline{\phi} (s)$ is given by 
\begin{eqnarray}
\delta (\overline{\phi }(t)\overline{\phi} (s))=\overline{\phi _0}\left (\frac{1}{2}\overline{\phi _0^{\prime \prime}}\delta D(t,t)+\overline{\phi _0^{\prime }}\delta m(t)+\frac{1}{2}\overline{\phi _0^{\prime \prime}}\delta D(s,s)+\overline{\phi _0^{\prime }}\delta m(s)\right ),
\end{eqnarray}
we have the correlation matrix for $(1+\partial _t)^{-1}\eta (t)$,
\begin{eqnarray}
\hspace{-1cm}V(t,s)&\overset{\mathrm{def}}{=}& (1+\partial _t)^{-1}(1+\partial _s)^{-1}\left (\widetilde{C} (t,s)-\overline{\phi }(t)\overline{\phi }(s)\right )\nonumber \\
&=&\left (\sqrt{V_0} +v(t)\right )\left (\sqrt{V_0} +v(s)\right )+O(|\mathbf{p}|^2),\label{first_order_representation_of_V_given_delta_m}
\end{eqnarray}
where we define $V_0=\frac{1}{2\sigma _0^2}D_0-\overline{\phi } _0^2$ and 
\begin{eqnarray}
\hspace{-0.5cm}v(t)=\frac{1}{2\sigma _0^2\sqrt{V_0}}\left \{ \left (1-\frac{2\sigma _0^2\overline{\phi _0}\overline{\phi _0^{\prime \prime }}}{a_1+2a_2}\right )r(t)+2\sigma _0^2\left (\frac{\overline{\phi _0}\overline{\phi _0^{\prime \prime }}a_3}{a_1+2a_2} -\overline{\phi _0}\overline{\phi _0^{\prime }}\right )(1+\partial _t)^{-1}\delta m(t)+\frac{2\sigma _0^2\overline{\phi _0}\overline{\phi _0^{\prime \prime }}}{a_1+2a_2}(1+\partial _t)^{-1}p_1(t)\right \} . \label{definition_v_fp}
\end{eqnarray}
Noting that $V(t,s)=V_0$ and $v(t)=v(s)=0$ holds for $t,s\leq 0$, and that $V(t,s)$ is the correlation matrix that determines the realization probability of $(1+\partial _t)^{-1}\eta (t)$, we have 
\begin{eqnarray}
(1+\partial _t)^{-1}\eta (t)&=&\frac{m_{\mathrm{f}}}{\sqrt{2} g_0}\left (1+\frac{v(t)}{\sqrt{V_0}}\right ) +\delta \eta ^{(2)}(t)+O(|\mathbf{p}|^2),  \label{eta_equation_1st_fp}
\end{eqnarray}
where $\delta \eta ^{(2)}(t)$ represents $O(|\mathbf{p}|)$ fluctuations due to the higher-order response in $D(t,s)$. In the case with finely-tuned synaptic weights, the correlation matrix for $(1+\partial _t)^{-1}\eta (t)$ is given by $D(t,s)/2\sigma _0^2$ and we have the same representation with $V_0=D_0/2\sigma _0^2$ and $v(t)=r(t)/2\sigma _0^2\sqrt{V_0}$.

Next, we consider the second-order response in $D(t,s)$, which is denoted by $\delta D^{(2)}(t,s)$. Differentiating Eq.(\ref{1st_order_linear_response}) once again, we obtain
\begin{eqnarray}
(1+\partial _t)(1+\partial _s)\delta D^{(2)}(t,s)&=&a_1\delta D^{(2)}(t,s)+a_2(\delta D^{(2)}(t,t)+\delta D^{(2)}(s,s))+b_1\delta D^{(1)}(t,s)^2\nonumber \\
&&\hspace{-2cm}+b_2(\delta D^{(1)}(t,t)^2+\delta D^{(1)}(s,s)^2)+b_3\delta D^{(1)}(t,t)\delta D^{(1)}(s,s)+b_4(\delta D^{(1)}(t,t)+\delta D^{(1)}(s,s))\delta D^{(1)}(t,s)\nonumber \\
&&+b_5(\delta m(t)^2+\delta m(s)^2)+b_6\delta m(t)\delta m(s)+\sum _{i}p_{2,i}(t)p_{2,i}(s), \label{2nd_order_linear_response} 
\end{eqnarray}
where we define $b_1=2\sigma _0^2\langle \phi ^{\prime \prime }(h)\phi ^{\prime \prime }(h)\rangle _0$, $b_2=\frac{\sigma _0^2}{2}\langle \phi ^{\prime \prime \prime \prime}(h)\phi (h)\rangle _0$, $b_3=\sigma _0^2\langle \phi ^{\prime \prime }(h)^2\rangle _0$, $b_4=2\sigma _0^2\langle \phi ^{\prime \prime \prime }(h)\phi ^{\prime }(h)\rangle _0$, $b_5=2\sigma _0^2\langle \phi ^{\prime \prime }(h)\phi (h)\rangle _0$ and $b_6=4\sigma _0^2\langle \phi ^{\prime }(h)\phi ^{\prime }(h)\rangle _0$. We also define the random part of the input, $p_{2,i}(t)$.

The solution for the above non-homogeneous linear equation is obtained as the superposition of special solutions for the equations with each of the non-homogeneous terms on the right-hand side plus a solution for the homogeneous equation. Noting $\delta D^{(1)}(t,s)=r(t)+r(s)$, the non-homogeneous terms are rewritten as: 
\begin{eqnarray}
\hspace{-1.5cm} (b_1+4b_2+2b_4)(r(t)^2+r(s)^2)+b_5(\delta m(t)^2+\delta m(s)^2)+(2b_1+4b_3+4b_4)r(t)r(s)+b_6\delta m(t)\delta m(s)+\sum _{i}p_{2,i}(t)p_{2,i}(s). \label{non-homogeneous_terms_fp}
\end{eqnarray}
We ignore the first two terms above, because they only yield responses of $O(|\mathbf{p}|^2)$ magnitude in $\eta (t)$. This can be seen by checking that these two terms only make corrections to Eq.(\ref{first_order_representation_of_V_given_delta_m}) of the following form:
\begin{eqnarray}
V(t,s)\approx (\sqrt{V_0}+v(t)+O(|\mathbf{p}|^2))(\sqrt{V_0}+v(s)+O(|\mathbf{p}|^2))+o(|\mathbf{p}|^2).
\end{eqnarray}
Also note that the response in $\overline{\phi }(t)\overline{\phi }(s)$ due to $\delta D^{(2)}(t,s)$ is negligible for the same reason.  
Thus, we are interested in the non-homogeneous equations of the following form: 
\begin{eqnarray}
(1+\partial _t)(1+\partial _s)\delta D^{(2)}(t,s)=a_1\delta D^{(2)}(t,s)+a_2(\delta D^{(2)}(t,t)+\delta D^{(2)}(s,s))+q_0(t)q_0(s). \label{typical_second_order_equation_fp}
\end{eqnarray}
In solving this equation, we first ignore the unknown quantities on the right-hand side and obtain the following approximate solution: 
\begin{eqnarray}
\delta D^{(2)}_{q,1}(t,s)=q_1(t)q_1(s)\overset{\mathrm{def}}{=}\left \{ (1+\partial _t)^{-1}q_0(t)\right \}\left \{(1+\partial _s)^{-1}q_0(s)\right \}.
\end{eqnarray}
With this solution, residual error $a_1q_1(t)q_1(s)+a_2(q_1(t)^2+q_1(s)^2)$ remains on the right-hand side of Eq.(\ref{typical_second_order_equation_fp}). We then solve Eq.(\ref{typical_second_order_equation_fp}) by regarding the residual-error term as a new non-homogeneous term and by ignoring the two unknown terms on the right-hand side again. Then, we obtain the following two corrections: 
\begin{eqnarray}
\delta D^{(2)}_{q,2}(t,s)&=&q_2(t)q_2(s)\overset{\mathrm{def}}{=}\left \{ \sqrt{a_1}(1+\partial _t)^{-1}q_1(t)\right \} \left \{ \sqrt{a _1}(1+\partial _s)^{-1}q_1(s)\right \} ,\label{second_order_recursion_fixed_point} \\
\delta D^{(2)\prime }_{q,2}(t,s)&=&a_2\left \{ (1+\partial _t)^{-1}(q_1^2)(t)+(1+\partial _s)^{-1}(q_1^2)(s)\right \} .
\end{eqnarray}
The second term in the above is again negligible because of its $O(|\mathbf{p}|^2)$ contribution to $\eta $. We can then recursively obtain $\delta D^{(2)}_{q,j}$, ($j\geq 2$) in the same manner as above and obtain the relevant part of the solution for Eq.(\ref{2nd_order_linear_response}) as $\sum _{1\leq j<\infty }\delta D^{(2)}_{q,j}(t,s)$. If we have $a_1<1$, this series is convergent because we have,
\begin{eqnarray}
\| \widehat{q_{j+1}}\| _1=\left \|\sqrt{a_1}\widehat{(1+\partial _t)^{-1}q_j}\right \| _1\leq \sqrt{a_1}\| \widehat{q_j}\| _1.
\end{eqnarray}
Here, $\widehat{\cdot }$ denotes the Fourier transform, and $\| \cdot \| _1$ denotes the $L^1$-norm.
From the above analysis, we now have the first- and second-order responses in $D(t,s)$, which yields
\begin{eqnarray}
\hspace{-1cm}D(t,s)\approx D_0+r(t)+r(s)+\sum _{j, 1\leq \ell <\infty }d_{2,j\ell }(t)d_{2,j\ell }(s)+\sum _{j,1\leq \ell <\infty }d_{3,j\ell }(t)d_{3,j\ell }(s). \label{representation_of_D_given_delta_m}
\end{eqnarray}
In the above, we obtain $d_{2,j\ell}(t)$ by applying the recursion relation in Eq.(\ref{second_order_recursion_fixed_point}) with $q_{\ell }(t)=d_{2,j\ell}(t)$ and $q_{0}(t)=d_{2,1,0}(t)=\sqrt{2b_1+4b_3+4b_4}r(t)$ or $d_{2,2,0}(t)=\sqrt{b_6}\delta m(t)$. We also define $d_{3,j\ell }(t)$ by applying the recursion relation in Eq.(\ref{second_order_recursion_fixed_point}) with $q_{\ell }(t)=d_{3,j\ell}(t)$ and $q_{0}(t)=d_{3,j,0}(t)=p_{2,j}(t)$. Note that $v(t)v(s)$ needs to be suitably subtracted from these terms to compensate the corresponding term in Eq.(\ref{first_order_representation_of_V_given_delta_m}), which modifies the definitions of $d_{2,j1}$ and $d_{3,j1}$ described above. We omit the precise expressions of these terms, because they do not affect our conclusion.

Noting that the second-order response in $D(t,s)$ does not evoke a response in $\overline{\phi }(t)\overline{\phi }(s)$ which leads to $O(|\mathbf{p}|)$ response in $\eta $, we have the relevant part of the response in $V(t,s)$, 
\begin{eqnarray}
\hspace{-1cm}V(t,s)\approx \left (\sqrt{V_0} +v(t)\right )\left (\sqrt{V_0} +v(s)\right )+\frac{1}{2\sigma _0^2}\left \{ \sum _{j, 1\leq \ell <\infty }d_{2,j\ell }(t)d_{2,j\ell }(s)+\sum _{j,1\leq \ell <\infty }d_{3,j\ell }(t)d_{3,j\ell }(s)\right \} .\label{expansion_of_V_ts}
\end{eqnarray}   
Putting $d_1(t)=v(t)/\sqrt{V_0}$, we now have $\delta \eta ^{(2)}$ in Eq.(\ref{eta_equation_1st_fp}) and obtain
\begin{eqnarray}
(1+\partial _t)^{-1}\eta (t)&=&\frac{m_{\mathrm{f}}}{\sqrt{2}g_0}\left (1+d_1(t)\right )+\frac{1}{\sqrt{2}\sigma _0}\left (\sum _{j,\ell }\xi _{j\ell } d_{2,j\ell }(t)+\sum _{j,\ell }\xi _{j\ell }^{\prime }d_{3,j\ell }(t)\right )+O(|\mathbf{p}|^2), \label{self_consistent_fp}
\end{eqnarray}
where we define {\it i.i.d.} unit Gaussian random variables, $\{ \xi _{j\ell }\} _{j,\ell}$ and $\{ \xi _{j\ell }^{\prime }\} _{j,\ell}$. Here, note that the outer-product representation in Eq.(\ref{expansion_of_V_ts}) gives a transform of the form of $(1+\partial _t)^{-1}\eta =U\xi $, $UU^T=V$. This gives a representation of $\eta $ with unit Gaussian variables, $\xi $, as we discussed in Appendix \ref{appendix:O}. Together with the equation for $m(t)$ for $t>0$,
\begin{eqnarray}
(1+\partial _t)\delta m(t)=\sqrt{2}g_0\delta \eta (t)+p_0(t), \  \  \  \delta \eta (t)=\eta (t)-\frac{m_{\mathrm{f}}}{\sqrt{2}g_0}, \label{delta_m_equation}
\end{eqnarray}
Eq.(\ref{self_consistent_fp}) gives a self-consistent equation that the first-order responses in $m(t)$ must satisfy. Here, we define the uniform component of the input, $p_0(t)$.

From the above relation, we obtain the first-order response in $m(t)$ for a given set of values of $\{ \xi _{j\ell }\} $ and $\{ \xi _{j\ell }^{\prime }\} $ by further iteration. Initially, calculating $d_{1}(t)$ and $d_{3,j\ell }(t)$ for $\delta m(t)=0$ [denoted by $d_{1}^{(0)}(t)$ and $d_{3,j\ell }^{(0)}(t)$, respectively], we have,
\begin{eqnarray}
\delta m^{(0)}(t)=(1+\partial _t)^{-1}p_0(t)+m_{\mathrm{f}} d_1^{(0)}(t) +\frac{g_0}{\sigma _0}\sum _{j,\ell }\xi _{j\ell }^{\prime }d_{3,j\ell }^{(0)}(t).
\end{eqnarray}
The solution for the above equation needs the following correction to the right-hand side of the first of Eq.(\ref{delta_m_equation}): 
\begin{eqnarray}
\sqrt{2}g_0\delta \eta ^{(1)}(t)=(1+\partial _t)\left \{ m_{\mathrm{f}}d_1^{(1)}(t) + \frac{g_0}{\sigma _0} \sum _{j,\ell }\xi _{j\ell }d_{2,j\ell } ^{(1)}(t)\right \} +O(|\mathbf{p}|^2), \label{computing_residual_error_fixed_point}
\end{eqnarray}
where $d_{1}^{(1)}(t)$ and $d_{2,j\ell }^{(1)}(t)$ are the corrections to $d_{1}(t)$ and $d_{2,j\ell }(t)$ due to the change $\delta m^{(0)}(t)$. We then recursively obtain $\delta m^{(j)}$ and $\delta \eta ^{(j)}$ by alternately correcting the errors in Eqs.(\ref{self_consistent_fp}) and (\ref{delta_m_equation}). The sum of the series thus obtained gives the desired first-order solution, $\delta m(t)=\sum _{j\geq 0}\delta m^{(j)}(t)$, and the sum of this series converges with a non-zero probability if we have $a_1<1$ and 
\begin{eqnarray}
\left \| \widehat{d_{1}^{(k)}}\right \| _{1}\leq \theta _1\| \widehat{\delta m^{(k)}}\|  _{1},
\end{eqnarray}
with $m_{\mathrm{f}}\theta _1<1$. From Eqs.(\ref{1st_order_perturbation_in_D}) and (\ref{definition_v_fp}), we find 
\begin{eqnarray}
\theta _1=\frac{1}{2\sigma _0^2V_0}\sup _{\omega }\left | \frac{a_3}{1-a_1-2a_2+\mathrm{i}\omega }\left (1-\frac{2\sigma _0^2\overline{\phi _0}\overline{\phi _0^{\prime \prime }}}{a_1+2a_2}\right )+\frac{2\sigma _0^2}{1+\mathrm{i}\omega }\left ( \frac{a_3\overline{\phi _0}\overline{\phi _0^{\prime \prime }}}{a_1+2a_2}-\overline{\phi _0}\overline{\phi _0^{\prime}}\right ) \right | . \label{theta_1_definition}
\end{eqnarray} 
Here, note that the norm of $d_{2,j\ell }^{(k)}$ exponentially decreases as index $\ell $ increases and the norm of $d_{2,j0}^{(k)}$ is bounded by a certain multiple of the norm of $\delta m^{(k)}$. This implies that, for any positive $\theta _2$,
\begin{eqnarray}
\left \| \sqrt{2} g_0\sum _{j,\ell }\xi _{j\ell }\widehat{d_{2,j\ell }^{(k)}}\right \| _{1}<\theta _2\| \widehat{\delta m^{(k)}}\| _{1}
\end{eqnarray}
holds with a non-zero probability. The convergence in 1-norm in the frequency domain implies the uniform convergence in the time domain, which asserts that $p(t)\rightarrow 0$ implies $\delta m(t)\rightarrow 0$ as $t\rightarrow \infty $ with the aid of the dominated convergence theorem. Hence we have proved the linear stability with a non-zero probability. Also note that a single configuration of the random connectivity of the network corresponds to a single set of values for $\{ \xi _{j\ell }\} $. Otherwise, the above solution is not consistent with the linearity of the response.

Besides the above stability result, we can also examine how the response dynamics diverge depending on the configuration. For example, suppose that the random coefficients $\xi _{j\ell}$ take large values for small $\ell $, and that all the inputs initially take constant values for a sufficiently long time. Then, the constructed solution is likely to diverge at some $t$, because the divergence of $d_{2,j\ell }^{(k)}$ for small $\ell $ and for $k\rightarrow \infty$ cannot be compensated by the other terms. However, even for such a divergent solution, if we restrict the domain of the solution to a short time interval, $t\in [0, t_0]$, the supremum of $d_{2,j\ell}^{(k)}$ can be controlled to decay as index $k$ increases. Here, recall that $d_{2,j\ell }^{(k)}$ is obtained by recursively applying $(1+\partial _t)^{-1}$ to $d_{2,j0}^{(k)}$ and suppose to choose a very short time interval compared with the unit decay constant of $(1+\partial _t)^{-1}$. This consideration indicates that, for any values of $\xi _{j\ell }$ and $\xi _{j\ell }^{\prime }$, the constructed solution converges over a short time interval.

Also, suppose the case in which the equality, $m_{\mathrm{f}}=\sqrt{2}g_0\eta (0)$, is slightly violated and that the mean activity is initially clamped to $m_{\mathrm{f}}$ by an external input until the clamping input is removed at $t=0$. Then, the inequality introduces a small constant, uniform, input term to the above self-consistent equation for $\delta m$. Up to the first-order, this results in the convergence of $\delta m$ and $\eta -m_{\mathrm{f}}/\sqrt{2} g_0$ to non-zero values for $t\rightarrow \infty$, with the same non-zero probability as that for the above stability to an external input. This indicates that a non-zero interval of values of $\eta (0)$ around $m_{\mathrm{f}}/\sqrt{2} g_0$ results in convergence to nearby fixed points. Thus, we reasonably expect the existence of fixed points with a non-zero probability, although a rigorous proof of this needs careful evaluation of all higher-order terms of the perturbative expansion.

\section{Perturbative stability analysis of regular oscillatory solutions} \label{perturbative_expansion_oscillation}
In this section, we extend the perturbation analysis developed in Appendix \ref{perturbative_expansion_fixed_point} and analyze the stability of the regular oscillations observed in the main text. Although this analysis is along the same lines as that in Appendix \ref{perturbative_expansion_fixed_point}, much more complex calculations are required for the present case. The complexity of these calculations would make it harder to grasp the basic idea behind it. Thus, we recommend readers to first check Appendix \ref{perturbative_expansion_fixed_point} and to familiarize themselves with the basic idea before reading this section. In what follows, we present a theoretical framework of the analysis first and then show the values of the derived bounding constants for concrete cases at the end of the section.

Suppose that the mean activity of a network is initially set to a periodic orbit $m_{\mathrm{o}}(t)$ for $t\leq 0$ and that temporary external perturbative inputs, collectively denoted by $\mathbf{p}$, are applied for $t>0$. Also suppose that the neuronal fluctuations are coherent and phase locked to the oscillation for $t\leq 0$. As we have seen in the main text, such coherent neuronal fluctuations are found by iteratively applying the self-consistent equation: 
\begin{eqnarray}
(1+\partial _t)(1+\partial _s)D(t,s)=2\sigma _0^2\widetilde{C}(t,s),
\end{eqnarray}
with the fixed mean activity $m_{\mathrm{o}}(t)$. Concretely, we first transform the autocorrelation function on the right-hand side to the frequency domain: 
$\widehat{\widetilde{C}} (\omega _1,\omega _2)=\frac{1}{2\pi }\int e^{-\mathrm{i}\omega _1t}e^{\mathrm{i}\omega _2s}\widetilde{C}(t,s)\mathrm{d} t\mathrm{d} s$. Then, the left-hand side is obtained by a simple algebraic computation in the frequency domain. After transforming this back to the time domain, we update the values of $\widetilde{C}(t,s)$ using Eq.(\ref{recovering_C_again}). Although this type of iterative approach is thought to be a heuristic method for finding a solution, the convergence argument we make below provides criteria for judging whether the solutions thus obtained are stable. In the unperturbed dynamics, we have the following eigen-decomposition of the correlation matrix for $(1+\partial _t)^{-1}\eta (t)$: $V(t,s)\overset{\mathrm{def}}{=}(1+\partial _t)^{-1}(1+\partial _s)^{-1}(\widetilde{C} (t,s)-\overline{\phi }(t)\overline{\phi }(s))$$=\sum _i\lambda _iv_i(t)\overline{v} _i(s)$ with real eigenvectors $v_i(t)$. If the microscopic part of the dynamics is coherent, the eigenvectors can be expanded with the Fourier basis, $\mathbf{e} _{\ell}(t)=e^{\mathrm{i} \ell \omega _0t}$, for the basic frequency $\omega _0$, as $v_i(t)=\sum _{\ell }v_{i\ell }\mathbf{e} _{\ell }(t)$. In the initial unperturbed dynamics, the driving-force term $\eta (t)$ is given by
\begin{eqnarray}
(1+\partial _t)m_{\mathrm{o}}(t)=\sqrt{2} g_0\eta (t)+g_0\widetilde{\phi }(t),\  \  \  \eta (t)=(1+\partial _t)\sum _{i}\sqrt{\lambda _i}\xi _{\mathrm{o}, i}v_i(t),\  \  \  (t\leq 0), \label{eigendecomp_of_examined_dynamics}
\end{eqnarray}
for a suitable set of values for $\{ \xi _{\mathrm{o} ,i}\} _i$. Otherwise, the sample path for $\eta (t)$ is never realized [cf. Eq.(\ref{probability_density_simplified_eta})]. Here, the term $\widetilde{\phi }$ is the external driving-force term defined in Eq.(\ref{MFT_ext}). For the model without external inputs, we set $\widetilde{\phi }(t)$ to zero. For $t>0$, the mean activity obeys the following dynamical equation: 
\begin{eqnarray}
(1+\partial _t)\delta m(t)=\sqrt{2} g_0\eta (t)-(1+\partial _t)m_{\mathrm{o}}(t)+g_0\delta \widetilde{\phi } (t)+p_0(t), \label{initial_delta_m_equation_osc}
\end{eqnarray}
where we define $\delta m(t)=m(t)-m_{\mathrm{o}}(t)$ and define $\delta \widetilde{\phi }(t)$ as the deviation of the function $\widetilde{\phi } (t)$ from the initial periodic orbit. The last term, $p_0(t)$, represents the effect of the uniform component of the input.

To analyze the above response dynamics, we first examine how a change in $m(t)$ evokes a response in $D(t,s)$ for $t,s\geq 0$. The first-order response in $D(t,s)$ is denoted by $\delta D^{(1)}(t,s)$ and obeys
\begin{eqnarray}
\hspace{-1cm} (1+\partial _t)(1+\partial _s)\delta D^{(1)}(t,s)&=&a_1(t,s)\delta D^{(1)}(t,s)+a_2(t,s)\delta D^{(1)}(t,t)+a_2(s,t)\delta D^{(1)}(s,s)\nonumber \\
&&+a_3(t,s)\delta m(t)+a_3(s,t)\delta m(s)+\sum _i\sqrt{\lambda _i}((1+\partial _t) v_i(t)p_{1,i}(s) +p_{1,i}(t)(1+\partial _s)v_i(s)), \label{1st_order_linear_response_osc}
\end{eqnarray} 
where we define $a_1(t,s)=2\sigma _0^2\langle \phi ^{\prime }(h(t))\phi ^{\prime}(h(s))\rangle _0$. $a_2(t,s)=\sigma _0^2\langle \phi ^{\prime \prime }(h(t))\phi (h(s))\rangle _0$, and $a_3(t,s)=2\sigma _0^2\langle \phi ^{\prime}(h(t))\phi (h(s))\rangle _0$. For the model with finely-tuned synaptic weights, the above coefficient functions must be suitably replaced. The coefficients originate from the differentiation of $\widetilde{C}(t,s)$ and $C(t,s)$ with respect to $D(t,s)$ and $m(t)$, which is summarized in Appendix \ref{differentiation_of_C}. The term $p_{1,i}$ represents the effects of the component of the input that is correlated with the $i$-th mode of the unperturbed neuronal fluctuations.

We also note that the first-order response in $V(t,s)$ is given by 
\begin{eqnarray}
\hspace{-2cm} (1+\partial _t)(1+\partial _s)\delta V^{(1)}(t,s)&=&\frac{1}{2\sigma _0^2}\left (a_1(t,s)\delta D^{(1)}(t,s)+\widetilde{a_2}(t,s)\delta D^{(1)}(t,t)+\widetilde{a_2}(s,t)\delta D^{(1)}(s,s)\right. \nonumber \\
&&\left. +\widetilde{a} _3(t,s)\delta m(t)+\widetilde{a} _3(s,t)\delta m(s)+\sum _i\sqrt{\lambda _i}((1+\partial _t) v_i(t)p_{1,i}(s) +p_{1,i}(t)(1+\partial _s)v_i(s))\right ) ,
\end{eqnarray}  
where we define $\widetilde{a} _2(t,s)=2\sigma _0^2(\langle \phi ^{\prime \prime }(h(t))\phi (h(s))\rangle _0-\overline{\phi ^{\prime \prime }}_0(t)\overline{\phi }_0(s))$, and $\widetilde{a} _3(t,s)=2\sigma _0^2(\langle \phi ^{\prime}(h(t))\phi (h(s))\rangle _0-\overline{\phi ^{\prime}}_0(t)\overline{\phi }_0(s))$. For the model with finely-tuned synaptic weights, the response dynamics for $D(t,s)$ and $V(t,s)$ are the same up to a multiplication constant. 

A solution of the form $\delta V^{(1)}(t,s)=\sum _i\sqrt{\lambda _i}(v_i(t)\overline{\delta v} _i(s)+\delta v _i(t)\overline{v} _i(s))$ is obtained from the above equations. By initially ignoring the first three terms on the right-hand side of this equation and using the expansion $a_3(t,s)=\sum _{ij}a_{3,ij}\mathbf{e} _i(t)\overline{\mathbf{e}} _{j}(s)$, we obtain the following approximate solution: 
\begin{eqnarray}
\hspace{-1.5cm}\delta D_0^{(1)}(t,s)=\sum _i\sqrt{\lambda _i}(v_i(t)\overline{\delta d} _{0,i}(s)+\delta d _{0,i}(t)\overline{v} _i(s)),\  \  \  \delta d_{0,i}(t)=\sum _j\overline{M} _{3,ji}(1+\partial _t)^{-1}(\mathbf{e} _j(t)\delta m(t))+(1+\partial _t)^{-1}p_{1,i}(t), \label{compute_delta_d_0}
\end{eqnarray}
where we define $\mathbf{e} _{\ell }(t)=\sum _ie_{\ell i}v_i(t)$ and $M_{3,ji}=\sum _{\ell }\overline{a} _{3,j\ell }e_{\ell i}/\sqrt{\lambda _i}(1+\mathrm{i}\ell \omega _0)$. We also have 
\begin{eqnarray}
\hspace{-1.5cm}\delta V_0^{(1)}(t,s)=\sum _i\sqrt{\lambda _i}(v_i(t)\overline{\delta v} _{0,i}(s)+\delta v _{0,i}(t)\overline{v} _i(s)),\  \  \  \delta v_{0,i}(t)=\frac{1}{2\sigma _0^2}\sum _j\overline{\widetilde{M} } _{3,ji}(1+\partial _t)^{-1}(\mathbf{e} _j(t)\delta m(t))+(1+\partial _t)^{-1}p_{1,i}(t), \label{compute_delta_v_0}
\end{eqnarray}
where we define $\widetilde{M} _{3,ji}=\sum _{\ell }\overline{\widetilde{a} } _{3,j\ell }e_{\ell i}/\sqrt{\lambda _i}(1+\mathrm{i}\ell \omega _0)$ with $\widetilde{a} _3(t,s)=\sum _{ij}\widetilde{a}_{3,ij}\mathbf{e} _i(t)\overline{\mathbf{e}} _{j}(s)$.

Since this solution leaves a residual error on the right-hand side of Eq.(\ref{1st_order_linear_response_osc}), we make corrections recursively by solving 
\begin{eqnarray}
(1+\partial _t)(1+\partial _s)\delta D_{k+1}^{(1)}(t,s)=a_{1}(t,s)\delta D_k^{(1)}(t,s)+a_2(t,s)\delta D_k^{(1)}(t,t)+a_2(s,t)\delta D_k^{(1)}(s,s), \  \  \  (k\geq 0). \label{recursion_first_order_oscillation}
\end{eqnarray}
It is easily seen that each additional response can be represented as $\delta D_k^{(1)}(t,s)=\sum _i\sqrt{\lambda _i}(v_{i}(t)\overline{\delta d} _{k,i}(s)+\delta d _{k,i}(t)\overline{v} _{i}(s))$, with 
\begin{eqnarray}
\hspace{-1.5cm} \delta d_{k+1, i}(t)=\frac{1}{\sqrt{\lambda _i}} \sum _{j,\ell}M_{12,j\ell i}(1+\partial _t)^{-1}(\mathbf{e} _j(t)\delta d_{k,\ell}(t)), \  \  M_{12,j\ell i}=\sqrt{\lambda _{\ell }}\sum _{m,n}\left \{ \frac{\overline{v} _{\ell m}a_{1,jn}\overline{e} _{n+m, i}}{1-\mathrm{i}(n+m)\omega _0} +2\sum _{m,n}\frac{v_{\ell m}a_{2,j-m,n}\overline{e} _{ni}}{1-\mathrm{i}n\omega _0}\right \} . \label{first_order_response_oscillation_12}
\end{eqnarray}
In the derivation of this equation, note that $v_{\ell }(t)$ and $\delta d_{k,\ell }(t)$ are real. We also have the following similar equation for corrections to $\delta V^{(1)}(t,s)$, $\delta V^{(1)}_k(t,s)=\sum _i\sqrt{\lambda _i}(v_{i}(t)\overline{\delta v} _{k,i}(s)+\delta v _{k,i}(t)\overline{v} _{i}(s))$: 
\begin{eqnarray}
\hspace{-1.5cm} \delta v_{k+1, i}(t)=\frac{1}{2\sigma _0^2\sqrt{\lambda _i}} \sum _{j,\ell}\widetilde{M} _{12,j\ell i}(1+\partial _t)^{-1}(\mathbf{e} _j(t)\delta d_{k,\ell}(t)), \  \  \widetilde{M} _{12,j\ell i}=\sqrt{\lambda _{\ell }}\sum _{m,n}\left \{ \frac{\overline{v} _{\ell m}a_{1,jn}\overline{e} _{n+m, i}}{1-\mathrm{i}(n+m)\omega _0} +2\sum _{m,n}\frac{v_{\ell m}\widetilde{a} _{2,j-m,n}\overline{e} _{ni}}{1-\mathrm{i}n\omega _0}\right \} . \label{first_order_response_oscillation_12_tilde}
\end{eqnarray}

The first-order response in $V(t,s)$ is then given by
\begin{eqnarray}
\delta V^{(1)}(t,s)=\sum _{k=0}^{\infty }\delta V_k^{(1)}(t,s)=\sum _{k,i}\sqrt{\lambda _i}(v_{i}(t)\overline{\delta v} _{k,i}(s)+\delta v _{k,i}(t)\overline{v} _{i}(s))=\sum _{i}\sqrt{\lambda _i}(v_{i}(t)\overline{\delta v} _{i}(s)+\delta v _{i}(t)\overline{v} _{i}(s)), 
\end{eqnarray}
if the series of $\delta V_k^{(1)}$ converges. Here, we define $\delta v_i(t)=\sum _k\delta v_{k,i}(t)$.

To check the convergence and magnitude of the above first-order response, we iterate the recursion relation from different initial values and check the 1-norm of the final result. More precisely, we numerically examine $\{ \delta v_i\} _i$ for initial input $\delta m(t)=e^{\mathrm{i} \omega t}$ with different values of $\omega $. From this calculation, we estimate the value of the bounding constant $\theta _1$, for $\sqrt{2}g_0\sum _{i} |\xi _{\mathrm{o}, i}|\| \widehat{\delta v} _i\| _1\leq \theta _1\| \widehat{\delta m}\| _1+\sum _i\theta _{1,i}^{\prime }\| \widehat{p} _{1,i}\| _1$. For use below, we also estimate the value of the bounding constant $\theta _2$, for $\sum _{i}\| \widehat{\delta v} _i\| _1\leq \theta _2\| \widehat{\delta m}\| _1+\sum _i\theta _{2,i}^{\prime }\| \widehat{p} _{1,i}\| _1$. Here, note that the 1-norm is evaluated as the sum of the discrete components over the multiples of the basic frequency $\omega _0$ and the continuous component over the other frequencies.

From the convergent solution, we obtain
\begin{eqnarray}
V(t,s)=\sum _i(\sqrt{\lambda _i} v_i(t)+\delta v_{i}(t))(\sqrt{\lambda _i}\overline{v} _i(s)+\overline{\delta v} _i(s))+O(|\mathbf{p}|^2).
\end{eqnarray} 
The first-order response in $V(t,s)$ results in 
\begin{eqnarray}
\eta (t)=(1+\partial _t)\left \{ \sum _i\xi _{\mathrm{o}, i}(\sqrt{\lambda _i}v_i(t)+\delta v_i(t))\right \} +\delta \eta ^{(2)}(t), \  \  \  (t\geq 0), \label{first_order_part_of_eta_osc}
\end{eqnarray}
where $\delta \eta ^{(2)}(t)$ is the response in $\eta (t)$ due to the higher-order responses in $D(t,s)$.

Next, we examine the second-order response in $D(t,s)$, which is denoted by $\delta D^{(2)}(t,s)$ and obeys the following equation: 
\begin{eqnarray}
\hspace{-1cm} (1+\partial _t)(1+\partial _s)\delta D^{(2)}(t,s)&=&a_1(t,s)\delta D^{(2)}(t,s)+a_2(t,s)\delta D^{(2)}(t,t)+a_2(s,t)\delta D^{(2)}(s,s)\nonumber \\
&&+b_1(t,s)\delta D^{(1)}(t,s)^2+b_2(t,s)\delta D^{(1)}(t,s)\delta D^{(1)}(t,t)+b_2(s,t)\delta D^{(1)}(t,s)\delta D^{(1)}(s,s)\nonumber \\
&&+b_3(t,s)\delta D^{(1)}(t,t)^2+b_3(s,t)\delta D^{(1)}(s,s)^2+b_4(t,s)\delta D^{(1)}(t,t)\delta D^{(1)}(s,s)\nonumber \\
&&+b_5(t,s)\delta D^{(1)}(t,s)\delta m(t)+b_5(s,t)\delta D^{(1)}(t,s)\delta m(s)+b_6(t,s)\delta D^{(1)}(t,t)\delta m(t)+b_6(s,t)\delta D^{(1)}(s,s)\delta m(s)\nonumber \\
&&+b_7(t,s)\delta D^{(1)}(t,t)\delta m(s)+b_7(s,t)\delta D^{(1)}(s,s)\delta m(t)\nonumber \\
&&+b_8(t,s)\delta m(t)^2+b_8(s,t)\delta m(s)^2+b_9(t,s)\delta m(t)\delta m(s) +\sum _ip_{2,i}(t)p_{2,i}(s),\label{2nd_order_linear_response_osc}
\end{eqnarray}
where coefficient functions $b_1(t,s)$---$b_9(t,s)$ are suitably defined and the random component of the input is denoted by $\{ p_{2,i}\} _i$. Similarly to Appendix \ref{perturbative_expansion_fixed_point}, we see that the second-order response consist of two components: the first, denoted by $\delta D^{(2)\prime }(t,s)$, involves second-order products of either $\delta m(t)$ or $\delta m(s)$ and contributes to $O(|\mathbf{p}|^2)$ response in $\eta $; the second, denoted by $\delta D^{(2)\prime \prime }(t,s)$, involves cross-terms such as $\delta m(t)\delta m(s)$ and newly generates $O(|\mathbf{p}|)$ fluctuations in $\eta$. Then, our aim is to show that the latter part of the second-order response in $D(t,s)$ can be represented as the sum of a series of outer products of functions of time with exponentially decreasing magnitudes, as we did in Appendix \ref{perturbative_expansion_fixed_point}. Note that the relevant part of the second-order response in $V(t,s)$ is the same as that in $D(t,s)/2\sigma _0^2$.

The approximate special solution $\delta D_0^{(2)}(t,s)$ for Eq.(\ref{2nd_order_linear_response_osc}) is obtained by ignoring the first three terms on the right-hand side. Let us assume that the component of this solution that contributes to $O(|\mathbf{p}|)$ response in $\eta (t)$ can be represented as the sum of outer products with a certain convergence property. Concretely, we assume 
\begin{eqnarray}
\delta D_0^{(2)}(t,s)=\sum _iq_{0,i}(t)q_{0,i}(s), \label{compute_p_0}
\end{eqnarray}
where $q_{0,i}(t)$ is a linear transformation of $\delta m(t)$ or input components. We assume that the magnitude of $q_{0,i}$ is bounded as $\sum _i\| \widehat{q_{0,i}}\| _1\leq c\| \widehat{\delta m} \| _1, c\sum _j\| \widehat{p} _{2,j}\| _1$ for some finite constant $c$. This assumption is reasonable because the non-homogeneous terms on the right-hand side of Eq.(\ref{2nd_order_linear_response_osc}) should represent a positive semi-definite correlation matrix for newly generated fluctuations, as we have observed in the case for fixed points.

We recursively make corrections to this solution by solving 
\begin{eqnarray}
(1+\partial _t)(1+\partial _s)\delta D _{\ell +1}^{(2)}(t,s)=a_{1}(t,s)\delta D _{\ell }^{(2)}(t,s)+a_2(t,s)\delta D _{\ell }^{(2)}(t,t)+a_2(s,t)\delta D_{\ell }^{(2)}(s,s). \label{recursion_second_order_oscillation}
\end{eqnarray}
Since the latter two terms on the right-hand side of this equation yield only $O(|\mathbf{p}|^2)$ contributions in $\eta (t)$, we ignore these terms. Then, for $\delta D_{\ell }^{(2)}(t,s)=\sum _{j}q_{\ell ,j}(t)q_{\ell ,j}(s)$, we have
\begin{eqnarray}
\delta D _{\ell +1}^{(2)}(t,s)&=&\sum _{i,j}q_{\ell +1,i,j}^{\prime }(t)q_{\ell +1,i,j}^{\prime }(s), \nonumber \\
q_{\ell +1,i,j}^{\prime }(t)&=&(1+\partial _t)^{-1}\sum _{j}\sqrt{\rho _{1,j}}u_{1,j}(\tau )q_{\ell ,i}(t), \label{recursion_of_p}
\end{eqnarray}
with eigen-decomposition, $a_1(t,s)=\sum _{j}\rho _{1,j}u_{1,j}(t)u_{1,j}(s)$. Thus, the iteration keeps the solution positive semi-definite.

If the solutions for this recursion equation converges, we obtain the relevant part of the second-order response as $\delta D ^{(2)}(t,s)=\sum _{\ell }\delta D_{\ell }^{(2)}(t,s)$. To evaluate the magnitude of the solution, we iteratively solve the recursion equation and see that, after a sufficient number of iterations, the solution satisfies
\begin{eqnarray}
\delta D_{\ell +1}^{(2)}(t,s)\approx \theta _3\delta D_{\ell }^{(2)}(t,s).
\end{eqnarray}
If we have $\theta _3<1$, the series converges.

Using the representation in terms of outer products, $\delta D_{\ell }^{(2)}(t,s)=\sum _{i}q_{\ell ,i}(t)q_{\ell ,i}(s)$, we obtain the response in $\eta $. Combined with Eq.(\ref{first_order_part_of_eta_osc}), the overall linear response in $\eta (t)$ reads
\begin{eqnarray}
\eta (t)=(1+\partial _t)\left \{ \sum _i\xi _{\mathrm{o}, i}(\sqrt{\lambda _i}v_i(t)+\delta v_i(t))+\frac{1}{\sqrt{2} \sigma _0}\sum _{\ell ,j}\xi _{\ell ,j}q_{\ell ,j}(t)\right \} ,
\end{eqnarray}
where $\{ \xi _{\ell ,j}\} _{\ell ,j}$ are $i.i.d.$ unit Gaussian random variables. Substituting the above into Eq.(\ref{initial_delta_m_equation_osc}), we obtain a self-consistent equation,
\begin{eqnarray}
\delta m(t)=\sqrt{2} g_0\sum _i \xi _{\mathrm{o}, i}\delta v_i(t)+\frac{g_0}{\sigma _0}\sum _{\ell ,j}\xi _{\ell ,j} q_{\ell ,j}(t)+g_0(1+\partial _t)^{-1}\delta \widetilde{\phi }(t)+(1+\partial _t)^{-1}p_0(t). \label{random_self_consistent_equation_osc}
\end{eqnarray}
The norms of $\delta v_i(t)$ and $q_{\ell ,j}(t)$ were evaluated by comparison with the norm of $\delta m(t)$, as described above. For $\delta \widetilde{\phi }(t)$, by putting $\langle \phi ^{\prime \prime }(h_j^{(E)}(t))\rangle _{j\notin \mathcal{S}}-\langle \phi ^{\prime \prime }(h_j^{(E)}(t))\rangle _{j\in \mathcal{S}}=\sum _{\ell }\phi ^{(2)}_{\ell }\mathbf{e} _{\ell }(t)$ and $\langle \phi ^{\prime}(h_j^{(E)}(t))\rangle _{j\notin \mathcal{S}}-\langle \phi ^{\prime}(h_j^{(E)}(t))\rangle _{j\in \mathcal{S}}=\sum _{\ell }\phi ^{(1)}_{\ell }\mathbf{e} _{\ell }(t)$, we have 
\begin{eqnarray}
\delta \widetilde{\phi }(t)=\sum _{\ell } \phi ^{(2)}_{\ell }\mathbf{e} _{\ell }(t)\delta D^{(1)}(t,t)+\sum _{\ell }\phi ^{(1)}_{\ell }(t)\mathbf{e} _{\ell }(t)\delta m(t)+O(|\mathbf{p}|^2), \  \  \  \| \widehat{(1+\partial _t)^{-1}\delta \widetilde{\phi }}\| _1\leq \theta _4\| \widehat{\delta m} \| _1, \label{compute_d_phi}
\end{eqnarray}
with
\begin{eqnarray}
\theta _4=\theta _2\sup _{i,\omega ^{\prime }}\left |\sum _{\ell ,n}\frac{\sqrt{\lambda _i} \phi ^{(2)}_{\ell }v_{in}}{1+\mathrm{i}((\ell +n)\omega +\omega ^{\prime })}\right |+\sup _{\omega ^{\prime }}\left |\sum _{\ell }\frac{\phi ^{(1)}_{\ell }}{1+\mathrm{i} (\omega ^{\prime }+\ell \omega )}\right | .
\end{eqnarray}
From this equation, we obtain the overall response in $m(t)$ by recursive computation. We first calculate $\delta m^{(0)}(t)$ from Eq.(\ref{random_self_consistent_equation_osc}), on the right-hand side of which $\delta v_i$ and $q_{\ell ,j}$ are computed just from the input $\mathbf{p}$ by setting $\delta m=0$. Then, by calculating functions $\delta v_i^{(k)}$, $q_{\ell ,j}^{(k)}$ and $\delta \widetilde{\phi }^{(k)}(t)$ from $\delta m^{(k)}(t)$, ($k\geq 0$) according to Eqs.(\ref{compute_delta_v_0}), (\ref{first_order_response_oscillation_12}), (\ref{compute_p_0}), (\ref{recursion_of_p}), and (\ref{compute_d_phi}), we make corrections recursively: 
\begin{eqnarray}
\delta m^{(k+1)}(t)=\sqrt{2}g_0\sum _i\xi _{\mathrm{o}, i}\delta v_i^{(k)}(t)+\frac{g_0}{\sigma _0}\sum _{\ell ,j}\xi _{\ell ,j} q_{\ell ,j}^{(k)}(t) +g_0(1+\partial _t)^{-1}\delta \widetilde{\phi }^{(k)}(t).
\end{eqnarray}
We finally obtain $\delta m(t)=\sum _{k=0}^{\infty }\delta m^{(k)}(t)$, which satisfies Eq.(\ref{random_self_consistent_equation_osc}), if $\| \delta m^{(k+1)}\| _1\leq \theta \| \delta m^{(k)}\| _1$ holds for some constant $\theta <1$. This convergence condition is satisfied with a non-zero probability if  
\begin{eqnarray}
\theta _1+g_0\theta _4<1, \  \  \theta _3<1.
\end{eqnarray}
 Similarly to Appendix \ref{perturbative_expansion_fixed_point}, this convergence implies the linear stability of the regular oscillation under examination for a non-zero fraction of configurations of the random connectivity. A single configuration of the random connectivity corresponds to a single set of values for $\{ \xi _{\ell j}\} _{\ell,j}$.

We numerically calculate the bounding constants in the above condition for convergence, using the orbits of the mean activity observed in Figs.\ref{zerosum_autonomous_oscillation}(c), \ref{figure_single_EI_pair_nonzerosum}(g) and \ref{figure_external_input}(c) for $m_\mathrm{o}(t)$. To reduce the computational cost, we ignore eigen-modes $\{ v_i(t)\} _i$ except for those for the 48 largest eigenvalues. The calculated values are summarized in table \ref{bounding_constant_oscillation}, which suggest that the regular oscillations observed in Fig.\ref{figure_single_EI_pair_nonzerosum}(g) and Fig.\ref{figure_external_input}(c) are stable, while those observed in Fig.\ref{zerosum_autonomous_oscillation}(c) are not.

\begin{table}
\begin{tabular}{|lr|c|c|c|c|c|c|}
\hline
model and orbit& & $\theta _1$ & $\theta _2$ & $\theta _3$ & $\theta _4$ & $\theta _1+g_0\theta _4$\\
\hline \hline 
untuned w/o input  & Fig.\ref{figure_single_EI_pair_nonzerosum}(g) & 0.159 & -- & 0.487 & 0 & 0.159\\
\hline 
finely-tuned w/o input & Fig.\ref{zerosum_autonomous_oscillation}(c) & $3.81\times 10^{-3}$ & -- & 1.02 & 0 & $3.81\times 10^{-3}$ \\
\hline
finely-tuned with input & Fig.\ref{figure_external_input}(c) & $4.36\times 10^{-3}$ & $5.83\times 10^{-4}$ & 0.896 & $3.32\times 10^{-5}$ & $4.39\times 10^{-3}$ \\
\hline 
\end{tabular}
\caption{The values of the bounding constants that determine stability of regular oscillatory solutions are numerically estimated and summarized in a table. } \label{bounding_constant_oscillation}
\end{table}

\section{Price theorem and differentiation of correlation matrix $C$ and $\widetilde{C}$} \label{differentiation_of_C}
We have used the differentiation of correlation matrix $\widetilde{C}$ with respect to $D$ and $m$ in Appendix \ref{perturbative_expansion_chaos}--\ref{perturbative_expansion_oscillation}. This is given by the following formula: 
\begin{eqnarray}
\begin{array}{llll}
\frac{\partial \widetilde{C}(t_1,t_2)}{\partial D_{11}}=\frac{1}{2} \langle \phi ^{\prime \prime }(z_1)\phi (z_2)\rangle , & \frac{\partial \widetilde{C}(t_1,t_2)}{\partial D_{12}}=\langle \phi ^{\prime }(z_1)\phi ^{\prime }(z_2)\rangle ,& \frac{\partial \widetilde{C}(t_1,t_2)}{\partial D_{22}}=\frac{1}{2} \langle \phi (z_1)\phi ^{\prime \prime }(z_2)\rangle ,& \frac{\partial \widetilde{C}(t_1,t_2)}{\partial m_{1}}=\langle \phi ^{\prime }(z_1)\phi (z_2)\rangle ,\\
\frac{\partial \widetilde{C}(t_1,t_2)}{\partial m_{2}}=\langle \phi (z_1)\phi ^{\prime }(z_2)\rangle ,& \frac{\partial \overline{\phi }(t_1)}{\partial D_{11}}=\frac{1}{2} \langle \phi ^{\prime \prime }(z_1)\rangle ,& \frac{\partial \overline{\phi }(t_1)}{\partial D_{12}}=\frac{\partial \overline{\phi }(t_1)}{\partial D_{22}}=0, & \\
\frac{\partial \overline{\phi }(t_1)}{\partial m_{1}}=\langle \phi ^{\prime}(z_1)\rangle ,& \frac{\partial \overline{\phi }(t_1)}{\partial m_{2}}=0,  & \frac{\partial \overline{\phi }(t_2)}{\partial m_{2}}=\langle \phi ^{\prime}(z_2)\rangle , & \frac{\partial \overline{\phi }(t_2)}{\partial m_{1}}=0.  \label{derivatives_of_C}
\end{array}
\end{eqnarray}
The variables $\{ z_\alpha \} _{\alpha =1,2}$ are Gaussian random variables that have the same first and second-order moments as $\{ h_{j}^{(\ell )}(t_{\alpha })\} _{\alpha =1,2}$ ($\ell =E,I$), similarly to Eq.(\ref{recovering_C_again}). Also note that the above partial derivatives with respect to $D_{\alpha \beta }$ are not total differentials with respect to $D(t_{\alpha }, t_{\beta })$ but are derivatives with respect to the corresponding variables that appear in Eq.(\ref{recovering_C_again}). The twice differentiation is performed similarly and easily inferred from the above results. The differentiation of $C$ is easily obtained from the results for the differentiation of $\widetilde{C}$ and $\overline{\phi }$. The above results can be derived by the differentiation of both sides of Eq.(\ref{recovering_C_again}) followed by integration by parts, but are obtained more easily by performing the differentiation in the frequency domain in a manner similar to the derivation of Price theorem \cite{probability1991stochastic} (also see \cite{Schuecker:2016uj} for a simpler example). We show the outline of how it is performed below.

Let us rewrite the representation of $\widetilde{C}$ in Eq.(\ref{recovering_C_again}) as 
\begin{eqnarray}
\widetilde{C}(t_1,t_2)&=&\int _{\mathbf{R} ^3}  \phi (f_1(\{ D_{\alpha \beta }\} )y_1+f_2(\{ D_{\alpha \beta}\} )w+m_1)\phi (f_3(\{ D_{\alpha \beta}\} )y_2+f_4(\{ D_{\alpha \beta}\} )w+m_2) \mathrm{d} \mathcal{N}(y_1) \mathrm{d} \mathcal{N}(y_2)  \mathrm{d} \mathcal{N}(w) \nonumber \\
&=& \frac{1}{2\pi }\int _{\mathbf{R} ^5} \widehat{\phi } (\omega _1)\widehat{\phi } (\omega _2)e^{\mathrm{i} \omega _1((f_1(\{ D_{\alpha \beta }\} )y_1+f_2(\{ D_{\alpha \beta}\} )w+m_1)+\mathrm{i} \omega _2 (f_3(\{ D_{\alpha \beta}\} )y_2+f_4(\{ D_{\alpha \beta}\} )w+m_2)} \mathrm{d} \mathcal{N}(y_1) \mathrm{d} \mathcal{N}(y_2)  \mathrm{d} \mathcal{N}(w) \mathrm{d} \omega _1\mathrm{d} \omega _2\nonumber \\
&=& \frac{1}{2\pi }\int _{\mathbf{R} ^2}\widehat{\phi } (\omega _1)\widehat{\phi } (\omega _2)e^{-\frac{1}{2} \omega _1^2f_1(\{ D_{\alpha \beta }\} )^2-\frac{1}{2}\omega _2^2f_3(\{ D_{\alpha \beta}\} )^2-\frac{1}{2} (\omega _1f_2(\{ D_{\alpha \beta}\} )+\omega _2f_4(\{ D_{\alpha \beta}\} ))^2+\mathrm{i}\omega _1m_1+\mathrm{i} \omega _2m_2}\mathrm{d} \omega _1\mathrm{d} \omega _2 ,
\end{eqnarray}
where functions $f_1$--$f_4$ of $\{ D_{\alpha \beta }\} _{\alpha ,\beta =1,2}$ are suitably defined. The Fourier transform is denoted by $\widehat{\cdot }$. From the second line to the third line, we have used the characteristic function of a unit Gaussian measure, $e^{-\frac{1}{2}\omega ^2}=\int e^{\mathrm{i}\omega X} \mathrm{d} \mathcal{N} (X)$.
Differentiating the last line with respect to, {\it e.g.} $D_{12}$, and rearranging terms, we have,
\begin{eqnarray}
\frac{\partial \widetilde{C} (t_1,t_2)}{\partial D_{12}} &=&-\frac{1}{2\pi }\int _{\mathbf{R} ^2} \omega _1\omega _2\widehat{\phi } (\omega _1)\widehat{\phi } (\omega _2)e^{-\frac{1}{2} \omega _1^2f_1(\{ D_{\alpha \beta }\} )^2-\frac{1}{2}\omega _2^2f_3(\{ D_{\alpha \beta}\} )^2-\frac{1}{2} (\omega _1f_2(\{ D_{\alpha \beta}\} )+\omega _2f_4(\{ D_{\alpha \beta}\} ))^2+\mathrm{i}\omega _1m_1+\mathrm{i} \omega _2m_2}\mathrm{d} \omega _1\mathrm{d} \omega _2\nonumber \\
&=&\frac{1}{2\pi }\int _{\mathbf{R} ^2} \widehat{\phi ^{\prime }}(\omega _1)\widehat{\phi ^{\prime }} (\omega _2)e^{-\frac{1}{2} \omega _1^2f_1(\{ D_{\alpha \beta }\} )^2-\frac{1}{2}\omega _2^2f_3(\{ D_{\alpha \beta}\} )^2-\frac{1}{2} (\omega _1f_2(\{ D_{\alpha \beta}\} )+\omega _2f_4(\{ D_{\alpha \beta}\} ))^2+\mathrm{i}\omega _1m_1+\mathrm{i} \omega _2m_2}\mathrm{d} \omega _1\mathrm{d} \omega _2 \nonumber \\
&=&\int _{\mathbf{R} ^3}  \phi ^{\prime }(f_1(\{ D_{\alpha \beta }\} )y_1+f_2(\{ D_{\alpha \beta}\} )w+m_1)\phi ^{\prime }(f_3(\{ D_{\alpha \beta}\} )y_2+f_4(\{ D_{\alpha \beta}\} )w+m_2) \mathrm{d} \mathcal{N}(y_1) \mathrm{d} \mathcal{N}(y_2)  \mathrm{d} \mathcal{N}(w)\nonumber \\
&=&\langle \phi ^{\prime }(z_1)\phi ^{\prime }(z_2)\rangle 
\end{eqnarray} 
From the first line to the second line, we have used the formula for the Fourier transform of derivatives, $\widehat{\phi ^{\prime}}(\omega )=\mathrm{i}\omega \widehat{\phi } (\omega )$. Other differentiations in Eq.(\ref{derivatives_of_C}) are calculated similarly. 
\end{widetext}
\bibliography{ref.bib}

\end{document}